\documentclass[prb,superscriptaddress,floatfix,letterpaper,twocolumn,aps,showpacs]{revtex4-1}
\usepackage[utf8]{inputenc}
\usepackage[usenames,dvipsnames,svgnames,table]{xcolor}
\usepackage{natbib,amsmath,amsbsy,amssymb,scrextend,graphicx,color,placeins,array,siunitx}
\setcitestyle{square, comma, numbers}
\usepackage{hyperref}

\begin{document}
\title{Strong pinning theory of thermal vortex creep in type II superconductors}
\author{M.\ Buchacek}
\affiliation{Institute for Theoretical Physics, ETH Zurich, 8093 Zurich
Switzerland}
\author{R.\ Willa}
\affiliation{Materials Science Division, Argonne National Laboratory, 
Lemont, IL 60439, USA}
\author{V.B.\ Geshkenbein}
\affiliation{Institute for Theoretical Physics, ETH Zurich, 8093 Zurich
Switzerland}
\author{G.\ Blatter}
\affiliation{Institute for Theoretical Physics, ETH Zurich, 8093 Zurich
Switzerland}
\date{\today}
\begin{abstract}
We study thermal effects on pinning and creep in type-II superconductors where
vortices interact with a low density $n_p$ of strong point-like defects with
pinning energy $e_p$ and extension $\xi$, the vortex core size. Defects are
classified as strong if the interaction between a single pin and an individual
vortex leads to the appearance of bistable solutions describing pinned and
free vortex configurations. Extending the strong pinning theory to account for
thermal fluctuations, we provide a quantitative analysis of vortex depinning
and creep.  We determine the thermally activated transitions between bistable
states using Kramer's rate theory and find the non-equilibrium steady-state
occupation of vortex states.  The latter depends on the temperature $T$ and
vortex velocity $v$ and determines the current--voltage (or force--velocity)
characteristic of the superconductor at finite temperatures.  We find that the
$T=0$ linear excess-current characteristic $v \propto (j-j_c) \,
\Theta(j-j_c)$ with its sharp transition at the critical current density
$j_c$, keeps its overall shape but is modified in three ways due to thermal
creep: a downward renormalization of $j_c$ to the thermal depinning current
density $j_\mathrm{dp}(T) < j_c$, a~smooth rounding of the characteristic around
$j_\mathrm{dp}(T)$, and the appearance of thermally assisted flux flow (TAFF) ${v
\propto j \exp(-U_0/k_{\rm \scriptscriptstyle B} T)}$ at small drive $j \ll
j_c$, with the activation barrier $U_0$ defined through the energy landscape
at the intersection of free and pinned branches.  This characteristic
emphasizes the persistence of pinning of creep at current densities beyond
critical.

\end{abstract}
%
\maketitle


\section{Introduction}\label{sec:intro} 

The properties of numerous materials are determined by the presence of
topological excitations in the ordered states of matter; examples include
vortices in type-II superconductors \cite{Blatter1994,Brandt1995}, domain
walls in ferroic materials \cite{Kleemann2007,Gorchon2014}, or dislocations in
metals \cite{Kassner2015,Zhou2015}. The motion of such objects within the host
material has a significant effect on its response, e.g., the onset of finite
resistivity in superconductors or the loss of coercivity in magnets.
Immobilizing these excitations, usually by pinning onto material defects, is
thus of great technological relevance. The dynamics of topological objects
then exhibits a transition between a static or pinned phase and a sliding or
unpinned phase upon exceeding a threshold or critical force $F_c$.
Understanding the pinned-to-sliding transition, optimizing the pinning
threshold, and stabilizing it against thermal fluctuations present vital tasks
at the cross-roads of disordered statistical physics and non-equilibrium
phenomena.  The complete description of the material's response is captured by
the force--velocity ($F-v$) characteristic of topological excitations; here,
we extend the strong pinning theory to include effects of finite temperatures
and calculate the response characteristic of vortex matter in type-II
superconductors subject to a low density of point-like strong defects.

\begin{figure}[b]
\centering 
\includegraphics[scale=1]{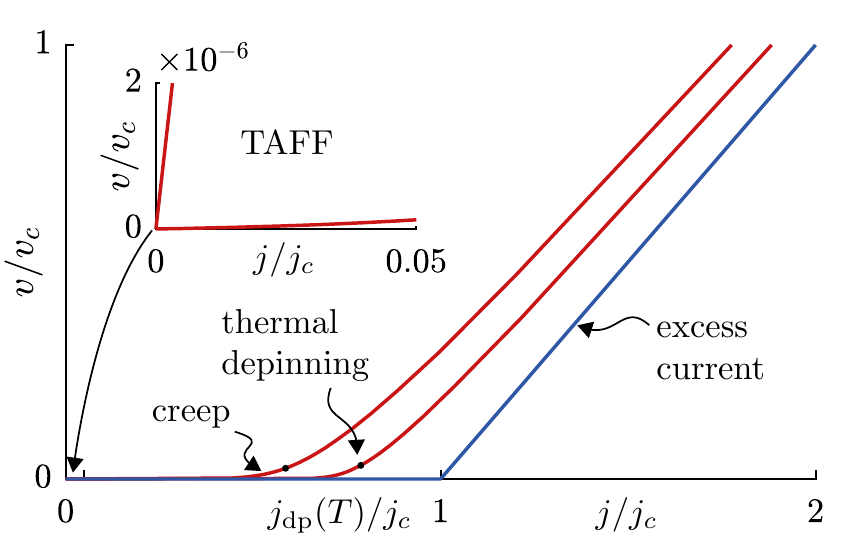}
\caption{Current--velocity characteristic of a type II superconductor with
strongly pinned vortices and including effects of thermal fluctuation.  The
$T=0$ linear excess-current characteristic (in blue) with its critical current
density $j_c$ is modified due to thermal creep (red curves with ${T/e_p =
(0.5,1)\times 10^{-2}}$): i) finite temperatures $T > 0$ shift the linear
branch to the left as described by a thermal reduction of $j_c$ to the
depinning current density $j_\mathrm{dp}(T) < j_c$, ii) the characteristic is
rounded at the onset of vortex motion near $j_\mathrm{dp}(T)$, and iii) the
characteristic turns ohmic at small current densities $j \to 0$, see inset, an
effect commonly known as thermally assisted flux flow or TAFF.}
\label{Fig:IV_schematic}
\end{figure}

Vortex pinning is described by either of two frameworks: within weak pinning
theory \cite{Larkin1979,Blatter1994}, the pinning force due to an individual
defect vanishes and it is the collective action of many defects which
generates the average pinning-force density. On the contrary, for strong
pinning \cite{Labusch1969,Larkin1979}, individual defects induce substantial
deformations that lead to bistable behavior and generate an average non-zero
pinning force on the vortex lattice that scales linearly in the (small)
density $n_p$ of defects.  Weak collective pinning has been fully developed in
the wake of the high-$T_c$ discovery \cite{Blatter1994,
NattermannScheidl2000}, although results have remained qualitative. On the
other hand, the theory of strong pinning provides quantitative results, but
its development is less advanced.  The critical currents
\cite{IvlevOvchinnikov91}, current--voltage characteristics
\cite{Thomann2012,Thomann2017}, $ac$-response \cite{Willa2015,Willa2016,Willa2015B}, and
the overall pinning diagram \cite{Blatter2004,Koopmann2004} have been analyzed
and augmented by numerical simulations \cite{KoshelevKolton2011,Koshelev2016,
Willa2018}. However, so far no systematic theory including thermal
fluctuations has been developed, although important understanding of the creep
mechanism can be derived from the work on charge-density wave pinning by
Brazovskii, Larkin, and Nattermann \cite{BrazovskiiLarkin1999,
BrazovskiiNattermann2004}, see also early work by Fisher
\cite{Fisher1983,Fisher1985}.

In this paper, we adopt the strong pinning paradigm and present quantitative
results on the pinning and creep of vortices in type II superconductors in the
presence of thermal fluctuations. Such vortices result from a magnetic field
$\mathbf{H}$ penetrating the superconductor \cite{Shubnikov1937,Abrikosov1957,
Tinkham1996}, each vortex carrying a quantum of flux $\phi_0 = hc/2e$ and
together forming a lattice of density $a_0^{-2} = B/\phi_0$ inducing the
average magnetic field $\mathbf{B}$.  The resulting vortex matter is pushed by
the current density $\mathbf{j}$ via the Lorentz-force density
$\mathbf{F}_{\rm\scriptscriptstyle L} = \mathbf{j}\times\mathbf{B}/c$.  The
resulting force--velocity characteristic follows from the dissipative
force-balance equation
\begin{align}\label{eq:force_balance}
   \eta v = F_{\rm \scriptscriptstyle L} (j)-F_\mathrm{pin}(v,T),
\end{align}
where $\eta \approx B H_{c2}/\rho_n c^2$ denotes the Bardeen-Stephen viscosity
(per unit volume, $H_{c2}$ is the upper-critical field and $\rho_n$ the
normal-state resistivity) and $F_\mathrm{pin}(v,T)$ is the average pinning force
density, the quantity of central importance in this paper.

The weak pinning approach provides estimates for the dynamical properties of
vortices: the pinning force density $F_\mathrm{pin}$ has been calculated by
Larkin and Ovchinnikov \cite{Larkin1974} and by Schmid and Hauger
\cite{Schmid1973} using a high-velocity perturbative expansion, while the
depinning dynamics around $j_c$ has been studied via renormalization group
techniques by Narayan and Fisher \cite{Narayan1992} and by Chauve {\it et al.}
\cite{Chauve2000}.  On the contrary, the strong pinning scenario produces
quantitative results, provided that the density $n_p$ of defects is small such
that they act independently, i.e., the pinning-force density $F_\mathrm{pin}
\propto n_p$ is linear in the density of defects (but scales non-trivially in
the force $f_p \sim e_p/\xi$ of individual defects).  Pinning is strong if the
largest (negative) curvature of the pinning potential $\partial_R^2 e_p(R)$
wins over the effective vortex stiffness $\bar{C}$; this is expressed in the
Labusch criterion \cite{Labusch1969} $\kappa = \max_R |\partial_R^2
e_p(R)|/\bar{C} > 1$ characterizing strong pins.  The vortex deformation due
to an individual pin then exhibits bistable solutions defining pinned and free
vortex branches (Fig.\ \ref{Fig:energy_landscape}).  Their asymmetric
occupation is at the origin of a finite average pinning-force density
$F_\mathrm{pin}$ exerted on the vortex lattice by the randomly positioned
defects. The determination of its maximal value provides the critical force
density \cite{Labusch1969,Larkin1979,Blatter2004} $F_c$. The calculation of
its dynamical variant $F_\mathrm{pin}(v)$ in the absence of thermal
fluctuations \cite{Thomann2012,Thomann2017} gives access to the full $T=0$
characteristic; this turns out to be of an excess-current form, i.e., the
linear flux-flow characteristic of the defect-free superconductor is shifted
by a finite critical current density $j_c$, see Fig.\ \ref{Fig:IV_schematic}.
Note that, although pinning by an individual defect is strong, the small
defect density $n_p$ results in small or moderate pinning forces; hence,
strong pinning theory is not necessarily the theory producing highest critical
current densities.

In order to account for thermal fluctuations in the average pinning-force
density $F_\mathrm{pin}(v,T)$, we follow Brazovskii and Larkin
\cite{BrazovskiiLarkin1999} and use Kramer's rate theory \cite{Kramers1940} to
describe transitions between the pinned and free vortex branches and determine
the branch occupation at finite temperatures and velocities. At large
velocities but below the (thermal) velocity $v_\mathrm{th} \sim \kappa (T/e_p)
v_p$, finite temperature assists the motion of vortices, diminishes the
asymmetry in the vortex branch occupation, and thus reduces the pinning-force
density to lie below the critical value $F_c$; here, the depinning velocity
$v_p = f_p/\eta a_0^3$ characterizes the dissipative motion in the defect
potential. The critical current density $j_c$ is reduced to a depinning
current density $j_\mathrm{dp}(T)$ separating flat and steep regions of the
characteristic, see Fig.\ \ref{Fig:IV_schematic}; to leading order, the
relative shift depends on temperature as $1-j_\mathrm{dp}(T)/j_c \propto
(T/e_p)^{2/3}$ and is logarithmically dependent on the density $n_p$ of
defects, see Sec.\ \ref{sec:high_velocities}. Beyond depinning, we find a weak
dependence of $F_\mathrm{pin}(v,T)$ on the velocity $v$ and thus recover a
close to linear excess-current characteristic, shifted downward in current
with respect to the $T = 0$ result. Hence, contrary to usual expectations, a
large pinning-force density as well as thermal creep remain present far beyond
the critical current density $j_c$, see Ref.\ [\onlinecite{Buchacek2018}].  Finally,
writing the vortex velocity in the form $v = v_\mathrm{th}e^{-U(j)/T}$,
reminiscent of its thermal origin with $U(j)$ the activation barrier, we find
a decreasing activation barrier $U(j < j_c) \propto (1-j/j_c)^{3/2}$ when
approaching the depinning region from below.  However, the barrier persists
well beyond $j_c$, where it is characterized by a slow logarithmic variation
with the current density $j$, $U(j \gtrsim j_c) \approx U(j_c) -
T\ln[1+(j-j_c) / (j_c-j_\mathrm{dp}(T))]$, consistent with a linear
force--velocity characteristic.

Weak drives $j$ are characterized by a nearly symmetric occupation of
branches, more precisely, an occupation that is shifted linearly in $j$ with
respect to the thermal equilibrium occupation. This results in an ohmic
response with exponentially small velocities $v$, usually known as TAFF,
thermally assisted flux-flow \cite{Kes1989}, a specific form of vortex creep
at low drive. As implied by its name, the resistivity is thermally assisted,
i.e., $\rho_{\rm \scriptscriptstyle TAFF} \propto \rho_\mathrm{ff} \exp(-U_0
/k_{\rm \scriptscriptstyle B}T)$, $\rho_\mathrm{ff}$ the flux-flow
resistivity, with the finite activation barrier $U_0$ derived directly from
the bistable solutions, see Sec.\ \ref{sec:low_velocities} below. As a result,
within the framework of strong pinning, the superconductor loses its defining
property of dissipation-free current transport. This is quite different as
compared with the result of weak collective pinning theory, where barriers
diverge $U(j\to 0) \propto j^{-\mu}$, thereby establishing a truly
superconducting (glass) state at low drives $j$.

Below, we start with a brief review of the strong pinning formalism and show
how the interaction of independent defects with the vortex lattice is reduced
to a single-pin--single-defect problem involving the effective vortex
elasticity $\bar{C}$, see Sec.\ \ref{subsec:strong-pinning}. In Sec.\
\ref{sec:thermal_creep}, we extend the analysis to include thermal
fluctuations; we discuss creep effects at large drives and velocities in Sec.\
\ref{sec:high_velocities} and find the depinning current density
$j_{\mathrm{dp}} (T)$ and the relevant creep barriers $U(j)$ in its vicinity.
In section \ref{sec:low_velocities}, we focus on small drives and low
velocities and find the ohmic TAFF characteristics with a quantitative
prediction of the activation barrier. Finally, in the appendices we provide details of the energy landscape in the marginally- and very strong pinning regime and other technical details of calculations omitted in the main text. In our analytic work, we focus on the
relevant limiting cases, small ($j \to 0$) and large ($j \sim j_c$) drives as
well as the limits of marginally strong pinning with $\kappa \gtrsim 1$ and
very strong pinning, $\kappa \gg 1$. The new insights on the persistence of
pinning and creep beyond the critical drive has been published in a short
format in Ref.\ [\onlinecite{Buchacek2018}].


\section{Formalism}\label{sec:formalism}

\subsection{Strong Pinning}\label{subsec:strong-pinning}
In the absence of defects (pins) and thermal fluctuations, our vortex array,
aligned along the $z$-axis, is arranged in a two-dimensional (2D) hexagonal
lattice with equilibrium positions $\mathbf{R}_\mu = (x_\mu,y_\mu)$. The
presence of strong defects results in deformations of the lattice described by
the planar displacement field $\mathbf{u}_{\mu}(z)$. We consider a
representative single defect placed in the origin and characterised by a
radially symmetric bare pinning potential $e_p(R)\, \delta(z)$, with $e_p(R)$
decaying on the length $\sigma$ and $e_p$ the maximal pinning energy. For a
point-like defect, the pinning potential extends over a distance $R \sim \xi$;
for a defect of size $\sigma \sim \xi$, the energy $e_p$ is determined by the
condensation energy, $e_p \sim H_c^2 \xi^3$, see Ref.\ [\onlinecite{Willa2016}] for
more details.  Furthermore, we consider a situation where the repulsion
between vortices prevents two of them from occupying the same defect \cite{Willa2018PRB}, limiting
the interaction between vortices and the defect to the single reference vortex
$\mu_0 \equiv 0$ closest to the origin. Such a situation is realized at small
and intermediate fields with $a_0\gg \xi$ and a not too large pinning energy
$e_p$.  The free energy of this setup then takes the form
\begin{align}\label{eq:free_energy}
\begin{split}
   F[{\bf u}] &= \int d z\, e_p[\mathbf{R}_{0}+\mathbf{u}_{0}(z)]\delta(z)\\
   &+\frac{1}{2}\int\frac{d^3 \mathbf{k}}{(2\pi)^3}u_\alpha(\mathbf{k})
   \phi_{\alpha\beta}(\mathbf{k})u_{\beta}(-\mathbf{k}),
\end{split}
\end{align}
where the first term describes the vortex--defect interaction and the second
contributes the elastic energy, expressed through the (symmetric and real)
reciprocal-space elastic matrix \cite{Labusch1969} $\phi_{\alpha\beta}
(\mathbf{k})$.  Here, $\mathbf{u}_{0}(z=0) \equiv \mathbf{u}_{0}$ denotes the
tip position of our reference vortex pinned to the defect at the origin. The
displacement fields in real and reciprocal space are related through [we
decompose $\mathbf{r}_\mu = (\mathbf{R}_\mu,z)$ and $\mathbf{k} =
(\mathbf{K},k_z)$]
\begin{align}{\label{eq:u_FT}}
   \mathbf{u}(\mathbf{k}) &= a_0^2\int d z\,\sum_{\mu}e^{-i\mathbf{k}
   \cdot\mathbf{x}_\mu}\mathbf{u}_\mu(z),\\
   \mathbf{u}_\mu(z) &= \int \frac{d^3\mathbf{k}}{(2\pi)^3}\,
   e^{i\mathbf{k}\cdot\mathbf{x}_\mu}\mathbf{u}(\mathbf{k}).
\end{align}
The integration over $\mathbf{K}$ is restricted to the 2D Brillouin zone of
the vortex lattice, $K_{\rm \scriptscriptstyle BZ} = 4\pi/a_0^2$ in the
circularized approximation, while $k_z$ is subject to the cutoff
$|k_z|<\pi/\xi$.

The variation of Eq.\ \eqref{eq:free_energy} with respect to the displacement
field $\delta \mathbf{u}$ provides us with the response
\begin{align}\label{energy_variation1}
\begin{split}
   \delta F &= -f_{p,\alpha}[\mathbf{R}_{0}+\mathbf{u}_{0}]\,
   \delta u_{0,\alpha}\\
   &+\int\frac{d^3 \mathbf{k}}{(2\pi)^3}\delta u_\alpha(\mathbf{k})
   \phi_{\alpha\beta}(\mathbf{k})u_{\beta}(-\mathbf{k}),
\end{split}
\end{align}
where $f_{p,\alpha} = -\partial e_p/\partial x_\alpha$ denotes the pinning
force. Expressing the real-space perturbation $\delta{u}_{0,\alpha}$ in the
first term through the Fourier modes $\delta u_\alpha(\mathbf{k})$, this reads
\begin{align}\nonumber
   \delta F =\int\frac{d^3 \mathbf{k}}{(2\pi)^3}\,\delta u_\alpha(\mathbf{k})
   &\left\lbrace -f_{p,\alpha}[\mathbf{R}_{0}+
   \mathbf{u}_{0}]e^{i\mathbf{K}\cdot\mathbf{R}_{0}}\right.\\ 
   \label{eq:energy_variation2}
   &\,\,\,+\left.\phi_{\alpha\beta}(\mathbf{k})u_{\beta}(-\mathbf{k})\right\rbrace.
\end{align}
For a lattice moving with a steady drift velocity $\mathbf{v}$, the asymptotic
vortex positions are given by $\mathbf{R}_\mu(t) =
\mathbf{R}_{\mu}(0)+\mathbf{v}t$.  The full dynamical response of the vortex
lattice including time-dependent displacements has to be calculated from the
dissipative dynamical equation of motion $\eta \dot{\mathbf{u}} = -\delta
F/\delta \mathbf{u}$ and is addressed in Refs.\
\cite{Thomann2012,Thomann2017}. In the present work, we neglect dynamical
effects and assume that the drift velocity is sufficiently small such that the
vortex displacement field locally minimizes the energy \eqref{eq:free_energy}
at any moment of time, $\delta F/\delta \mathbf{u} = 0$.  The displacement
field ${\bf u}$ then depends on time only through the boundary condition, the
asymptotic position $\mathbf{R}_{0}$ of the reference vortex $\mu_0 = 0$, and
relates to the pinning force ${\bf f_p}$ via
\begin{align}\label{eq:equilibrium_fourier}
   u_\alpha(\mathbf{k}) = G_{\alpha\beta}(\mathbf{k})f_{p,\beta}[\mathbf{R}_{0}
   +\mathbf{u}_{0}]\, e^{-i\mathbf{K}\cdot\mathbf{R}_{0}}
\end{align}
with the Green's function $G_{\alpha\beta}(\mathbf{k}) = [\phi^{-1}
(\mathbf{k})]_{\alpha \beta}$. Using Eq.\ \eqref{eq:equilibrium_fourier}, we
can first solve for $u_{0,\alpha}$, the $\alpha$-component of
$\mathbf{u}_{0}$, and then express the complete displacement field ${\bf
u}({\bf k})$ through the tip position $\mathbf{u}_{0}$ of the reference
vortex.  After transformation back to real space, we obtain
\begin{align}\label{eq:equilibrium_vector}
   u_{0,\alpha} = f_{p,\beta}[\mathbf{R}_{0}+\mathbf{u}_{0}]
   \int \frac{d^3\mathbf{k}}{(2\pi)^3}G_{\alpha\beta}(\mathbf{k}).
\end{align}
We express the last integral through the effective elasticity $\bar{C}$,
\begin{align}
   \bar{C}^{-1} = \frac{1}{2}\int
   \frac{d^3\mathbf{k}}{(2\pi)^3}G_{\alpha\alpha}(\mathbf{k}),
\end{align}
(with summation over $\alpha$ implied) and make use of the self-consistent
solution of Eq.\ (\ref{eq:equilibrium_vector}) to obtain the displacement
field expressed through the amplitude $u_{0,\beta}$,
\begin{align}\label{eq:u-field}
   u_{\alpha}(\mathbf{k}) = \bar{C} \, G_{\alpha\beta}(\mathbf{k}) u_{0,\beta} 
   \, e^{-i \mathbf{K}\cdot \mathbf{R}_{0}}.
\end{align}
Inserting this result back into Eq.\ \eqref{eq:free_energy}, we obtain a
simple expression for the free energy $F[{\bf u}] \to
e_\mathrm{pin}(\mathbf{R}_{0}, \mathbf{u}_{0})$ for our specific configuration
with the tip at $z=0$ of the vortex $\mu_0 = 0$ displaced by $\mathbf{u}_{0}$
from its asymptotic position $\mathbf{R}_{0}$ due to the action of the defect,
\begin{align}
   &e_\mathrm{pin}(\mathbf{R}_{0},\mathbf{u}_{0}) =
   e_p[\mathbf{R}_{0}+\mathbf{u}_{0}]\\ \nonumber
   &+\frac{\bar{C}^2}{2}\int\frac{d^3 \mathbf{k}}{(2\pi)^3} G_{\alpha\beta}(\mathbf{k})
   u_{0,\beta} \, \phi_{\alpha\gamma}(\mathbf{k})G_{\gamma\delta}(-\mathbf{k})
   u_{0,\delta}\\
   \label{eq:free_energy_effective}
   &=e_p[\mathbf{R}_{0}+\mathbf{u}_{0}] + \frac{\bar{C}}{2} \mathbf{u}_{0}^2.
\end{align}

\begin{figure}[h]
\centering \includegraphics[scale=1]{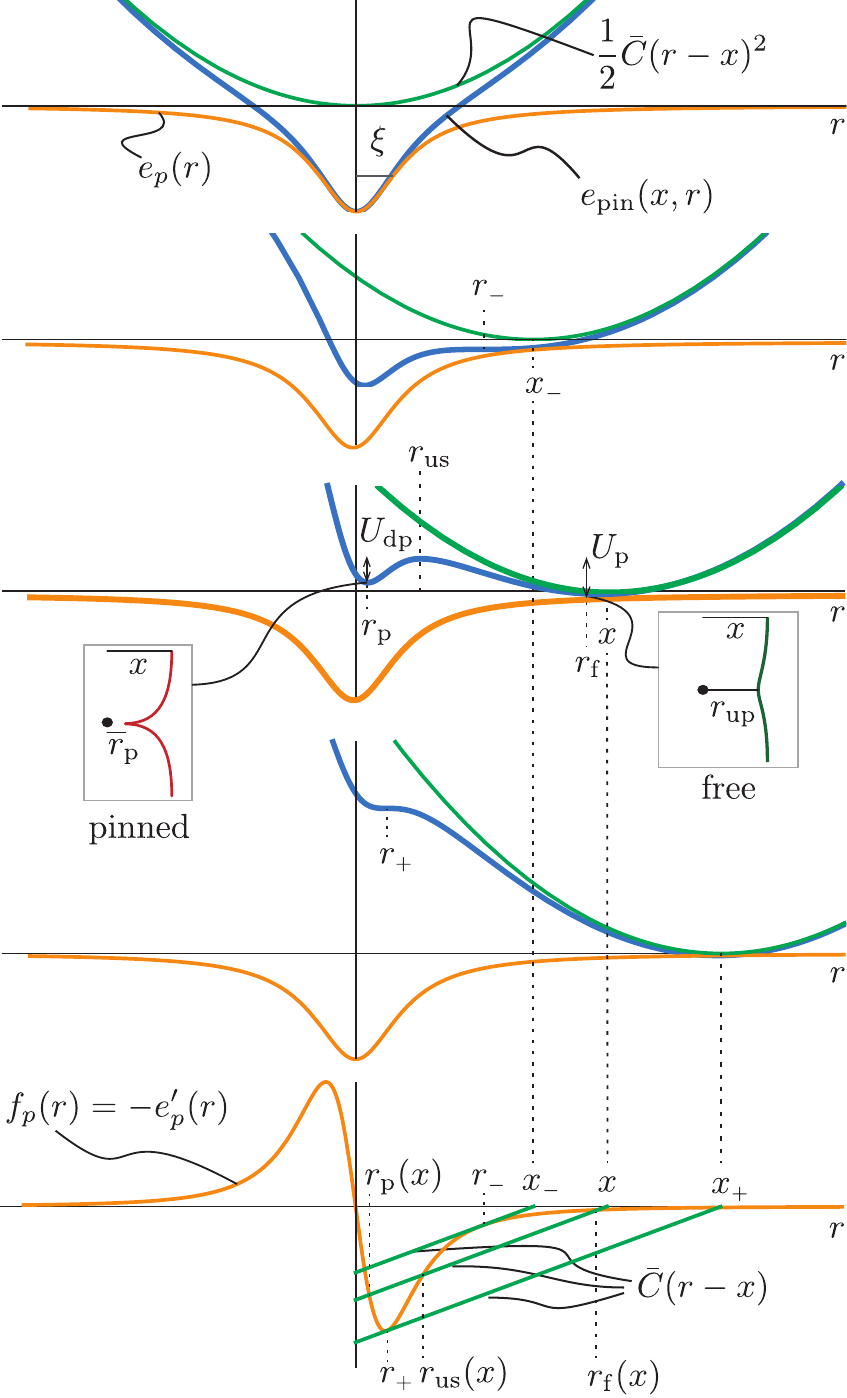}
\caption{Effective pinning energy $e_\mathrm{pin}(x;r)$ versus $r$ for four
different asymptotic vortex positions $x$, $0 \leq x \leq
x_{\scriptscriptstyle +}$. We assume a Lorentzian defect with a potential
$e_p(R) = -e_p/(1+R^2/2\xi^2)$ and a Labusch parameter $\kappa = 5$. Two local
minima at $r_\mathrm{p}$ and $r_\mathrm{f}$ appear for $x_{\scriptscriptstyle
-} < x < x_{{\scriptscriptstyle +}}$ and are separated by the local maximum at
$r_\mathrm{us}$ defining the barrier $U_\mathrm{dp}(x)$ for depinning (the
barrier to escape the pin) and the barrier $U_\mathrm{p}(x)$ for pinning (or
jumping into the pin). Bottom: Minimizing $e_\mathrm{pin}(x;r)$ with respect
to the tip position $r$ at fixed asymptotic position $x$ corresponds to
solving the self-consistency equation $\bar{C} (r-x) = f_p(r)$, see
\eqref{eq:equilibrium}, here done graphically. Multiple solutions $r_i(x)$,
$(i=\mathrm{p,~f,~us})$, show up if the slope $\bar{C}$ is smaller than the
maximum slope of $f_p(r)$, what corresponds to a Labusch parameter $\kappa >
1$.}
\label{Fig:metastable_states} 
\end{figure}

Next, we choose the vortex position along the $x$-axis, $\mathbf{R}_{0} = x\,
\mathbf{e}_x$ and assume a radially symmetric pin, implying that
$\mathbf{u}_{0} = u\,\mathbf{e}_x$; expressing the effective free energy in
Eq.\ \eqref{eq:free_energy_effective} in terms of the vortex tip coordinate $r
= x+u$, we arrive at the simplified effective pinning energy
\begin{align}\label{eq:effective_energy}
   e_\mathrm{pin}(x;r) = \frac{1}{2}\bar{C} (r-x)^2+e_p(r).
\end{align}
The effective pinning energy $e_\mathrm{pin}(x;r)$ involves the bare pinning
potential $e_p(r)$ augmented by the elastic deformation energy of the vortex
in the form of a parabolic potential centered at $r = x$, see Fig.\
\ref{Fig:metastable_states}, with a curvature given by the effective vortex
lattice elasticity $\bar{C}$. The latter can be expressed through the
compression, tilt and shear elastic moduli known from elasticity theory
\cite{Blatter1994, Willa2016}, $\bar{C} = \gamma (a_0^2/\lambda) [c_{66}\,
c_{44}({\bf k} = 0)]^{1/2}$; the numerical $\gamma$ depends on the chosen
approximations, see Refs.\ \cite{Thomann2017,Willa2018PRB}.  The simple
estimate $\bar{C} \sim \sqrt{\varepsilon_0\varepsilon_\ell}/a_0$, where
$\varepsilon_\ell = \varepsilon_0 \ln(a_0/\xi)$, is governed by the value of
the vortex line energy $\varepsilon_0 = (\phi_0/4\pi \lambda)^2$.

If pinning is sufficiently strong, i.e., $e_p(r)$ is sufficiently deep, the
total free energy $e_\mathrm{pin}(x;r)$ has two minima within a finite range
$|x|\in [x_{{\scriptscriptstyle -}},x_{{\scriptscriptstyle +}}]$ of asymptotic
vortex positions $x$. Minima appear or vanish whenever the total pinning
energy develops an inflection point, ${\partial^2 e_\mathrm{pin}/\partial r^2
= \bar{C}-f_p'(r)=0}$. This requires that the condition
\begin{align}\label{eq:Labusch}
   \kappa = \frac{\max_r f_p'(r)}{\bar{C}}>1,
\end{align}
the so-called Labusch criterion\cite{Labusch1969}, is fulfilled, see bottom of
Fig.\ \ref{Fig:metastable_states}. The value $\kappa=1$ marks the transition
between weak pinning with a unique vortex configuration and the strong-pinning
situation where the vortex can choose between two alternative configurations,
a pinned and a free one, whenever its asymptotic position $|x|$ resides in the
interval $[x_{\scriptscriptstyle -},x_{\scriptscriptstyle +}]$.  The two local
minima $r_\mathrm{p}(x)$ and $r_\mathrm{f}(x)$ are obtained from minimizing
$e_\mathrm{pin}(x;r)$ with respect to $r$ at fixed $x$,
\begin{align}\label{eq:equilibrium}
   \bar{C}(r-x) = f_p(r).
\end{align}
The local maximum at $r_\mathrm{us}(x)$ is an unstable solution that plays an
important role in the context of creep, see below.

Fig.\ \ref{Fig:energy_landscape} shows the multi-valued energy landscape with
the three branches $e_\mathrm{pin}^\mathrm{p}(x)$,
$e_\mathrm{pin}^\mathrm{f}(x)$, and $e_\mathrm{pin}^\mathrm{us}(x)$
corresponding to the extremal solutions, $e_\mathrm{pin}^i(x) =
e_\mathrm{pin}[x;r_i(x)]$; three of them coexist in the two intervals where
$|x|\in [x_{{\scriptscriptstyle -}},x_{{\scriptscriptstyle +}}]$, while outside those regions only one solution is realized. The total
pinning force exerted on a moving vortex is derived from this energy
landscape, with the pinning force acting on a vortex given by
$f_\mathrm{pin}(x) = -d e_{\mathrm{pin}}(x)/d x$. As this force differs when
evaluated in the pinned and free branches, it is the occupation of these
branches that determines the total pinning-force density acting on the vortex
system. The pinned and free branches in the pinning-energy landscape are
separated by an energy barrier.  The \textit{depinning} barrier
$U_\mathrm{dp}(x) = e_\mathrm{pin}^\mathrm{us}(x) -
e_\mathrm{pin}^\mathrm{p}(x)$ has to be overcome for transitions to the free
branch, while the \textit{pinning} barrier $U_\mathrm{p}(x) =
e_\mathrm{pin}^\mathrm{us}(x)-e_\mathrm{pin}^\mathrm{f}(x)$ is relevant for
jumps into the pin, i.e., the transitions to the pinned branch.

\begin{figure}
\centering \includegraphics{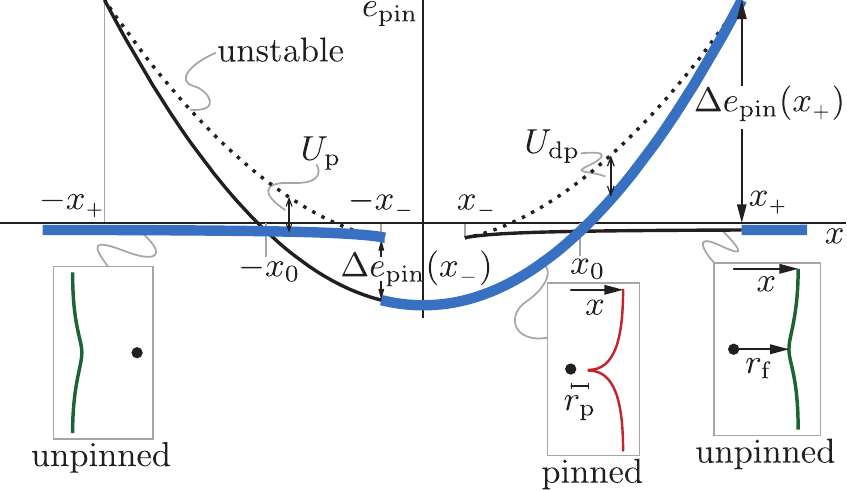}
\caption{Multivalued energy landscape for a strong defect. Within the regions
$|x|\in [x_{\scriptscriptstyle -},x_{{\scriptscriptstyle +}}]$, pinned and
free equilibrium positions show up with different locations $r_\mathrm{p}(x)$
and $r_\mathrm{f}(x)$ of the vortex tips either close to or away from the
defect; they define the energies $e_\mathrm{pin}^\mathrm{p}(x)$ and
$e_\mathrm{pin}^{\mathrm{f}}(x)$ of the pinned (nearly parabolic) and free
(nearly flat) branches. In addition, an unstable extremum at
$r_\mathrm{us}(x)$, see Fig.\ \ref{Fig:metastable_states}, maximises the
pinning energy (dotted) unstable branch $e_\mathrm{pin}^{\mathrm{us}}$ and
defines the energy barriers connecting pinned and free branches.  Thick blue
curves denote the occupation of branches in the critical state realised for $T
= 0$.  The pinning force is proportional to the sum of jumps $\Delta
e_\mathrm{pin}(x_{{\scriptscriptstyle +}})$, $\Delta
e_\mathrm{pin}(x_{\scriptscriptstyle -})$ in the energy landscape.}
\label{Fig:energy_landscape}
\end{figure}

Unfortunately, no closed expressions for the branches $e_\mathrm{pin}^i(x)$
can be given since the equilibrium equation \eqref{eq:equilibrium} fixing
$r(x)$ is in general not solvable analytically. Progress can be made in the
limits of marginally strong pinning with $\kappa-1\ll 1$ or for very strong
pinning $\kappa\gg 1$. In the first case the slope $\bar{C}$ is close to the
maximum slope of $f_p(r)$ and the energy branches can be derived from a cubic
expansion of $f_p(r)$ around the point $r_m$ of maximum slope, $f_p'(r_m) =
\kappa \bar{C}$.  For very strong pinning, the slope $\bar{C}$ is small
compared to $f_\mathrm{p}'(r_m)$ and the pinned and unstable solutions are
obtained by analyzing the tail of $f_p(r)$.  We will assume an algebraically
decaying potential, $e_p(r\gg\xi)\sim e_p(\xi/r)^{n}$ such that the pinning
force $f_p(r\gg \xi)\sim f_p(\xi/r)^{n-1}$ with $f_p \sim e_p/\xi$, in order
to make analytic progress in this situation.

The boundaries $x_{\scriptscriptstyle\pm}$ of the multi-valued interval are
found by first determining the critical tip positions $r_{\scriptscriptstyle
\pm}$ from the condition $f'_p(r_{\scriptscriptstyle \pm}) = \bar{C}$ for the
appearance of inflection points in $e_\mathrm{pin}(x;r)$ and then deriving the
associated asymptotic position $x_{\scriptscriptstyle \pm}$ from the
equilibrium condition Eq.\ \eqref{eq:equilibrium}, see also Fig.\
\ref{Fig:metastable_states}. For marginally strong pinning, we find that (see
Ref.\ [\onlinecite{Blatter2004}], Appendix \ref{sect:APP_mod_strong} and
\ref{sect:APP_very_strong} for the derivation)
\begin{align}\label{eq:rpm_mp}
   r_{\scriptscriptstyle \pm} - r_m &\sim \mp (\kappa - 1)^{1/2} \xi, \\
   \label{eq:xpm_mp}
   x_{\scriptscriptstyle \pm} - x_m &\sim \pm (\kappa - 1)^{3/2} \xi, 
\end{align}
where $r_m \lesssim x_m \sim \xi$. Note that $x_m$ coincides with the branch
crossing point $x_0$, $x_m = x_0$. For very strong pinning,
$r_{\scriptscriptstyle {\scriptscriptstyle -}}$ resides on the tail of the
pinning potential, while $r_{\scriptscriptstyle +}$ is located near the
maximum force,
\begin{align}\label{eq:rpm_sp}
   r_{\scriptscriptstyle -} \sim \kappa^{1/(n+2)} \xi, \qquad
   r_{\scriptscriptstyle +} \sim \xi. 
\end{align}
The associated asymptotic positions are largely different, see also
Fig.\ \ref{Fig:metastable_states},
\begin{align}\label{eq:xpm_sp}
   x_{\scriptscriptstyle -} \sim \kappa^{1/(n+2)} \xi, \qquad 
   x_{\scriptscriptstyle +} \sim \kappa \xi.
\end{align}
Finally, the branch crossing point is located at a position where the free and
pinned branches have the same energies, $\bar{C} x_0^2/2 \approx  e_p$,
implying that $x_0\approx \sqrt{2 e_p/\bar{C}} \sim \kappa^{1/2}\xi$.

The free-energy landscape in Fig.\ \ref{Fig:energy_landscape} has much in
common with the one appearing in the phenomenological theory of a first-order
phase transition in thermodynamic systems, e.g., the Gibb's energy $g(p,T)$ of
the Van der Waal's theory of the gas--liquid transition or the energy $g(h,T)$
of a magnetic transition. In developing this analogy, we can identify $\bar{C}
r$ with the volume $V$ and $\bar{C}$ (or the inverse Labusch parameter
$1/\kappa$) with the reduced temperature $\tau = T/T_c$. Expanding
$e_\mathrm{pin}(x;r) = e_p(r)+\bar{C} r^2/2-\bar{C} r x +\bar{C} x^2/2$, we
can identify the energy $e_p(r)+\bar{C} r^2/2$ with the free energy
$f(V,\tau)$. If $x$ is identified with pressure $p$, then
$e_\mathrm{pin}(x;r)$ is (up to the constant term $\bar{C} x^2/2$) equivalent
to the Gibb's energy $g=f-pV$. Minimizing $e_\mathrm{pin}$ with respect to $r$
for fixed $x$ and $\bar{C}$ corresponds to minimizing $g$ with respect to $V$
for fixed $p$ and $T$ and provides the (metastable) equilibrium states (note
that $\bar{C} r$ and $V$ play the roles of constraint parameters). The barrier
separating the minima in the thermodynamic system are relevant in the
description of the hysteretic transition and nucleation phenomena---here, the
analogous barriers describe thermal transitions between pinned and free states
and thus are relevant in the description of thermal creep.

\subsection{Pinning force}\label{subsect:formalism_pinning_force}

The average pinning force per defect acting on the vortex system is obtained
by position-averaging the force between a defect and its nearest vortex while
accounting for the random positions of the defects in the material.  Driving
the vortices in the positive $x$-direction results in an average force
$-\langle f_\mathrm{pin} \rangle < 0$ per defect (in accordance with
\eqref{eq:force_balance}, we choose $\langle F_\mathrm{pin} \rangle$ and hence
$\langle f_\mathrm{pin} \rangle$ to be positive).  The instantaneous force
acting on a vortex with asymptotic position $x$ is different for pinned and
free states.  Let $p(x)$ be the occupation probability of the pinned branch;
the occupation probability for the free branch then is $1-p(x)$.  For a vortex
passing centrally through the defect, the average pinning force is given by
the position and occupation average
\begin{align}\label{eq:f_pin}
   \langle f_\mathrm{pin}\rangle = -\frac{1}{a_0}\int_{-a_0/2}^{a_0/2}
   d x\, \bigl[p\, f_\mathrm{pin}^\mathrm{p} 
   + (1-p)f_\mathrm{pin}^{\mathrm{f}}\bigr](x),
\end{align}
with $f_\mathrm{pin}^{\mathrm{p},\mathrm{f}}(x) = -d
e_\mathrm{pin}^{\mathrm{p},\mathrm{f}}(x)/d x$ denoting the pinning forces on
the pinned and free branches. For $|x|<x_{\scriptscriptstyle -}$ only the
pinned branch is available, while for $|x|>x_{\scriptscriptstyle +}$ the
occupation is restricted to the free branch; hence, we set $p(x) = 1$ and
$p(x) = 0$, respectively, in those two regions.  The integration is restricted
to $|x|<a_0/2$ due to the vortex lattice periodicity. The antisymmetric force
with $f^{\mathrm{p},\mathrm{f}}_\mathrm{pin}(x) =
-f^{\mathrm{p},\mathrm{f}}_\mathrm{pin}(-x)$ allows to write the previous
equation in the form
\begin{align}\label{eq:f_pin2}
   \langle f_\mathrm{pin}\rangle = -\frac{1}{a_0}\int_{I_\mathrm{mv}}
   d x\, p(x)\, \Delta f_\mathrm{pin}(x)
\end{align}
with $\Delta f_\mathrm{pin} = f^\mathrm{p}_\mathrm{pin} -
f_\mathrm{pin}^{\mathrm{f}}$ and the integration restricted to the multivalued
intervals $I_\mathrm{mv} = [-x_{{\scriptscriptstyle +}},-x_{\scriptscriptstyle
-}]\cup[x_{\scriptscriptstyle -},x_{{\scriptscriptstyle +}}]$.

The $T = 0$ branch occupation for vortices driven along the positive $x$-axis
is shown in Fig.\ \ref{Fig:energy_landscape}, see blue solid lines. An
individual vortex approaches the defect on the free branch and remains there
until $-x_{\scriptscriptstyle -}$, even though the pinned branch becomes
energetically more favorable at the branch crossing point $-x_0 <
-x_{\scriptscriptstyle -}$. The vortex becomes pinned at
$-x_{\scriptscriptstyle -}$ when the \textit{pinning} barrier $U_\mathrm{p}$
vanishes and stays on the pinned branch until $x_{\scriptscriptstyle +}$,
where it jumps again to the unpinned branch, this time due to the vanishing of
the \textit{depinning} barrier $U_\mathrm{dp}$. The occupation then can be
written through the Heaviside step function $\Theta(x)$,
\begin{align}\label{eq:p_c}
   p_\mathrm{c}(x) = \Theta(x+x_{\scriptscriptstyle -})-
   \Theta(x-x_{\scriptscriptstyle +}),
\end{align}
and using $f_\mathrm{pin} = -\partial_x e_\mathrm{pin}$ in Eq.\
\eqref{eq:f_pin}, we find that
\begin{align}\label{eq:f_pin_av}
   \langle f_\mathrm{pin}\rangle = \frac{\Delta e_{c}}{a_0}
\end{align}
with $\Delta e_{c} = \Delta e_\mathrm{pin}(x_{\scriptscriptstyle +})-\Delta
e_\mathrm{pin}(-x_{\scriptscriptstyle -})$ and $\Delta e_\mathrm{pin} =
e_\mathrm{pin}^\mathrm{p}-e_\mathrm{pin}^{\mathrm{f}}$.  Note that $\Delta
e_\mathrm{pin}(-x_{\scriptscriptstyle -})=\Delta e_\mathrm{pin}
(x_{\scriptscriptstyle -})<0$ and $\Delta e_c$ thus corresponds to the sum of
the energy jumps in the multivalued energy landscape evaluated at the end
points of the multivalued intervals.

Estimates for the jumps $\Delta e_\mathrm{pin}(x_{\scriptscriptstyle \pm})$
are derived in Appendix \ref{sect:APP_mod_strong} and
\ref{sect:APP_very_strong}, see also Refs.\ \cite{Labusch1969,Blatter2004}.
For marginally strong pinning, one finds that
\begin{align}\label{eq:de_pin_mp}
   \Delta e_\mathrm{pin}(x_{\scriptscriptstyle \pm}) \sim \bar{C} \xi^2 (\kappa - 1)^2
\end{align}
while for very strong pinning
\begin{align}\label{eq:de_pin_sp}
   \Delta e_\mathrm{pin}(x_{\scriptscriptstyle \pm}) 
   \approx \frac{\bar{C}}{2} x^2_{\scriptscriptstyle \pm},
\end{align}
in particular, using Eq.~\eqref{eq:xpm_sp}, we find that $\Delta
e_\mathrm{pin}(x_{\scriptscriptstyle {\scriptscriptstyle +}}) \sim
\bar{C}\kappa^2 \xi^2\sim \kappa e_p$ is large compared to $\Delta
e_\mathrm{pin}(x_{\scriptscriptstyle {\scriptscriptstyle -}})\sim
\kappa^{1/(n+2)}e_p $.
\begin{figure}[t]
\centering \includegraphics[scale=1]{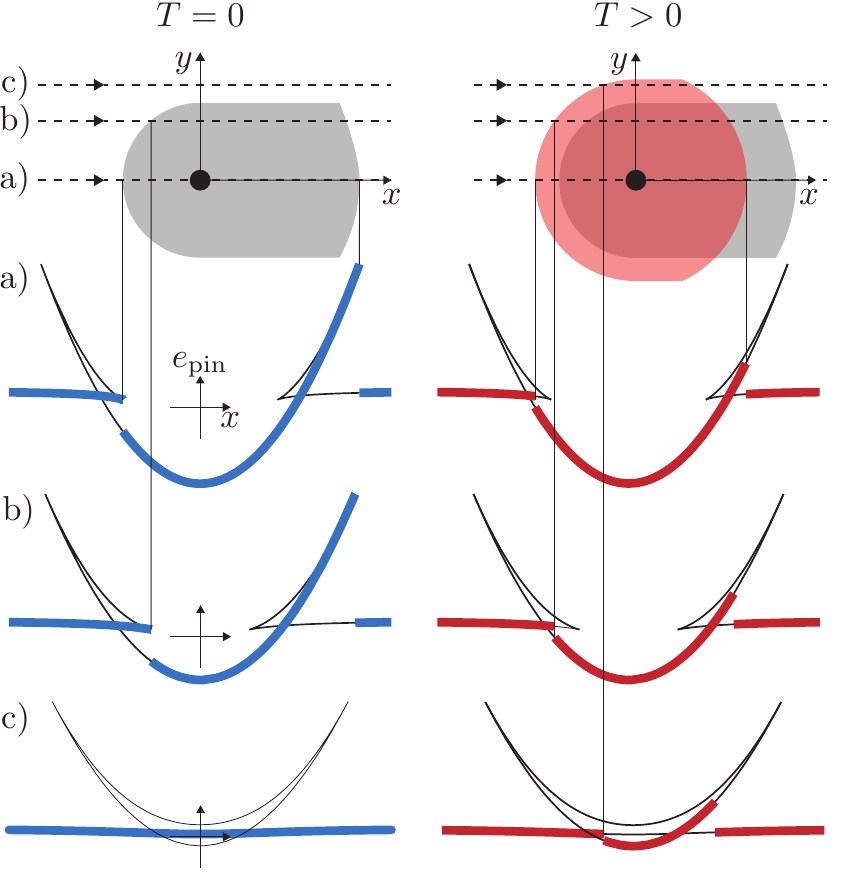}
\caption{Pinning of vortices passing the defect at transverse distances $y =0$
(a), $y =0.8x_{\scriptscriptstyle-}$ (b), and $y = 1.2 x_{\scriptscriptstyle
-}$ (c) for $T = 0$ (blue, left) and finite temperatures (red, right). The
corresponding pinning energy landscapes $e_\mathrm{pin}^i(x,y)$ are plotted
for the Lorentzian pinning potential and $\kappa = 5$. At $T = 0$, vortices
are pinned whenever they enter the defect trapping area (along the grey half
circle of radius $x_{\scriptscriptstyle -}$, where the free and unstable
branches merge and disappear such that $U_\mathrm{p}(x,y) = 0$.  In case (c)
the vortex trajectory lies outside the trapping area and vortices always stay
on the free branch, which is available all the way along the vortex
trajectory; the pinning force vanishes in this case.  At finite temperature $T
> 0$, the transitions between the free and pinned branches are realized at a
radial distance $x^\mathrm{jp}_{\scriptscriptstyle \pm} >
x_{\scriptscriptstyle -}$ from the pinning centre.  Due to thermal activation,
vortices can overcome the barrier separating free and pinned branches and
occupy the pinned branch even for $y > x_{\scriptscriptstyle -}$, resulting in
a non-vanishing average pinning force in the extended region $y <
x_{\scriptscriptstyle -}^\mathrm{jp}$ (light red).}
\label{Fig:energy_landscape_transverse}
\end{figure}

Above, we have considered the situation where the vortex impacts straight on
the defect center.  For those vortices passing the defect at a finite
transverse distance $y$ (Fig.~\ref{Fig:energy_landscape_transverse}), the
effective pinning energy is given by $e_\mathrm{pin}(\mathbf{R}, \mathbf{r}) =
\frac{1}{2}\bar{C} (\mathbf{R}-\mathbf{r})^2+e_p(\mathbf{r})$, where
$\mathbf{R}$ and $\mathbf{r}$ denote the asymptotic- and tip-position of the
vortex line, see Eq.\ \eqref{eq:free_energy_effective} (we drop the index
$\mu_0 = 0$).  The equilibrium condition $\nabla_\mathbf{r} e_\mathrm{pin} =
0$ yields the solutions $\mathbf{r}_i(\mathbf{R})$, $i = \mathrm{f},
\mathrm{p}, \mathrm{us}$ for the free, pinned, and unstable branches. For a
radially symmetric pinning potential, we have $e_p(\mathbf{r}) = e_p(r)$ and
the equilibrium condition is satisfied for the radial geometry $\mathbf{r}
\parallel \mathbf{R}$. The energy is then brought to the same form as in Eq.\
\eqref{eq:effective_energy}, albeit with the replacement $x \to |\mathbf{R}| =
\sqrt{x^2+y^2}$. Evaluating $e_\mathrm{pin}^i(\mathbf{R}) =
e_\mathrm{pin}[\mathbf{R}, \mathbf{r}_i(\mathbf{R})]$ provides us with the
energies of the various branches in the multivalued energy landscape, $i =
\mathrm{f}, \mathrm{p}, \mathrm{us}$,
\begin{align}\label{eq:e_pin_xy}
   e_\mathrm{pin}^i(\mathbf{R}) = e_\mathrm{pin}^i(x,y) 
   = e_\mathrm{pin}^i\left(\sqrt{x^2+y^2},0\right).
\end{align}
The energy landscape is plotted in Fig.\ \ref{Fig:energy_landscape_transverse}
for three impact parameters $y = 0$, $y = 0.8x_{\scriptscriptstyle -}$, and $y
= 1.2x_{\scriptscriptstyle -}$. Eq.\ \eqref{eq:e_pin_xy} shows that the shape
of the energy landscape at finite impact parameter $y$ is similar to the one
at $y = 0$ with an excluded region $|x|<y$. In particular, the minimal energy
of the pinned branch satisfies $e_\mathrm{pin}^\mathrm{p}(0,y) =
e_\mathrm{pin}^\mathrm{p}(y,0) > e_\mathrm{pin}^\mathrm{p}(0,0)$. For a large
impact parameter $y > x_{\scriptscriptstyle -}$ (the situation with $y =
1.2x_{\scriptscriptstyle -}$ is shown in Fig.\
\ref{Fig:energy_landscape_transverse}) the free branch never terminates,
implying that such a vortex is never trapped at $T=0$.  On the other hand,
vortices hitting the defect with a finite impact parameter $y <
x_{\scriptscriptstyle -}$ are trapped onto the defect at the radial distance
$R = x_{\scriptscriptstyle -}$ and released back to the free branch at a
radial distance $R = x_{\scriptscriptstyle +}$, hence the vortex remains
pinned over the finite interval $x \in [-\sqrt{x_{\scriptscriptstyle -}^2 -
y^2}, \sqrt{x_{\scriptscriptstyle +}^2 - y^2}]$, with $x$ the direction of
drive, and for all $y < t_\perp = x_{\scriptscriptstyle -}$; we call $t_\perp$
the transverse trapping length. The average pinning force along the
$x$-direction is once more given by Eq.\ \eqref{eq:f_pin}, but with the
pinning forces replaced by $f^{\mathrm{p},\mathrm{f}}_\mathrm{pin}(x) \to
f^{\mathrm{p},\mathrm{f}}_\mathrm{pin}(x,y) = -\partial_x
e_\mathrm{pin}^{\mathrm{p},\mathrm{f}}(x,y) = -\partial_x
e_\mathrm{pin}^{\mathrm{p},\mathrm{f}} (\sqrt{x^2+y^2},0)$ and the jumps in
the occupation \eqref{eq:p_c} now appearing at $\sqrt{x_{\scriptscriptstyle
-}^2 - y^2}$ and $\sqrt{x_{\scriptscriptstyle +}^2 - y^2}$. A simple
calculation then shows, that the average force $\langle
f_\mathrm{pin}(y)\rangle$ contributed by a vortex with an impact parameter $y
< x_{\scriptscriptstyle -}$ is identical with the result \eqref{eq:f_pin_av}.
While vortices passing the defect at larger distances cannot get trapped at
zero temperature, fluctuations at finite temperature will render such
processes statistically possible, see Sec.\
\ref{subsubsect:large_drives_pinning_force} below.

Combining the above results, we can determine the average pinning force
density for a finite density of defects by multiplying the average pinning
force \eqref{eq:f_pin_av} with the fraction $2x_{\scriptscriptstyle -}/a_0$ of
trajectories that are trapped by one defect and the density $n_p$ of
independently acting defects; including a minus sign in order to respect our
definition of pinning force density in the equation of motion
\eqref{eq:force_balance}, we obtain the critical force density
\begin{align}\label{eq:F_crit}
   F_c = n_p\frac{2x_{\scriptscriptstyle -}}{a_0}\frac{\Delta e_{c}}{a_0}.
\end{align}
Collecting the various factors from above, we obtain the estimates
\begin{align}\label{eq:F_crit_mp}
   F_c \sim (\xi/a_0)^2 n_p f_p (\kappa-1)^2  
\end{align}
and 
\begin{align}\label{eq:F_crit_sp}
   F_c  \sim (\xi/a_0)^2 n_p f_p \kappa^{(n+3)/(n+2)}
\end{align}
in the marginally strong and very strong pinning limits, respectively. Quite
often, these results are written through the trapping
area\cite{IvlevOvchinnikov91,Blatter2004} $S_\mathrm{trap}$,
\begin{align}\label{eq:S_trap}
   S_\mathrm{trap} = 2 x_{\scriptscriptstyle -} (x_{\scriptscriptstyle +} 
   + x_{\scriptscriptstyle -}),\qquad 
   F_c \sim \frac{S_\mathrm{trap}}{a_0^2} n_p f_p,
\end{align}
which assumes values $S_\mathrm{trap} \sim \xi^2$ and $S_\mathrm{trap} \sim
\kappa^{(n+3)/(n+2)} \xi^2$ at marginally strong and very strong pinning.  For
a rapidly decaying pinning potential (with a large value of $n$), the trapping
area, critical force density, and critical current density scale like
$S_\mathrm{trap} \propto 1/\sqrt{B}$, $F_c \propto \sqrt{B}$, and $j_c \propto
1/\sqrt{B}$; such a field dependence (cut off at small fields when strong
pinning becomes 1D, single-vortex type, see Ref.\ [\onlinecite{Blatter2004}]) is often
taken as a signature for strong pinning.

The critical state occupation $p_\mathrm{c}(x)$ in \eqref{eq:p_c} is the one
maximizing the pinning force. If the applied force density
$F_{\rm\scriptscriptstyle L}$ exceeds $F_c$, the vortex lattice moves with
drift velocity as given through the dissipative force balance equation $\eta v
= F_{\rm\scriptscriptstyle L}-F_c$.  The dynamical pinning force
$F_\mathrm{pin}(v)$ changes on a scale $v_p = f_p/a_0^3\eta \gg v_c =
F_c/\eta$ and has been calculated at $T=0$ in Refs.\
\cite{Thomann2012,Thomann2017}; below, we focus on the calculation of
$F_\mathrm{pin}(v,T)$ at finite temperatures $T$ but small velocities $v \ll
v_p$, where the dynamical motion of the vortex through the pin can be
neglected, and derive the thermally renormalized force--velocity
characteristic.

\section{Thermal creep}\label{sec:thermal_creep}

We start with a short qualitative overview of thermal creep effects at large
and small velocities before deriving precise expressions for the two limits.

\subsection{Qualitative overview}
At finite temperatures $T > 0$, one has to account for thermal fluctuations in
the determination of the branch occupation as vortices can jump between
branches by overcoming the activation barrier; the same physics appears in the
context of pinned charge density waves, see Refs.\
\cite{BrazovskiiLarkin1999,BrazovskiiNattermann2004}.  We find the pinned
branch occupation $p(x)$ through solving the rate equation derived from
Kramers' theory \cite{Kramers1940} (we set $k_\mathrm{B} = 1$ from now on),
\begin{align}\label{eq:rate}
   \frac{d p}{d t} = v\frac{d p}{d x}
   = -\omega_\mathrm{p}\, e^{-U_\mathrm{dp}/T}p
   +\omega_\mathrm{f}\,e^{-U_\mathrm{p}/T}(1-p).
\end{align}
This rate equation accounts for the depinning of vortices via the activation
barrier $U_\mathrm{dp}(x)$ as well as the filling of the pinned branch due to
transitions over the barrier $U_\mathrm{p}(x)$. The steady-state probabilities
depend on the time $t$ only through the coordinate $x$ and thus we have
replaced the total derivative by $d/d t = v\,d/d x$. The frequencies
$\omega_\mathrm{p}(x)$ and $\omega_\mathrm{f}(x)$ can be understood as the
number of attempts per unit of time made by a vortex to escape from its
current, pinned or free, state. The success probability of such attempts is
exponentially small in the activation barrier. We calculate the barriers and
attempt frequencies later in Sec.\ \ref{sec:high_velocities} from a
`microscopic' theory.

Focusing on the high- and low-velocity regimes with qualitatively distinct
solutions of the rate equation \eqref{eq:rate} provides us with a first
understanding of the problem. The velocity $v_\mathrm{th} =
\omega_\mathrm{p}\, T/ \partial_x U_\mathrm{dp}$ derived below, see Eqs.\
\eqref{eq:vth} and \eqref{eq:v_crit}, sets the scale below which thermal
effects modify the $T=0$ excess-current characteristic; above $v_\mathrm{th}$,
the $T>0$ characteristic smoothly joins the one at $T=0$.  The high-velocity
regime $v > v_\mathrm{th} e^{-U_0/T}$, with $U_0$ the maximal activation
barrier located at the branch crossing point $x_0$, see Fig.\
\ref{Fig:energy_landscape}, is characterized by an occupation $p(x)$ of a
shape similar to the one of the critical state, but with the transitions
between branches realized close to the thermally renormalized jump points
$x^\mathrm{jp}_{\scriptscriptstyle \pm}(v,T)$, see Fig.\
\ref{Fig:occupation_high_v}. Ignoring the finite width $\ell_\mathrm{p}$ and
$\ell_\mathrm{dp}$ of these jumps, we can write $p(x)\approx
\Theta(x-x^\mathrm{jp}_{\scriptscriptstyle
-})-\Theta(x-x^\mathrm{jp}_{\scriptscriptstyle +})$ and express the pinning
force density through the thermally renormalized jumps in the energy landscape
\begin{align}\label{eq:F_pin_large_v}
   F_\mathrm{pin}(v,T) = n_p\frac{2 x^\mathrm{jp}_{\scriptscriptstyle -}}{a_0}
   \frac{\Delta e_\mathrm{pin}^\mathrm{tot}(v,T)}{a_0},
\end{align}
where
\begin{align}\label{eq:Delta_e_pin_tot}
   \Delta e_\mathrm{pin}^\mathrm{tot}(v,T)
   =\Delta e_\mathrm{pin}(x^\mathrm{jp}_{\scriptscriptstyle +})
   -\Delta e_\mathrm{pin}(-x^\mathrm{jp}_{\scriptscriptstyle -})>0.
\end{align}
The jump location $x^\mathrm{jp}_{\scriptscriptstyle +}$ follows from the
following consideration (a corresponding analysis provides the location
$-x^\mathrm{jp}_{\scriptscriptstyle -}$, see below): Close to the jump at
$x^\mathrm{jp}_{\scriptscriptstyle +}$, the occupation dynamics is dominated
by the smaller depinning barrier $U_\mathrm{dp} < U_\mathrm{p}$ and the second
term on the right hand side of the rate equation \eqref{eq:rate} can be
ignored. The rate equation then takes the simple form $\partial_x p =
-p/\ell_\mathrm{dp}$, with
\begin{align}\label{eq:ell_dp}
   \ell_\mathrm{dp}(x) = \frac{v}{\omega_\mathrm{p}}e^{U_\mathrm{dp}(x)/T}
\end{align}
defining the depinning length at the position $x > x_0$, telling us over what
distance the vortex will transit from the pinned to the free branch. The
depinning length $\ell_\mathrm{dp}(x)$ is large near $x_0$ where the barrier
$U_\mathrm{dp}$ is large and decreases rapidly with increasing $x$ due to the
decreasing barrier $U_\mathrm{dp}$.  The transition to the lower (free) state
appears at the position $x_{{\scriptscriptstyle +}}^\mathrm{jp}$ where the
vortex can escape the pin while itself moving a distance
$\ell_\mathrm{dp}(x)$, implying that the relative change in
$\ell_\mathrm{dp}(x)$ over the distance $\ell_\mathrm{dp}(x)$ should be of
order unity.  With the help of Eq.\ \eqref{eq:ell_dp}, we can reexpress the
corresponding condition $|\partial_x \ell_\mathrm{dp}(x)
|_{x_{{\scriptscriptstyle +}}^\mathrm{jp}}| \approx 1$ in the form
\begin{align}\label{eq:x^jp_+}
   \ell_\mathrm{dp}\bigl( x_{{\scriptscriptstyle +}}^\mathrm{jp}\bigr) 
   \approx \frac{T}{U'_\mathrm{dp} \bigl(x_{{\scriptscriptstyle +}}^\mathrm{jp}\bigr)},
\end{align}
where we focus on the main $x$-dependence in the exponent and denote the space
derivative with a prime, $U'_\mathrm{dp} \equiv \partial_x U_\mathrm{dp}$.  At
the maximal value $x^\mathrm{jp}_{\scriptscriptstyle +} =
x_{\scriptscriptstyle +}$, the barrier $U_\mathrm{dp}(x_{\scriptscriptstyle
+})$ vanishes and we reach the maximal velocity $v_\mathrm{th}$,
\begin{align}\label{eq:vth}
  v_\mathrm{th} = \frac{\omega_\mathrm{p}\, T}{U'_\mathrm{dp}}
  \bigg|_{x_{\scriptscriptstyle +}},
\end{align}
where the thermal characteristic goes over into the $T=0$ excess-current
characteristic.  From the condition \eqref{eq:x^jp_+} and using Eq.\
\eqref{eq:vth}, we find that the relevant depinning barrier
$U_\mathrm{dp}(x^\mathrm{jp}_{\scriptscriptstyle +})$ can be written in the
form
\begin{align}\label{eq:Ux^jp_+}
   U_\mathrm{dp}(x^\mathrm{jp}_{\scriptscriptstyle +}) 
   \approx T \ln(v_\mathrm{th}/{v}),
\end{align}
where we have approximated
$[U'_\mathrm{dp}/\omega_\mathrm{dp}](x^\mathrm{jp}_{\scriptscriptstyle +})$ by
its value at $x_{\scriptscriptstyle +}$.  Similar results apply for the jump
at $- x^\mathrm{jp}_{\scriptscriptstyle -}$ and are quantitatively derived
below, see Sec.\ \ref{sec:high_velocities}. Given the barriers
$U_\mathrm{dp}(x)$ (and $U_\mathrm{p}(x)$) for a specific defect potential, we
can solve Eq.\ \eqref{eq:Ux^jp_+} for $x^\mathrm{jp}_{\scriptscriptstyle
+}(v)$ (and similar for $-x^\mathrm{jp}_{\scriptscriptstyle -}(v)$) and using
the results in the definition of the energy jump Eq.\
\eqref{eq:Delta_e_pin_tot} leads to the velocity- and temperature-dependent
pinning force density. We cast the final result (see
Sec.~\ref{sec:high_velocities} for details) in the form
\begin{align}\label{eq:F_pin_result}
   F_\mathrm{pin}(v,T) = F_c\Bigl[1- g(\kappa)\Bigl(\frac{T}{e_p}
   \log\frac{v_\mathrm{th}}{v}\Bigr)^{2/3}\Bigr],
\end{align}
with $g(\kappa)$ a factor of order unity that can be derived as a function of
pinning strength $\kappa$ for any given defect potential $e_p(r)$.

A different approach has to be used in solving the rate equation for small
velocities $v < v_{\rm \scriptscriptstyle TAFF} = v_\mathrm{th}e^{-U_0/T}$.
Starting from the above analysis and decreasing the velocity $v$, the jump
positions $x^\mathrm{jp}_{\scriptscriptstyle \pm}$ approach the branch
crossing point $x_0$ and the activation barriers increase towards their
maximum $U_0 = U_\mathrm{dp}(x_0)=U_\mathrm{p}(x_0)$, see Fig.\
\ref{Fig:energy_landscape}. At $x_0$, the renormalized energy jumps $\Delta
e_\mathrm{pin}(x^\mathrm{jp}_{\scriptscriptstyle \pm} \to x_0)$ vanish and
Eq.\ \eqref{eq:F_pin_large_v}, providing a vanishing pinning force, can no
longer be used.  In this limit, a good starting point for our analysis is the
equilibrium distribution obtained by setting $v = 0$ in the rate equation
\eqref{eq:rate}.
\begin{align}\label{eq:p_eq1}
   p_\mathrm{eq}(x) &= \frac{\omega_{\mathrm{f}}
   \,e^{-U_{\mathrm{p}}/T}}{\omega_{\mathrm{p}}
   \,e^{-U_{\mathrm{dp}}/T}\! +\omega_{\mathrm{f}}\,e^{-U_{\mathrm{p}}/T}} 
   = \frac{\ell_\mathrm{dp}(x)}{\ell_\mathrm{dp}(x) \! + \! \ell_\mathrm{p}(x)}
\end{align}
with $\ell_\mathrm{p} = (v/\omega_\mathrm{f})e^{U_p/T}$ defining the local
relaxation distance for the case of pinning. The expression Eq.\
\eqref{eq:p_eq1} is valid away from the endpoints of the multi-valued interval
where barriers vanish. The rate equation then can be cast into the form
\begin{align}\nonumber
   \frac{d p}{d x} &= \frac{1}{v}(p_\mathrm{eq} -p)
   \bigl(\omega_\mathrm{p}\,e^{-U_\mathrm{dp}/T}
   +\omega_\mathrm{f}\,e^{-U_p/T}\bigr)\\
   &=\frac{p_\mathrm{eq}(x)-p}{\ell_\mathrm{eq}(x)},\label{eq:rate_tau}
\end{align}
where the equilibrium relaxation distance $\ell_\mathrm{eq}$,
\begin{align}
   \ell_\mathrm{eq}(x) = [\ell_\mathrm{p}(x)^{-1} + \ell_\mathrm{dp}(x)^{-1}]^{-1},
   \label{eq:ell_eq}
\end{align}
includes processes that connect both pinned and free branches. Treating $v$ as
a small parameter, we find that the solution of the rate equation is given by
the shifted equilibrium distribution, $p(x)\approx p_\mathrm{eq}(x) -
\ell_\mathrm{eq}(x)p_\mathrm{eq}'(x)\approx p_\mathrm{eq}[x -
\ell_\mathrm{eq}(x)]$. Assuming similar scales $\omega_\mathrm{p} \sim
\omega_\mathrm{f}$ and $|U_\mathrm{dp}'|\sim U_\mathrm{p}'$, we obtain a
simple estimate for the equilibrium relaxation length in the form
$\ell_\mathrm{eq}(x_0) \sim (v/ \omega_\mathrm{p})\, e^{U_0/T}$ and the
condition $v \ll v_{\rm \scriptscriptstyle TAFF}$ defining the low-velocity
regime is then equivalent to $\ell_\mathrm{eq}(x_0) \ll T/|U_\mathrm{dp}'|$
implying that the shift $\ell_\mathrm{eq}(x)$ is small compared to the scale
of variations in $p_\mathrm{eq}(x)$.  Our low-velocity analysis improves on
the work by Brazovskii, Larkin, and Nattermann (BLN)
\cite{BrazovskiiLarkin1999,BrazovskiiNattermann2004} discussing thermal
effects on the pinning of charged density waves that exhibits similar bistable
solutions as found here. In their analysis, the smooth variation in the
equilibrium distribution is ignored, what results in a different shift scale
$\ell_\mathrm{eq}(x_0)$, see Sec.~\ref{sec:low_velocities} for further
details.

The equilibrium occupation $p_\mathrm{eq}(x)$ is symmetric and thus yields no
average pinning force, allowing us to rewrite Eq.~\eqref{eq:f_pin2} as
\begin{align}\label{eq:f_pin_small_v1}
   \langle f_\mathrm{pin}\rangle &= - \frac{1}{a_0}\int_{I_\mathrm{mw}}
   d x\, (p-p_\mathrm{eq})\, \Delta f_\mathrm{pin} \\
   &\approx  \frac{1}{a_0}\int_{I_\mathrm{mw}}d x\, \ell_\mathrm{eq}(x)\, 
   p_\mathrm{eq}'(x)\, \Delta f_\mathrm{pin}. \label{eq:f_pin_small_v2}
\end{align}
Hence, the pinning force depends linearly on $v$ for small velocities.  A
detailed analysis (see Sec.~\ref{sec:low_velocities}) shows that the average
pinning-force $\langle f_\mathrm{pin}\rangle$ has a non-trivial dependence on
the transverse distance, reaching its maximum value given by
Eq.~\eqref{eq:f_pin_small_v2} at $y = 0$ and vanishing at $y = x_0$. This
results in an additional numerical prefactor $\alpha = \pi/4$ in the formula
for the average pinning force density $F_\mathrm{pin}(v,T) = \alpha n_p
(2x_0/a_0) \langle f_\mathrm{pin} \rangle$.  Carrying out the integration in
Eq.\ \eqref{eq:f_pin_small_v2} yields a quantitative result for the
pinning-force density at small velocities,
\begin{align}\label{eq:F_pin_TAFF_result}
   F_\mathrm{pin}(v,T) = h(\kappa)\, n_p a_0 \xi^2\,e^{U_0/T}\eta \, v
\end{align}
with a $\kappa$-dependent factor $h(\kappa)$. The linear dependence
$F_\mathrm{pin} \propto v$ then immediately implies an ohmic characteristic at
small drive $j \to 0$; the exponential $\propto e^{U_0/T}$ leads to the
reduced flow velocity that is at the origin of the name TAFF, thermally
assisted flux flow \cite{Kes1989}.
\begin{figure}[t]
\centering \includegraphics[scale=1]{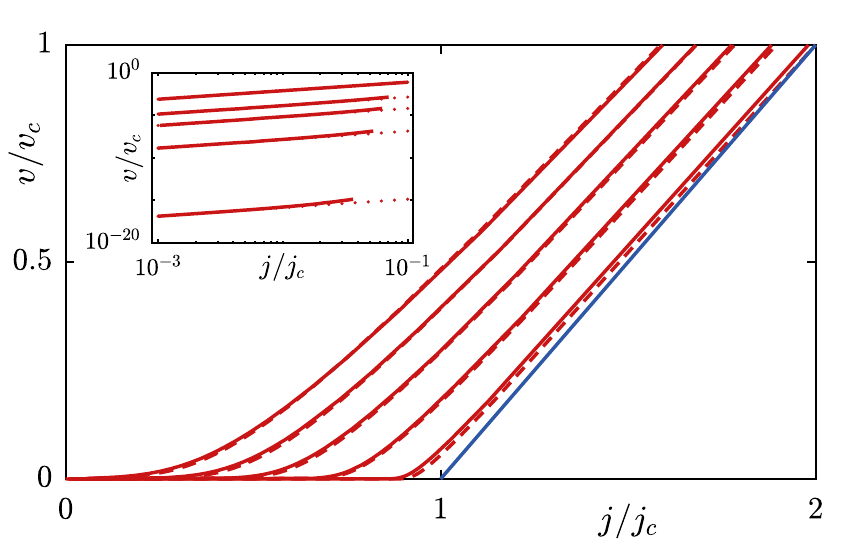}
\caption{The current--velocity characteristic derived from the dissipative
equation of motion \eqref{eq:force_balance} balancing the effect of the
driving force $F_{\rm\scriptscriptstyle L}(j)$ and the pinning-force density
$F_\mathrm{pin}(v,T)$. The right-most blue line represents the $T = 0$
excess-current characteristic with a constant pinning force density
$F_\mathrm{pin}(v,T) = F_c$. The creep characteristics (red) are shown for
temperatures ${T/e_p = (0.1,0.5,\,1,\,1.5,\,2.0)\times 10^{-2}}$, with the
left-most curve corresponding to the highest temperature. The pinning-force
densities for $T > 0$ are calculated numerically (solid lines) and the
resulting characteristics show good agreement with the analytic results
(dashed, see Eq.~\eqref{eq:F_pin_result}) in the regime of high and
intermediate velocities $v>v_{\rm \scriptscriptstyle TAFF}$. At low drives and
low velocities (logarithmic plot in the inset), the characteristics exhibit
the linear dependence on the drive $j/j_c$ as described by
Eq.~\eqref{eq:F_pin_TAFF_result}; solid and dotted lines refer to numerical
and analytic results, respectively.}\label{Fig:IV_numerics}
\end{figure}

The results outlined above can be compared with those obtained from a
numerical analysis.  Taking into account the $y$-dependence of the frequency
factors and barriers in the rate equation \eqref{eq:rate}, one can solve
numerically for the occupation probability $p(x,y)$. The average pinning-force
density then follows from a modified equation \eqref{eq:f_pin2},
\begin{align}\label{eq:F_pin_res}
	F_\mathrm{pin} = - n_p\int\frac{d xd y}{a_0^2}\,
	p(x,y)\Delta f_\mathrm{pin}(x,y)
\end{align}
with $\Delta f_\mathrm{pin}(x,y) = f_\mathrm{pin}^\mathrm{p}(x,y) -
f_\mathrm{pin}^\mathrm{f}(x,y)$ and the integration covering a unit cell of
the vortex lattice. We present the numerical results in the form of a
current--velocity characteristic in Fig.~\ref{Fig:IV_numerics}. The analytic
predictions introduced above and further elaborated in sections
\ref{sec:high_velocities} and \ref{sec:low_velocities} are in good agreement
with the numerical results: While Eq.~\eqref{eq:F_pin_result} describes the
characteristic for high and intermediate velocities $v>v_{\rm
\scriptscriptstyle TAFF}$, the linear response formula
Eq.~\eqref{eq:F_pin_TAFF_result} (including an accurate prefactor) gives a
precise result at low drives (see inset of Fig.~\ref{Fig:IV_numerics}).

\subsection{Large drives, high velocities}\label{sec:high_velocities}

In this section, we carry out the above program, that provides us with the
modification of the excess-current characteristic at large drives due to
thermal fluctuations, in particular, the thermal depinning current density
$j_\mathrm{dp}(T)$ and the barrier $U(j)$ determining the shape of the
characteristic in the vicinity of $j_\mathrm{dp}(T)$.

\subsubsection{Energy landscape}\label{subsect:pinning_landscape}

We first find the boundaries $x_{\scriptscriptstyle \pm}$ of the bistable
region where second minima appear and disappear at $r_{\scriptscriptstyle
\pm}$. The equilibrium condition \eqref{eq:equilibrium}, $\partial_r
e_\mathrm{pin}(x_{\scriptscriptstyle \pm};r)|_{r=r_{\pm}} = 0$, then has to be
simultaneously satisfied with the condition for an inflection point
$\partial^2_r e_\mathrm{pin}(x_{\scriptscriptstyle \pm};r)|_{r=r_{\pm}} = 0$,
see Fig.~\ref{Fig:metastable_states}.  The combination of these two equations
determines the end points $x_{\scriptscriptstyle \pm}$ of the multivalued
interval together with the associated tip positions $r_{\scriptscriptstyle
\pm}$ at pinning and depinning,
\begin{align}\label{eq:end_points}
   f_p'(r_{\scriptscriptstyle \pm}) = \bar{C},\qquad
   x_{\scriptscriptstyle \pm} = r_{\scriptscriptstyle \pm}
   -\frac{f_p(r_{\scriptscriptstyle \pm})}{\bar{C}}.
\end{align}
Note that the inflection points $r_{\scriptscriptstyle \pm}$ do not depend on
the asymptotic vortex position $x$ but are a property of the pinning potential
in relation to the effective elasticity.  Expanding the second equation away
from $x_{\scriptscriptstyle -}$ with $x = x_{\scriptscriptstyle -} + \delta
x_{\scriptscriptstyle -}$, we find the tip locations of the free and unstable
solutions near the onset of bistability, $r_\mathrm{f}(x) =
r_{\scriptscriptstyle -}+\delta r$ and $r_\mathrm{us}(x) =
r_{\scriptscriptstyle -}-\delta r$ with $\delta r = (\xi \delta
x_{\scriptscriptstyle -}/\kappa_{\scriptscriptstyle -})^{1/2}$ and
$\kappa_{\scriptscriptstyle -} = \xi |f_p''(r_{\scriptscriptstyle -})|/2
\bar{C}$.  Similarly, the tip locations of the pinned and unstable solutions
at $x = x_{\scriptscriptstyle +}-\delta x$ close to $x_{\scriptscriptstyle +}$
are given by $r_\mathrm{p}(x) = r_{\scriptscriptstyle +} +\delta r$ and
$r_\mathrm{us}(x) = r_{\scriptscriptstyle +} - \delta r$ with $\delta r = (\xi
\delta x_{\scriptscriptstyle +}/\kappa_{\scriptscriptstyle +})^{1/2}$ and
$\kappa_{\scriptscriptstyle +} = \xi f_p''(r_{\scriptscriptstyle
+})/2\bar{C}$. Simple estimates for $\kappa_{\scriptscriptstyle \pm}$ are (see Appendix
\ref{sect:APP_mod_strong})
\begin{align}\label{eq:x_pm_mp}
   \kappa_{\scriptscriptstyle \pm} 
   = \frac{\xi |f_p''(r_{\scriptscriptstyle \pm})|}{2\bar{C}} \sim \sqrt{\kappa - 1}
\end{align}
at marginally strong pinning (we use that $|f_p''(r_\pm)|\sim (f_p/\xi^2)|
(\kappa-1)^{1/2}$ and $f_p/\bar{C} \xi\sim \mathcal{O}(1)$) and (see
Appendix~\ref{sect:APP_very_strong})
\begin{align}\label{eq:x_pm_sp}
    \kappa_{{\scriptscriptstyle -}} \sim \kappa^{-1/(n+2)}, 
   \qquad \kappa_{\scriptscriptstyle +} \sim \kappa
\end{align}
at very strong pinning (we use that  $f_p''(r_{\scriptscriptstyle +}) \sim
f_p/ \xi^2$, $f_p/\bar{C}\xi\sim \kappa$, and $f_p''(r_{\scriptscriptstyle
-})\sim f_p/\kappa^\nu \xi^2$ with $\nu = (n+3)/(n+2)$).

Next, we discuss the frequencies $\omega_{\mathrm{p},\mathrm{f}}$ and barriers
$U_{\mathrm{dp},\mathrm{p}}$ in the rate equation \eqref{eq:rate}. The
deformation $u$ of the vortex tip extends a distance $z_\mathrm{tip}$ of order
the lattice constant $a_0$ along the $z$ direction. Given the viscosity
$\eta_l = \eta a_0^2$ for the motion of an individual vortex line, we
approximate the tip motion by the dissipative dynamics of a particle with a
friction coefficient $\eta_l z_\mathrm{tip} = \eta a_0^3$ in the effective
potential $e_\mathrm{pin}(x;r = x+u)$, $\eta a_0^3 \dot{u} = -\partial_u
e_\mathrm{pin}(x;x+u)$; the attempt frequencies then are given by the
expressions \cite{Kramers1940}
\begin{align}\label{eq:omega}
   \omega_\mathrm{p}(x) = \frac{\sqrt{\lambda_\mathrm{p}|\lambda_\mathrm{us}|}}
   {2\pi\eta a_0^3}, \qquad \omega_\mathrm{f}(x) = \frac{\sqrt{\lambda_\mathrm{f}
   |\lambda_\mathrm{us}|}}{2\pi\eta a_0^3},
\end{align}
where the curvatures $\lambda_i$, $i = \mathrm{p},\,\mathrm{f},\,\mathrm{us}$,
are to be evaluated at the local minima and at the maximum of the pinning
energy $e_\mathrm{pin}(x;r)$,
\begin{align}\label{eq:lambda}
   \lambda_i(x) = \partial^2_r e_\mathrm{pin}(x;r)\Big|_{r = r_i(x)} 
   \!\!\!\! = \bar{C}-f_p'[r_i(x)].
\end{align}
Since the curvatures $\partial^2_r e_\mathrm{pin}(x;r)$ vanish at the
inflection points $r_i(x_{\scriptscriptstyle \pm})$, we have
$\lambda_{\mathrm{p},\mathrm{us}}(x_{\scriptscriptstyle +}) =
\lambda_{\mathrm{f},\mathrm{us}}(x_{\scriptscriptstyle -}) = 0$.  Close to the
boundaries of the bistable regime, we obtain the expansions
\begin{align}\label{eq:lambda_onset_gen}
\begin{split}
   \lambda_\mathrm{f,us}(x_{\scriptscriptstyle -}\!\!+\delta x_{\scriptscriptstyle -})
   &= \pm 2\bar{C} (\kappa_{\scriptscriptstyle -}\, \delta x_{\scriptscriptstyle-}/\xi)^{1/2},\\
   \lambda_\mathrm{p,us}(x_{\scriptscriptstyle +}\!\!-\delta x_{\scriptscriptstyle +})
   &= \pm 2\bar{C} (\kappa_{\scriptscriptstyle +}\, \delta x_{\scriptscriptstyle +}/\xi)^{1/2}.
   \end{split}
\end{align}
Here, the $\pm$ signs refer to the free/pinned and unstable branches.  Simple
estimates for the attempt frequencies then are (we remind that $v_p = f_p
/\eta a_0^3$ provides the velocity scale for dissipative motion in the pinning
potential)
\begin{align}\label{eq:om_pf_mp}
   \omega_{\mathrm{p},\mathrm{f}}(\delta x_{\scriptscriptstyle \pm}) 
   \sim (v_p/\xi) (\kappa - 1)^{1/4} \sqrt{\delta x_{\scriptscriptstyle \pm}/\xi}
\end{align}
at marginally strong pinning and
\begin{align}\label{eq:om_p_sp}
   \omega_\mathrm{p}(\delta x_{\scriptscriptstyle +})
   &\sim (v_p/\xi) \sqrt{\delta x_{\scriptscriptstyle +}/\kappa\xi}, \\
   \label{eq:om_f_sp}
   \omega_\mathrm{f}(\delta x_{\scriptscriptstyle -})
   &\sim (v_p/\xi) \, \kappa^{-\nu/2} \sqrt{\delta x_{\scriptscriptstyle -}/\kappa\xi}
\end{align}
at very strong pinning (we remind that $\nu = (n+3)/(n+2)$).

Similarly, one finds for the onset of the barriers $U_\mathrm{p}(x) =
e_\mathrm{pin}(x; r_\mathrm{us}(x)) - e_\mathrm{pin}(x; r_\mathrm{f}(x))$ and
$U_\mathrm{dp}(x)$
\begin{align}\label{eq:U_onset_gen}
   &U_\mathrm{p}(x_{\scriptscriptstyle -}+\delta x_{\scriptscriptstyle -}) 
   = \frac{4 \bar{C}\xi^2}{3\sqrt{\kappa_{\scriptscriptstyle -}}}
   (\delta x_{\scriptscriptstyle -}/\xi)^{3/2},\\
   &U_\mathrm{dp}(x_{\scriptscriptstyle +}-\delta x_{\scriptscriptstyle +}) 
   = \frac{4 \bar{C}\xi^2}{3\sqrt{\kappa_{\scriptscriptstyle +}}}
   (\delta x_{\scriptscriptstyle +}/\xi)^{3/2}.
   \label{eq:U_onset_gen2}
\end{align}
For marginally strong pinning, the interval of bistability shrinks as $\propto
(\kappa-1)^{3/2} \xi$, see Eq.\ \eqref{eq:xpm_mp}, and the combination with
the factor $1/\sqrt{\kappa_{\scriptscriptstyle\pm}}$ produces barriers of size 
\begin{align}\label{eq:U_mp}
   U_{\mathrm{dp,p}} \sim e_p (\kappa-1)^2
\end{align}
(note that $\bar{C}\xi^2 \sim e_p/\kappa$). For very strong pinning, we find
the bistability extending over the region $x_{\scriptscriptstyle +} -
x_{\scriptscriptstyle -} \sim \kappa\xi$. For the depinning and pinning
barriers, we obtain
\begin{align}\label{eq:U_sp}
   U_\mathrm{p} \sim e_p \kappa^{\nu/2} (\delta x_{\scriptscriptstyle -}/ 
   \kappa \xi)^{3/2}, \quad
   U_\mathrm{dp} \sim e_p (\delta x_{\scriptscriptstyle +} /\kappa \xi)^{3/2}.
\end{align}

\begin{figure}
\centering \includegraphics[width=8truecm]{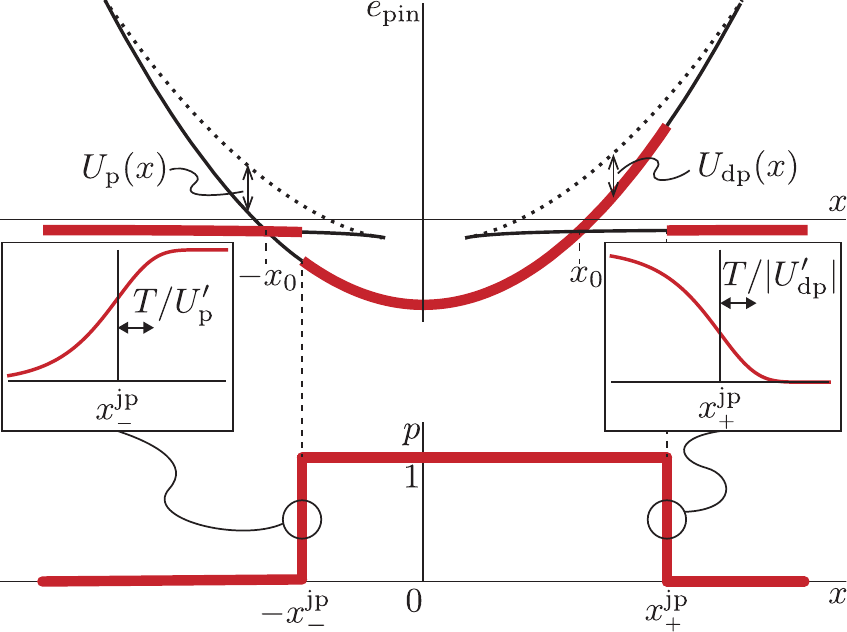}
\caption{Occupied branches (thick red) in the large velocity limit. The
pinned-branch occupation $p(x)$ is approximated by the top-hat function
(bottom) with two jump points at $-x^\mathrm{jp}_{\scriptscriptstyle -}(v,T)$
and $x^\mathrm{jp}_{\scriptscriptstyle +}(v,T)$. The insets show the
variations of $p(x)$ near the jump points occurring on scales
$T/|U_\mathrm{p}'(-x^\mathrm{jp}_{\scriptscriptstyle -})|$ and
$T/|U_\mathrm{dp}'(x^\mathrm{jp}_{\scriptscriptstyle +})|$ that are small
compared to the spatial dimensions of the energy landscape. }
\label{Fig:occupation_high_v}
\end{figure}

The expansions (\ref{eq:lambda_onset_gen})--(\ref{eq:U_onset_gen2}) break down
for $x$ close to the branch crossing point at $x_0$. However, one can still
show that the depinning barrier $U_\mathrm{dp}$ decreases monotonically in the
interval $[x_{\scriptscriptstyle -},x_{\scriptscriptstyle +}]$. Indeed,
$U'_\mathrm{dp}(x) = f_\mathrm{pin}^\mathrm{p}(x) -
f_\mathrm{pin}^\mathrm{us}(x) = \bar{C}(r_\mathrm{p}-r_\mathrm{us})<0$, see
Fig.~\ref{Fig:metastable_states}. The same way one shows that the pinning
barrier $U_\mathrm{p}$ is monotonically increasing.

\subsubsection{Solution of the rate equation}

The first-order differential rate equation \eqref{eq:rate} for the occupation
probability $p(x)$ gives rise to two initial-value problems, to be solved
separately in the multivalued intervals $[-x_{\scriptscriptstyle
+},-x_{\scriptscriptstyle -}]$ and $[x_{\scriptscriptstyle
-},x_{\scriptscriptstyle +}]$. Within the first interval, the initial
condition is $p(-x_{\scriptscriptstyle +}) = 0$ (as the pinned branch only
starts at $-x_{\scriptscriptstyle +}$), while for the second,
$p(x_{\scriptscriptstyle -}) = 1$ (as there is only a pinned branch just
before reaching $x_{\scriptscriptstyle -}$). Note that, in principle, the
solution can be discontinuous at the right end-points $-x_{\scriptscriptstyle
-}$ and $x_{\scriptscriptstyle +}$ of the bistable intervals.  Indeed, this is
the case for the $T=0$ critical state occupation $p_\mathrm{c}(x)$.

Focusing on the interval $[x_{\scriptscriptstyle -}, x_{\scriptscriptstyle
+}]$, we assume a free branch with an exponentially small occupation $1-p$ and
neglect transitions from this branch as described by the second term on the
right side of \eqref{eq:rate}.  The rate equation is then rewritten through
the depinning relaxation length, see Eq.\ \eqref{eq:ell_dp}, as
\begin{align}\label{eq:rate_large_v}
   \partial_x p = -p/\ell_\mathrm{dp}(x).
\end{align}
We define the jump point $x^\mathrm{jp}_{{\scriptscriptstyle +}}$ through the
condition $\partial_x^2 p|_{x^\mathrm{jp}_{\scriptscriptstyle +}} = 0$, i.e.,
as the inflection point of $p(x)$ or equivalently the point with the steepest
rate of change of the occupation of the pinned branch.  Taking the derivative
of Eq.~\eqref{eq:rate_large_v} with respect to $x$, we obtain the condition
$\partial_x^2 p(x) = [p/\ell_\mathrm{dp}(x)^2][1+\ell_\mathrm{dp}'(x)]$ and
the inflection point satisfies the relation $\ell_\mathrm{dp}'(x) = -1$.
Evaluating $\ell_\mathrm{dp}'(x)$ with the help of Eq.~\eqref{eq:ell_dp}, we
find that
\begin{align}\label{ell_dp_der}
   \ell_\mathrm{dp}'(x) = \Bigl[\frac{U_\mathrm{dp}'(x)}{T}-\frac{\omega_\mathrm{p}'(x)}
   {\omega_\mathrm{p}(x)}\Bigr]\ell_\mathrm{dp}(x).
\end{align}
At small temperatures $T\ll U_\mathrm{dp}(x_{\scriptscriptstyle
+}^\mathrm{jp})$ the second term can be dropped and we obtain the condition
for depinning in the form $\ell_\mathrm{dp}(x_{\scriptscriptstyle
+}^\mathrm{jp}) \approx T/|U_\mathrm{dp}'(x_{\scriptscriptstyle
+}^\mathrm{jp})|$. Finally, substituting back the definition
Eq.~\eqref{eq:ell_dp} of $\ell_\mathrm{dp}$ provides us with the condition
\begin{align}\label{eq:jump}
    v = \frac{\omega_\mathrm{p}(x^\mathrm{jp}_{\scriptscriptstyle +}) T}
    {|U_\mathrm{dp}'(x^\mathrm{jp}_{\scriptscriptstyle +})|}
   \exp\Bigl[-\frac{U_\mathrm{dp}(x^\mathrm{jp}_{\scriptscriptstyle +})}{T}\Bigr].
\end{align}
Eq.\ \eqref{eq:jump} is our quantitative condition determining the jump point
$x^\mathrm{jp}_{\scriptscriptstyle +}(v,T)$ out of the pin, with the
exponential providing the dominant factor. Since the barrier should be large
as compared to the temperature $T$ in order to validate Kramer's rate theory,
the above results apply for $x_{\scriptscriptstyle +}^\mathrm{jp}$ not too
close to $x_{\scriptscriptstyle +}$, i.e., $x_{\scriptscriptstyle +} -
x_{\scriptscriptstyle +}^\mathrm{jp} \gg (T/e_p)^{2/3} \kappa\xi$ at strong
pinning, see Eq.\ \eqref{eq:U_sp}.

With increasing velocity $v$, the barrier $U_\mathrm{dp}$ decreases and the
jump point approaches $x_{\scriptscriptstyle +}$ where vortices depin without
activation. The velocity $v$ for activated motion then is restricted by the
thermal velocity $v_\mathrm{th}$ that follows from Eq.\ (\ref{eq:jump}) in the
limit $x^\mathrm{jp}_{\scriptscriptstyle +} \to x_{\scriptscriptstyle +}$
where the barrier $U_\mathrm{dp} (x^\mathrm{jp}_{\scriptscriptstyle +})$
vanishes.  Using the expansions \eqref{eq:lambda_onset_gen} and
\eqref{eq:U_onset_gen2} for the barrier and the frequency factor
$\omega_\mathrm{p}$ near the point $x_{\scriptscriptstyle +}$, we find
\begin{align}\label{eq:v_crit}
   v_\mathrm{th} = \lim\limits_{x^\mathrm{jp}_{\scriptscriptstyle +}
   \to x_{\scriptscriptstyle +}} \frac{\omega_\mathrm{p}
   (x^\mathrm{jp}_{\scriptscriptstyle +}) T}
   {|U_\mathrm{dp}'(x^\mathrm{jp}_{\scriptscriptstyle +})|}  
   &=  \frac{Tf_p''(r_{\scriptscriptstyle +})}{4\pi \bar{C}\eta a_0^3}\\
   \nonumber
    &=  \frac{\kappa_{\scriptscriptstyle +}}{2\pi}\frac{T}{\eta a_0^3\xi}
   \sim \frac{T}{e_p}\kappa_{\scriptscriptstyle +} v_p,
\end{align}
where $v_p\sim f_p/\eta a_0^3$ is the velocity scale of dissipative motion in
the well above which dynamical effects become relevant in the depinning
process, see Ref.\ [\onlinecite{Thomann2012, Thomann2017}].  The thermal velocity
$v_{\mathrm{th}}$ separates two regimes, the small velocity regime $v <
v_{\mathrm{th}}$ where barriers are finite and creep is relevant, and the high
velocity region where the occupation $p(x)$ is given by the critical one,
$p(x) \approx p_c(x)$, and the pinning-force density is approximately given by
the critical value $F_c$, Eq.~\eqref{eq:F_crit}, as long as $v\ll v_p$. The
results in Refs.~\cite{Thomann2012, Thomann2017} describe the dynamical
situation at high velocities of order $v_p$ and beyond.

Finally, we can use the result for the thermal velocity $v_\mathrm{th}$ and
rewrite the jump condition Eq.\ \eqref{eq:jump} in the form
\begin{align}\label{eq:jump_simple}
   U_\mathrm{dp}(x_{\scriptscriptstyle +}^\mathrm{jp}) 
   \approx T\,\ln\frac{v_\mathrm{th}}{v}.
\end{align}
As the velocity $v$ decreases far below $v_\mathrm{th}$, the quantity
$\omega_\mathrm{p} T/|U_\mathrm{dp}'|$ in Eq.\ \eqref{eq:jump_simple} will
deviate from its value at $v_\mathrm{th}$, resulting in logarithmic
corrections which we neglect in comparison with the large ratio
$U_\mathrm{dp}(x_{\scriptscriptstyle +}^\mathrm{jp})/T$.

An analogous consideration applies to the interval $[-x_{\scriptscriptstyle
+},-x_{\scriptscriptstyle-}]$ and provides us with the condition for the
pinning barrier $U_\mathrm{p}(-x^\mathrm{jp}_{\scriptscriptstyle -})$
determining the jump location $-x^\mathrm{jp}_{\scriptscriptstyle -} > -x_0$
where transitions from the free to the pinned branch start to become
energetically favorable,
\begin{align}
   U_\mathrm{p}(-x^\mathrm{jp}_{\scriptscriptstyle -}) 
   = T\,\ln\frac{v_\mathrm{th}^-}{v},\quad v^-_\mathrm{th}
   = \frac{\kappa_{\scriptscriptstyle -}}{2\pi}\frac{T}{\eta a_0^3\xi}
   \sim \frac{T}{e_p}\kappa_{\scriptscriptstyle -} v_p.
\end{align}
The ratio of velocity scales $v_\mathrm{th}^{{\scriptscriptstyle
-}}/v_\mathrm{th} = \kappa_{\scriptscriptstyle -}/\kappa_{\scriptscriptstyle
+}$ is of order unity at marginally strong pinning and decays as
$\kappa^{-\nu}$ at very strong pinning. The activation barriers are thus
approximately related by
\begin{align}
   U_\mathrm{p}(-x^\mathrm{jp}_{\scriptscriptstyle -})
   \approx U_\mathrm{dp}(x^\mathrm{jp}_{\scriptscriptstyle +})
   -\nu T\ln\kappa.
\end{align}
In the following, we neglect the small difference between the effective
pinning and depinning barriers as both of them are supposed to be large
compared to $T$.  The conditions fixing the two jump points
$x^\mathrm{jp}_{\scriptscriptstyle \pm}$ as a function of $v$ and $T$ then can
be written in the simple form
\begin{align}\label{eq:jump_U}
   U_\mathrm{dp}(x^\mathrm{jp}_{{\scriptscriptstyle +}}) \approx 
   U_\mathrm{p}(-x^\mathrm{jp}_{{\scriptscriptstyle -}}) \approx U(v,T) \equiv 
   T\ln \frac{v_\mathrm{th}}{v}.
\end{align}

Next, we integrate Eq.~\eqref{eq:rate_large_v} with the boundary condition
$p(x_{\scriptscriptstyle -}) = 1$ in order to find the full functional
solution $p(x)$ of the rate equation inside the interval
$[x_{\scriptscriptstyle -},x_{\scriptscriptstyle +}]$,
\begin{align}\label{eq:rate_sol_large_v}
   p(x) &= \exp\biggl[-\frac{1}{v}\int_{x_{\scriptscriptstyle -}}^xd x'\,
   \omega_\mathrm{p}(x')\, e^{-U_\mathrm{dp}(x')/T}\biggr].
\end{align}
The depinning barrier $U_\mathrm{dp}(x')$ decreases with $x'$ such that at low
temperatures the integral is dominated by its contributions close to the upper
limit, while the lower limit $x_{\scriptscriptstyle -}$ is irrelevant. The
factor $e^{-U_\mathrm{dp}/T}$ entering the depinning distance
$\ell_\mathrm{dp}$ changes on the scale $T/|U'_\mathrm{dp}|$, while the change
in the frequency $\omega_\mathrm{p}$ is negligible on this scale.   Expanding
$U_\mathrm{dp}(x')$ to linear order near the upper boundary $x$ of the
integral and neglecting variations of $\omega_\mathrm{p}(x')$ then gives
\begin{align}\label{eq:p_large_v}
   p(x) &\approx \exp\biggl[-\frac{\omega_\mathrm{p}(x)}{v}
   \int_{-\infty}^x \!\!\!\!\!
   d x'\, e^{-[U_\mathrm{dp}(x)+U_\mathrm{dp}'(x) (x'-x)]/T}\biggr]\nonumber\\
   &= \exp\biggl[-\frac{\omega_\mathrm{p}(x) T}{v\, |U'_\mathrm{dp}(x)|}
   e^{-U_\mathrm{dp}(x)/T}\biggr].
\end{align}
Expanding around the jump point $x^\mathrm{jp}_{\scriptscriptstyle +}$ as
defined by the Eq.\ \eqref{eq:jump} and neglecting the changes of
$\omega_\mathrm{p}(x)$ and $|U_\mathrm{dp}'(x)|$ on the scale
$T/|U_\mathrm{dp}'(x^\mathrm{jp}_{\scriptscriptstyle +})|$, we find the
expansion of $p(x)$ near the jump point,
\begin{align}\label{eq:p_around_jump}
   p(x^\mathrm{jp}_{\scriptscriptstyle +}+\delta x)\approx
   \exp\left[-e^{|U_\mathrm{dp}'(x_{\scriptscriptstyle +}^\mathrm{jp})|\delta x/T}\right].
\end{align}
Indeed, the transition from $p(x) = 1$ to $p(x) = 0$ at $x_{\scriptscriptstyle
+}^\mathrm{jp}$ is realized on a scale
$\ell_\mathrm{dp}(x^\mathrm{jp}_{\scriptscriptstyle +}) =
T/|U_\mathrm{dp}'(x^\mathrm{jp}_{\scriptscriptstyle +})| \sim
[T/e_p]\,\kappa\xi$, that is small compared to the range $\kappa \xi$ of the
pinning landscape, see the inset of Fig.\ \ref{Fig:occupation_high_v}.

\subsubsection{Pinning force}\label{subsubsect:large_drives_pinning_force}

Given the thermally renormalized jumps at $\pm x_{\scriptscriptstyle
\pm}^\mathrm{jp}$, the average pinning force $\langle
f_\mathrm{pin}(v,T)\rangle$ acting on vortices straightly impacting on a
defect reads (see Eq.\ \eqref{eq:f_pin_av})
\begin{align}
   \langle f_\mathrm{pin}(v,T)\rangle = \frac{\Delta e_\mathrm{pin}^{\mathrm{tot}}(v,T)}{a_0},
\end{align}
with $\Delta e_\mathrm{pin}^\mathrm{tot}(v,T)=\Delta
e_\mathrm{pin}(x^\mathrm{jp}_{\scriptscriptstyle +}) -\Delta
e_\mathrm{pin}(-x^\mathrm{jp}_{\scriptscriptstyle -})>0$.  Again, we have to
generalize this result to the situation where vortices approach the defect at
arbitrary transverse distance $y$. Assuming a radially symmetric defect
potential, the same arguments can be made as for the $T=0$ situation, but with
the jump from the free to the pinned branch now determined by the condition
$U_\mathrm{p}(x,y) = U_\mathrm{p}[(x^2 + y^2)^{1/2},0] = U(v,T)$. The
condition for vortex pinning thus becomes $(x^2+y^2)^{1/2} =
x^\mathrm{jp}_{\scriptscriptstyle -}$, implying that vortices approaching the
pin at a tranverse distance $y< t_\perp = x^\mathrm{jp}_{\scriptscriptstyle
-}$ get trapped (note that the transverse trapping length is enhanced compared
to the $T = 0$ case). As a result, we find the finite-temperature
pinning-force density to be given by
\begin{align*}
   F_\mathrm{pin}(v,T) = n_p\frac{2 x^\mathrm{jp}_{\scriptscriptstyle -}}{a_0}
   \frac{\Delta e_\mathrm{pin}^\mathrm{tot}(v,T)}{a_0}.
\end{align*}

Below, we will make strong use of the scaled pinning-force density
\begin{align}\label{eq:force_ratio}
   \frac{F_\mathrm{pin}(v,T)}{F_c} 
   = \frac{x_{\scriptscriptstyle -}^\mathrm{jp}(v,T)}{x_{\scriptscriptstyle -}}
   \frac{\Delta e_\mathrm{pin}^\mathrm{tot}(v,T)}{\Delta e_c}
\end{align}
that depends only on the rescaled barrier $U(v,T)/e_p$: indeed, the activation
barrier $U(v,T)$ suffices to determine the position of the jumps $\pm
x^\mathrm{jp}_{\scriptscriptstyle \pm}$ as well as the magnitude of the jumps
in energy. Let us analyze this force ratio as a function of velocity.

At marginally high velocities close to $v_{\rm\scriptscriptstyle TAFF} =
v_\mathrm{th} e^{-U_0/T}$, vortices probe barriers close to the maximum
activation barrier $U_0$ and the jumps $x^\mathrm{jp}_{\scriptscriptstyle
\pm}$ in the energy landscape are realized close to the branch crossing point
$x_0$. At velocities beyond $v_{\rm\scriptscriptstyle TAFF}$, the barriers and
energy jumps scale linearly in the differences $\delta x_0 =
x^\mathrm{jp}_{\scriptscriptstyle +} - x_0$ (and
$-x^\mathrm{jp}_{\scriptscriptstyle -} + x_0$), resulting in a force ratio
that is linear in the activation barrier $U$ and that vanishes for $U = U_0$,
\begin{align}\label{eq:F_pin_interm_v}
  \frac{F_\mathrm{pin}(v,T)}{F_c} \! \approx \! 
   \varphi(\kappa)\frac{U_0-U(v,T)}{e_p} \!= \!\varphi(\kappa)\frac{T}{e_p}
   \ln\frac{v}{v_{\rm\scriptscriptstyle TAFF}}.
\end{align}
The exact expression for the slope $\varphi(\kappa)$ is given in Appendix
\ref{sect:APP_scal_fun}. The function $\varphi(\kappa)$ scales as $\sim
(\kappa-1)^{-2}$ for marginally strong pinning $\kappa\to 1$ and decays as
$\propto \kappa^{-{\nu'}}$ with the power $\nu' = (3n+4) / 2(n+1)(n+2)$ at
large $\kappa$, hence the function $\tilde{\varphi}(\kappa) = \varphi(\kappa)
(\kappa-1)^2  \kappa^{{\nu'}-2}$ is a slowly varying function in $\kappa$
ranging between $\tilde\varphi(\infty)\approx 3.1$ and
$\tilde\varphi(1)\approx 5.3$, see Fig.~\ref{Fig:scaling_functions}.

\begin{figure}
\centering
\includegraphics[width=8cm]{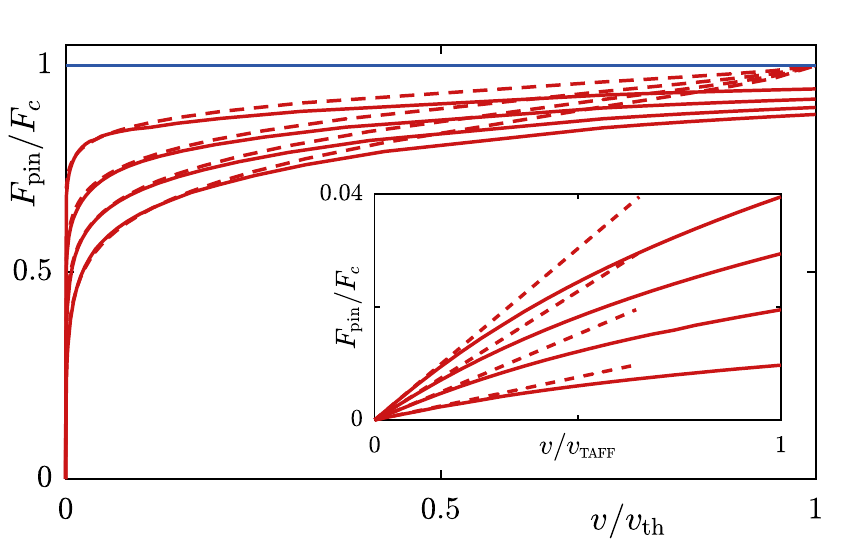}
\caption{Pinning-force density $F_\mathrm{pin}(v,T)$ calculated for the
Lorentzian pinning potential and a Labusch parameter $\kappa = 5$ at
temperatures $T/e_p = (0,\,0.5,\,1,\,1.5,\,2.0)\times 10^{-2}$ (top to bottom,
solid blue line corresponds to $T = 0$). The numerical result (solid line) as
obtained from integrating Eq.\ \eqref{eq:F_pin_res} is compared with the
analytical formula (dashed line) for high velocities,
Eq.~\eqref{eq:F_pin_expansion}, and low velocities (inset,
Eq.~\eqref{eq:F_pin_small_v}). The pinning-force density is reduced compared
to the critical force density $F_c$ due to thermal creep at velocities
$v<v_\mathrm{th}$.  While numerical and analytic results agree qualitatively,
they differ quantitatively close to $v_\mathrm{th}$. The inset shows the
crossover to the linear TAFF response relevant for small velocities
$v<v_{\scriptscriptstyle \mathrm{TAFF}}$; here the highest temperature
corresponds to the upper-most curve, case $T = 0$ is not shown.}
\label{Fig:vF_comparison}
\end{figure}

At large velocities $v \lesssim v_\mathrm{th}$, the departure from the
critical force $F_c$ is non-linear in $(T/e_p)\ln (v/v_\mathrm{th})$, a result
that is due to the non-linear scaling of the activation barrier with distance
away from the critical jumps at $x_{\scriptscriptstyle \pm}$, see Eq.\
\eqref{eq:U_onset_gen}.  Given the jumps at $x^\mathrm{jp}_{\scriptscriptstyle
+} = x_{\scriptscriptstyle +} - \delta x_{\scriptscriptstyle +}$ and
$x^\mathrm{jp}_{\scriptscriptstyle -} = -x_{\scriptscriptstyle -} - \delta
x_{\scriptscriptstyle -}$, we expand the total jump in energy as given by Eq.\
\eqref{eq:Delta_e_pin_tot} in the small quantities $\delta
x_{\scriptscriptstyle +}$ and $\delta x_{\scriptscriptstyle -}$,
\begin{align}\nonumber
   \Delta e_\mathrm{pin}^\mathrm{tot}(v,T) - \Delta e_c 
   = \Delta f_\mathrm{pin}(x_{\scriptscriptstyle +})
   \delta x_{\scriptscriptstyle +}\! -\Delta f_\mathrm{pin}(-x_{\scriptscriptstyle -})
   \delta x_{\scriptscriptstyle -},
\end{align}
where $\Delta f_\mathrm{pin}(x_{\scriptscriptstyle +})<0$ and $\Delta
f_\mathrm{pin}(-x_{\scriptscriptstyle -})=-\Delta f_\mathrm{pin}
(x_{\scriptscriptstyle -})>0$. The reduced jump in energy implies a reduction
in $\langle f_\mathrm{pin}(v,T)\rangle$ as compared to its $T=0$ value. On the
other hand, the trapping distance $x^\mathrm{jp}_{\scriptscriptstyle -}$ is
larger by $\delta x_{\scriptscriptstyle-}$ as compared to
$x_{\scriptscriptstyle -}$, hence, more vortices are trapped at $T>0$. These
two effects compete as expressed in the expansion of the force ratio Eq.\
\eqref{eq:force_ratio}, to linear order,
\begin{align}\label{eq:F_pin/F_c1}
   \frac{F_\mathrm{pin}}{F_c} &\approx 1\!
   +\!\frac{\delta x_{\scriptscriptstyle -}}{x_{\scriptscriptstyle -}}
   \!+\!\frac{\Delta f_\mathrm{pin}(x_{\scriptscriptstyle +})
   \delta x_{\scriptscriptstyle +}\! + \!\Delta f_\mathrm{pin}(x_{\scriptscriptstyle -})
   \delta x_{\scriptscriptstyle -}}{\Delta e_c},
\end{align}
where the first correction arises from the relative change in the trapping
distance $t_\perp$, see Fig.\ \ref{Fig:energy_landscape_transverse}, while the
second correction is the relative change in the pinning force exerted on the
vortex.

We use Eqs.\ \eqref{eq:U_onset_gen}--\eqref{eq:U_onset_gen2} to express
$\delta x_{\scriptscriptstyle \pm}$ through the activation barrier $U(v,T)$,
$\delta x_{\scriptscriptstyle \pm} = \xi (3 U \sqrt{\kappa_{\scriptscriptstyle
\pm}}/4 \bar{C} \xi^2)^{2/3}$, and arrive at the expansion
\begin{align}\label{eq:F_pin_expansion_barrier}
   \frac{F_\mathrm{pin}(v,T)}{F_c} &\approx 1-g(\kappa)[U(v,T)/e_p]^{2/3}\\
    &= 1 - g(\kappa)\Bigl(\frac{T}{e_p}\ln\frac{v_\mathrm{th}}{v}\Bigr)^{2/3},   
    \label{eq:F_pin_expansion}
\end{align}
with the coefficient
\begin{align}\label{eq:g}
   g(\kappa) &= - \frac{\xi \bar\kappa_{\scriptscriptstyle -}}{x_{\scriptscriptstyle -}}
   + \frac{\xi}{\Delta e_c}
   \bigl|\Delta f_\mathrm{pin}(x_{\scriptscriptstyle +}) \, 
   \bar\kappa_{\scriptscriptstyle +} \\
   \nonumber
   &\qquad\qquad\qquad 
   + \Delta f_\mathrm{pin}(x_{\scriptscriptstyle -}) \, 
   \bar\kappa_{\scriptscriptstyle -}\bigr|,
\end{align}
where $\bar\kappa_{\scriptscriptstyle \pm} = [3 e_p
\sqrt{\kappa_{\scriptscriptstyle \pm}}/4\bar{C}\xi^2]^{2/3}$; this factor is
of order $\sim (\kappa-1)^{1/6}$ for marginally strong pinning and
$\bar\kappa_{\scriptscriptstyle +} \sim \kappa$,
$\bar\kappa_{\scriptscriptstyle -} \sim \kappa^{(2n+3)/3(n+2)}$ for very
strong pinning.

The second term $\propto \xi/\Delta e_c$ in Eq.~\eqref{eq:g} is always
positive and dominates at marginally strong pinning, where $|\Delta
f_\mathrm{pin}(x_{\scriptscriptstyle \pm})| \sim (e_p/\xi)(\kappa-1)^{1/2}$
and $\Delta e_c \sim (\kappa-1)^2 e_p$, hence $g_{\scriptscriptstyle
+}(\kappa)\sim (\kappa-1)^{-4/3}$, see Appendix \ref{sect:APP_scal_fun}.  The
pinning force density $F_\mathrm{pin}(v,T)$ then is reduced compared to the
critical force density $F_c$ for sufficiently small $\kappa$. For very strong
pinning, the second term is of order $\kappa^0$, hence the function
$\tilde{g}(\kappa)= g(\kappa) (\kappa-1)^{4/3} \kappa^{-4/3}\approx 3.6$ is
slowly varying in this regime, see Fig.~\ref{Fig:scaling_functions}.  On the
other hand, at ultra-strong pinning, the first term originating from the
enhanced trapping distance is of order $\sim -\kappa^{2n/3(n+2)}$ and
eventually dominates over the positive second term, see Appendix
\ref{sect:APP_very_strong}. The function $g(\kappa)$ then turns negative for
$\kappa>\kappa_0$ and grows in magnitude with a power of $\kappa$, implying a
creep-enhanced pinning-force density beyond $F_c$, $F_\mathrm{pin} > F_c$, see
Eq.\ \eqref{eq:F_pin_expansion_barrier}.  However, the crossover value
$\kappa_0$ where $g$ turns negative is large and hardly accessible in a real
material: for a Lorentzian potential, we find that ${\kappa_0\approx 150}$,
while for the exponentially decaying pinning potential $e_p(r)=e_p/\cosh
(r/\xi)$, the crossover value is reduced but still large, $\kappa_0\approx
37$.

In Fig. \ref{Fig:vF_comparison}, we compare the numerical results derived from
integrating Eq.\ \eqref{eq:F_pin_res} for the pinning-force density with those
obtained from the analytic expression Eq.~\eqref{eq:F_pin_expansion}.  The
analytic solution predicts a somewhat larger pinning-force density at large
velocities $v\sim v_\mathrm{th}$.  This enhancement originates from assuming
sharp jumps in the branch occupation $p(x)$ at the points $\pm
x_{\scriptscriptstyle \pm}^\mathrm{jp}$ that is no longer valid at large
velocities $v\sim v_\mathrm{th}$: indeed for the depinning point, we have
$x_{\scriptscriptstyle +}^\mathrm{jp}\to x_{\scriptscriptstyle +}$ for $v\to
v_\mathrm{th}$ and the width $T/|U_\mathrm{dp}'
(x^\mathrm{jp}_{\scriptscriptstyle +})|\propto (x_{\scriptscriptstyle
+}-x_{\scriptscriptstyle +}^\mathrm{jp})^{-1/2}$ of the jump in
Eq.~\eqref{eq:p_around_jump} diverges.  In contrast to the analytic solution
assuming a fully occupied pinned branch up to $x_{\scriptscriptstyle +}$, the
regions of the pinned branch close to $x_{\scriptscriptstyle +}$ responsible
for large pinning forces become only partially occupied. The numerical
solution taking into account this partial occupation then predicts a smaller
pinning force.

Neither the analytic nor the numerical result is expected to be quantitatively
accurate for $v\sim v_\mathrm{th}$, as the assumption of large activation
barriers $U\gg T$ required by Kramer's rate theory is no longer satisfied.
Nevertheless, in the wide and important region of intermediate velocities
$v_{\rm\scriptscriptstyle TAFF}\ll v\ll v_\mathrm{th}$ where barriers are
large, both analytical and numerical results are reliable and show good
agreement.  As the velocity decreases, the pinning-force density
\eqref{eq:F_pin_expansion} valid at large drives where barriers scale $\propto
\delta x_{\scriptscriptstyle +}^{3/2}$ has to be replaced by
\eqref{eq:F_pin_interm_v} (where barriers scale linearly in $\delta x_0$); at
very small velocities $v < v_{\rm\scriptscriptstyle TAFF}$, the barrier
saturates at $x_0$ and we enter the linear response regime discussed in
Sect.~\ref{sec:low_velocities}.

\subsubsection{Current--velocity characteristic}\label{sec:IVchar}

Applying the equation of motion \eqref{eq:force_balance} and the results for
the force density ratio $F_\mathrm{pin}/F_c$ provides us with the
current--velocity characteristic of the superconductor. We first consider
velocities, $v_{\rm \scriptscriptstyle TAFF}\ll v < v_\mathrm{th}$ where the
expression \eqref{eq:F_pin_expansion} for the pinning force ratio holds and
assume a regular situation with $g(\kappa) > 0$, see Sec.\
\ref{sec:low_velocities} for a discussion of small velocities $v \ll v_{\rm
\scriptscriptstyle TAFF}$. Then the scaled equation of motion, to be solved
for the velocity $v$ at given drive $j$, takes the form
\begin{align}\label{eq:motion_expanded}
   \frac{v}{v_c} = \frac{j}{j_c}-1+
   g(\kappa)\left[\frac{T}{e_p}\ln\frac{v_\mathrm{th}}{v} \right]^{2/3}.
\end{align}
This expression is conveniently rewritten into the form
\begin{align}\label{eq:motion_expanded_2}
\frac{v}{v_\mathrm{th}} = \frac{1}{\mathcal{A}} \frac{\delta j}{j_c}
   +\frac{1}{\nu}
   \Bigl[\ln\Bigl(\frac{v_\mathrm{th}}{v}\Bigr) \Bigr]^{2/3}
\end{align}
with $\delta j = j-j_c$, $\mathcal{A}$ the ratio of thermal- and free-flow
velocities
\begin{align}\label{eq:v_th_v_c}
   \mathcal{A} \equiv \frac{v_\mathrm{th}}{v_c} 
   &= \frac{\kappa_{\scriptscriptstyle +}}{2\pi}\frac{T}{F_c a_0^2\xi} 
   = \frac{\kappa_{\scriptscriptstyle +}}{4\pi}\frac{T}{\Delta e_c}
   \frac{1}{n_p a_0 x_{\scriptscriptstyle -}\xi}\\
   &=\frac{T}{e_p} 
   \frac{a(\kappa)}{n_p a_0 \xi^2}
\end{align}
involving the $\kappa$-dependent scaling factor
\begin{align}
   a(\kappa) &= \frac{\kappa_{\scriptscriptstyle +}}{4\pi}
   \frac{\xi}{x_{\scriptscriptstyle -}}\frac{e_p}{\Delta e_c}\sim
   \left\lbrace\begin{array}{l l}
                 \kappa^{-1/(n+2)}, & 1 \ll \kappa,\\
                 (\kappa-1)^{-3/2}, & 1 \lesssim \kappa.
   \end{array}\right.\label{eq:a}
\end{align}
and the parameter
\begin{align}\label{eq:tau}
   \nu = \frac{\mathcal{A}}{g(\kappa)}\Bigl(\frac{e_p}{T}\Bigr)^{2/3}
   = \Bigl(\frac{T}{e_p}\Bigr)^{1/3} \frac{a(\kappa)}{n_p a_0 \xi^2 g(\kappa)}.
\end{align}
Given the above asymptotic behavior, the function $\tilde{a}(\kappa) =
a(\kappa)(\kappa-1)^{3/2} \kappa^{-(3n+4) /(2n+4)}$ is roughly constant and
ranges between $\tilde{a}(1) \approx 0.092$ and $\tilde{a}(\infty)\approx
0.22$, see Fig.~\ref{Fig:scaling_functions} in Appendix~\ref{sect:APP_scal_fun}.
The divergences in $\mathcal{A}\propto (\kappa-1)^{-3/2}$ and in $\nu \propto
(\kappa - 1)^{-1/6}$ as $\kappa\to 1$ are due to the vanishing of the
multi-stable solutions, in particular, the stabilizing barriers separating
free and pinned branches. At the same time, the condition for the activation
barriers being large compared to the temperature requires $T\ll
e_p(\kappa-1)^2$ at marginally strong pinning. This compensates the
divergences in $\mathcal{A}$ and $\nu$ and implies that the quantities
$(T/e_p)a(\kappa)$ and $(T/e_p)^{1/3} a(\kappa)/g(\kappa)$ are small
parameters. They compete with another small quantity $n_p a_0\xi^2$ (the 3D
bulk strong pinning regime requires $\kappa n_p a_0\xi^2<1$) and so the
parameters $\mathcal{A}$ and $\nu$ can assume both large and small values; we
discuss below the possible scenarios.

In a first characterization of thermal effects, we determine the velocity
ratio $v(j_c)/v_\mathrm{th}$ at the critical drive; with the finite $T$ and
$T=0$ characteristics joining at $v_\mathrm{th}$, this quantity tells us about
the importance of thermal fluctuations.

At moderately low temperatures such that $1\gg T/e_p\gg n_pa_0\xi^2
/a(\kappa)$, we have $\mathcal{A}\gg 1$ and therefore $v_\mathrm{th}\gg v_c$,
i.e., the characteristic joins the $T = 0$ excess current characteristic at
the current $(1+\mathcal{A})j_c$ far above $j_c$. Since $\nu/\mathcal{A} =
(e_p/T)^{2/3}/g(\kappa)\sim [{(\kappa-1)^2} e_p/T]^{2/3}\gg 1$, this implies
$\nu\gg 1$ as well. We derive $v(j_c)$ from Eq.~\eqref{eq:motion_expanded_2}
using the iterative scheme
\begin{align}\label{eq:iterations_v_jc}
   \begin{split}
   \frac{v^{(0)}(j_c)}{v_\mathrm{th}} &= \frac{1}{\nu},\\
   \frac{v^{(n+1)}(j_c)}{v_\mathrm{th}} 
   &= \frac{1}{\nu}\Bigl[\ln\frac{v_\mathrm{th}}{v^{(n)}(j_c)}\Bigr]^{2/3}.
   \end{split}
\end{align}
The velocity at critical drive can be formally expressed as $v(j_c) =
\lim_{n\to\infty}v^{(n)}(j_c)$. In the following, we will always ignore
corrections beyond $\ln[\ln(\cdots)]$ terms. For the velocity at critical
drive $j_c$, we approximate
\begin{align}\label{eq:v_jc}
   \frac{v(j_c)}{v_\mathrm{th}}
   \approx \frac{1}{\nu}\Bigl[\ln\frac{\nu}{(\ln\nu)^{2/3}}\Bigr]^{2/3}.
\end{align}

At very low temperatures, we enter the regime $\mathcal{A}\ll 1$ and hence
$v_\mathrm{th}\ll v_c$. The characteristic joins the $T = 0$ excess-current
characteristic only slightly above $j_c$ but the velocity $v(j_c)$ may scale
according to two different scenarios: If $\nu \gg 1$, i.e., $a(\kappa)
n_pa_0\xi^2\gg T/e_p\gg [g(\kappa)/a(\kappa)]^3(n_p a_0\xi^2)^3$, the
iteration \eqref{eq:iterations_v_jc} can be applied and $v(j_c)$ is
appreciably suppressed (by $\sim 1/\nu$) as compared to $v_\mathrm{th}$. If
the temperature is extremely low, $T/e_p\ll [g(\kappa)/a(\kappa)]^3(n_p
a_0\xi^2)^3$, we eventually have $\nu \ll 1$ and the  iteration procedure can
no longer be used (the convergence criterion is $\nu>(2e/3)^{2/3}\simeq 1.49$,
see Appendix \ref{sect:APP_IV}). For those small values of $\nu$, we use the
expansion $v(j_c)=v_\mathrm{th}-\delta v$ with $\delta v\ll v_\mathrm{th}$ in
Eq.~\eqref{eq:motion_expanded_2}, what yields the correction $\delta
v/v_\mathrm{th}\approx \nu^{3/2}$ and thus
\begin{align}\label{eq:approx_sol}
   \frac{v(j_c)}{v_\mathrm{th}} \approx 1 - \nu^{3/2}, \qquad \nu \ll 1,
\end{align}
i.e., $v(j_c)$ is very close to $v_\mathrm{th}$.

Next, we use the ratio $v(j_c)/v_c$ to define a depinning temperature
$T_\mathrm{dp}$ where thermal fluctuations lead to a substantial change in the
characteristic.  Substituting the ({\it ad hoc}) criterion
$[v(j_c)/v_c]_{T_\mathrm{dp}} \equiv 1/2$ to Eq.~\eqref{eq:motion_expanded_2}
provides us with the relation
\begin{align}
   \frac{1}{2} = g(\kappa)\Bigl[\frac{T_\mathrm{dp}}{e_p}
         \ln \frac{2a(\kappa)T_\mathrm{dp}}{n_p a_0\xi^2 e_p}\Bigr]^{2/3}
\end{align}
which we solve iteratively for $T_\mathrm{dp}$,
\begin{align}\label{eq:T_dp}
   \frac{T_\mathrm{dp}}{e_p} &\approx \frac{1}{[2g(\kappa)]^{3/2}}
          \frac{1}{\ln(\gamma/\ln\gamma)},\\ \nonumber
    \gamma &= \frac{a(\kappa)/\sqrt{2}}{g^{3/2}(\kappa)\,n_p a_0 \xi^2}\gg 1.
\end{align}
The depinning temperature $T_\mathrm{dp}$ vanishes as $(\kappa - 1)^2$ when
approaching the Labusch point and scales as $T_\mathrm{dp} \sim e_p$ for very
strong pinning.

\begin{figure}[t]
\centering \includegraphics[scale=1]{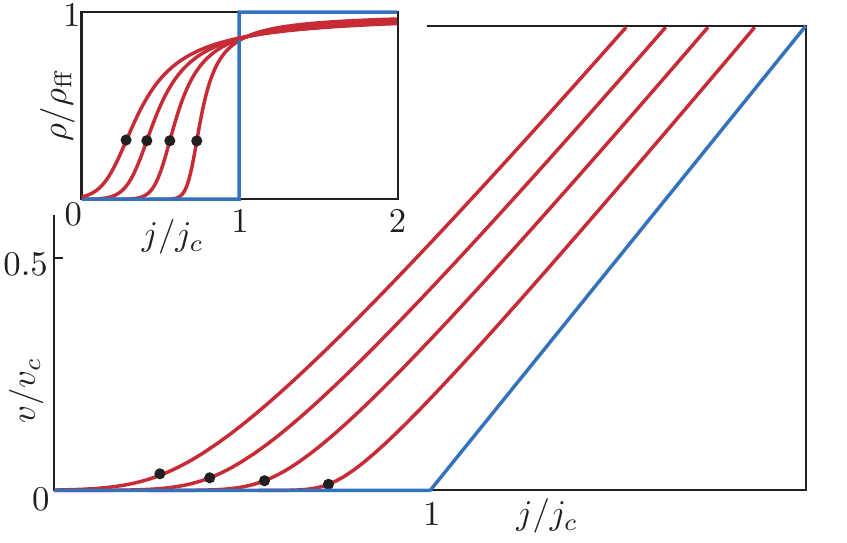}
\caption{Current--velocity characteristics at temperatures $T/e_p =
(0.50,\,1.0,\,1.5,\,2.0)\times 10^{-2}$ and for a small defect density $n_p
a_0\xi^2 = 10^{-4}$. We have chosen the Labusch parameter $\kappa = 5$
implying $g(\kappa)\approx 2.8$ and $a(\kappa)\approx 0.17$ for the Lorentzian
pinning potential, which gives $\mathcal{A}\approx 8.5$ for the lowest and
$\approx 34$ for the highest chosen temperature, respectively, guaranteeing
the applicability of Kramer's rate theory for the range of velocities shown.
The ratio of the chosen temperatures to the depinning temperature
$T_\mathrm{dp}\approx 2.0\times 10^{-2} e_p$ is $T/T_\mathrm{dp} =
\,0.25,\,0.50,\,0.75,\,1.0$. Thermal fluctuations lead to a downward shift of
the critical current density $j_c$ to the depinning current density
$j_\mathrm{dp}(T)$ (solid points) defined through the inflection point in the
differential resistivity (inset) where $\rho(j_\mathrm{dp})\approx
\rho_\mathrm{ff}/3$.  The characteristic steeply increases and follows roughly
parallel to the excess-current characteristic past $j_\mathrm{dp}$.  The
resistivity $\rho$ steeply increases towards the flux-flow resistivity
$\rho_\mathrm{ff}$ around the depinning current $j_\mathrm{dp}$; the increase
spreads around a wider region for higher temperatures.}\label{Fig:IV}
\end{figure}

Effects of thermal fluctuations are also conveniently described through the
differential resistivity $\rho$ rescaled by the flux-flow resistivity
$\rho_\mathrm{ff}$. Differentiating Eq.~\eqref{eq:motion_expanded_2}, we find
that
\begin{align}\label{eq:rho}
   \frac{\rho}{\rho_\mathrm{ff}} = \Bigl[\frac{\partial (j/j_c)}
   {\partial (v/v_c)}\Bigr]^{-1} = \Bigl[1+\frac{2}{3\nu}
   \frac{v_\mathrm{th}/v}{(\ln v_\mathrm{th}/v)^{1/3}}\Bigr]^{-1}.
\end{align}
As illustrated in Fig.~\ref{Fig:IV}, the differential resistivity assumes a
step-like form that is smeared and shifted to lower current densities as $T$
increases.  We define the depinning current density $j_\mathrm{dp}(T)$ as the
inflection point of $\rho(j)$, i.e., the point of the fastest rate of change
of the differential resistivity. Solving the condition $\partial^2
\rho/\partial j^2 = \partial^3 v/\partial j^3 = 0$ for $v$ leads to the
definition of depinning velocity $v_\mathrm{dp}$ (see Appendix
\ref{sect:APP_IV} for details),
\begin{align}\label{eq:v_dp}
   \frac{v_\mathrm{dp}}{v_\mathrm{th}} 
   \approx \frac{1}{3\nu}\frac{1}{(\ln[3\nu (\ln 3\nu)^{1/3}])^{1/3}}.
\end{align}
The corresponding depinning current density $j_\mathrm{dp}(T)$, where the
characteristic rises steeply (and thus assumes the role of the $T=0$ critical
current density $j_c$), is conveniently written as a reduction of $j_c$,
$\delta j_\mathrm{dp}(T) = j_\mathrm{dp}(T) - j_c < 0$ and reads
\begin{align}\label{eq:j_dp}
   \frac{\delta j_\mathrm{dp}(T)}{j_c} &\approx
   \frac{g(\kappa)(T/e_p)^{2/3}}{(\ln\,[3\nu (\ln 3\nu)^{1/3}])^{1/3}}
         \Bigl(\frac{1}{3}-\ln\,[3\nu (\ln 3\nu)^{1/3}]\Bigr).
\end{align}
For very large $\nu$ (note that $\nu \gtrsim 100$ for the range of
temperatures chosen in Fig.~\ref{Fig:IV}) the expression for the depinning
current simplifies to
\begin{align} \label{eq:j_dp_s}
   \frac{j_\mathrm{dp}(T)}{j_c} \approx 
   1 - g(\kappa)\Bigl(\frac{T}{e_p}\Bigr)^{2/3}(\ln 3\nu)^{2/3}.
\end{align}
Substituting $v_\mathrm{dp}$ as calculated to order $(\ln 3\nu)^{-4/3}$ to
Eq.~\eqref{eq:rho} (see Appendix \ref{sect:APP_IV}), we find that the
differential resistivity at depinning remains approximately constant, with
only a weak logarithmic temperature-dependence, ${\rho(j_\mathrm{dp})/
\rho_\mathrm{ff} \approx (1/3)[1-(3\ln 3\nu)^{-1}]}$.

%

Figure \ref{Fig:IV} shows both the current--velocity characteristic and the
differential resistivity in the range of velocities $v<v_c$. Our main finding
is the preservation of an excess-current characteristic also at finite
temperatures, demonstrating that pinning and creep both remain active beyond
$j_c$. Note the sharp rise of the characteristic at $j_\mathrm{dp}(T)$
(replacing $j_c$ at finite $T$) and the nearly parallel shift of the flux-flow
branch at large drives.  To have Kramer's rate theory valid throughout the
chosen range, we require that $U(v_c,T) = T\ln\mathcal{A} \gtrsim T$, which
provides us with the condition $\mathcal{A}\gg 1$. Furthermore, in applying
Kramer's rate theory, we have assumed that vortices reside in a local
equilibrium state at any time during their motion. This is true for velocities
$v$ that are small compared to the scale $v_p$ defined by the vortex dynamics
during depinning, $v_p \sim f_p/\eta a_0^3$.

Indeed, for velocities beyond $v_p$ the pinning force $F_\mathrm{pin}(v)$
decreases markedly, see Refs. \cite{Thomann2012, Thomann2017}. The velocity
scale $v_p$ is independent on $n_p$ and much larger than the flow velocity
$v_c$, $v_c/v_p \sim n_p a_0 \xi^2 (\Delta e_c/f_p\xi)(x_{\scriptscriptstyle
-}/\xi)\sim n_p a_0\xi^2 (\kappa-1)^2$ at moderate and $\sim n_p a_0\xi^2
\kappa^{(n+3)/(n+2)}$ for very strong pinning. This separation of scales
guarantees the simple shape of the excess-current characteristic over a large
regime $v < v_p$ including the free-flow velocity $v_c$ at $F_c$; the
inequality $n_p a_0 \xi^2 \kappa < 1$ is the condition for 3D bulk strong
pinning\cite{Blatter2004}. We then have to check that the thermal velocities
$v < v_\mathrm{th}$ below which finite temperatures modify the excess-current
characteristic remain below $v_p$.  With (we drop numericals and logarithmic
corrections)
\begin{align}
   \frac{v_\mathrm{th}}{v_p} \sim \frac{T}{e_p} \kappa_{\scriptscriptstyle +}
   \sim \frac{T}{T_\mathrm{dp}}\frac{\kappa_{\scriptscriptstyle +}}{g(\kappa)^{3/2}},
\end{align}
we find that this condition limits our temperature to a value below
$T_\mathrm{dp}/\kappa$ at very strong pinning, else dynamical effects have to be
accounted for. Note, however, that at temperatures $\sim T_\mathrm{dp}$ the
characteristic has already lost the essential signatures of pinning, hence
this limitation is easily satisfied. At marginally strong pinning, we have
$v_\mathrm{th}/v_p\sim (\kappa-1)^{5/2}(T/T_\mathrm{dp})$ and the condition
$v_\mathrm{th}<v_p$ is always satisfied for temperatures below $T_\mathrm{dp}$.

\subsubsection{Activation barriers}

The creep-type motion of vortices leads to an average velocity $v$ of the
vortex lattice as discussed in the previous chapters. The pinning and
depinning of individual vortices can be undestood as a thermal diffusion
process, with vortices undergoing transitions between the pinned and free
metastable states. Hence, the activation barriers play a central role in
determining the shape of the current--voltage characteristic. As explained in
Secs.\ \ref{sec:formalism} and \ref{sec:thermal_creep}, the barriers
separating the pinned and free states depend on the distance $x$ of the vortex
from the nearest defect, with the transitions between states taking place
close to the specific jump points $\pm x^\mathrm{jp}_{\scriptscriptstyle \pm}$.

In order to obtain expressions for the activation barriers $U(j)$ as a
function of drive $j$, we use Eq.\ \eqref{eq:jump_U} to express $U$ as
function of temperature and velocity and then combine the result with the
current-velocity characteristic $v(j)$, Eq.\ \eqref{eq:motion_expanded}; the
activation barrier $U(j) = U(v(j),T)$ provides us with the characteristic in
the Arrhenius form $v(j) = v_\mathrm{th}e^{-U(j)/T}$ typical for a thermally
activated motion.

A different behavior is expected for the creep barriers at small and large
drives $j$: As the driving current $j$ approaches $j_c$ from below, the
barrier is expected to vanish as \cite{Blatter1994} $U(j)\approx
U_c(1-j/j_c)^\alpha$ with an exponent $\alpha$ depending on the pinning
scenario. For small currents $j \to 0$, the barriers are usually discussed in
the context of weak pinning theory, predicting a glassy response of the vortex
lattice with a diverging activation barrier $U\approx U_0(j_0/j)^\mu$.

Within the framework of strong pinning that assumes independent action of
defects, the behavior of the activation barriers differs from this standard
expectation.  Starting at small drives $j\to 0$, the jump points
$x^\mathrm{jp}_{\scriptscriptstyle \pm}$ approach the branch crossing point
$x_0$ with an activation barrier $U_0 = U_\mathrm{p}(x_0) =
U_\mathrm{dp}(x_0)$ that remains finite, hence glassy response is replaced by
an ohmic one, see Sec.\ \ref{sec:low_velocities} for details. For larger
drives close to $j_c$, we discuss separately the two cases below and above
vortex depinning. Sufficiently below $j_c$, the vortex velocity is small and
the term $v/v_c$ at the left of Eq.~\eqref{eq:motion_expanded} can be
neglected (this is equivalent to neglecting the dissipative forces acting on
the vortex lattice). Using Eq.~\eqref{eq:jump_U}, we find the barrier
\begin{align}\label{eq:U_small_j}
   U(j) \approx U_c\Bigl(1-\frac{j}{j_c}\Bigr)^{3/2}, \qquad  
   U_c = \frac{e_p}{g^{3/2}(\kappa)}.
\end{align}
This expression remains valid for currents not too close to $j_c$, as we can
drop the term $v/v_c$ only provided that $v/v_c = \mathcal{A}e^{-U(j)/T} \ll
1-j/j_c$. As a result, Eq.\ \eqref{eq:U_small_j} remains valid if
\begin{align}\label{eq:condition_small_current}
   1- \frac{j}{j_c} \gtrsim g(\kappa)\Bigl(\frac{T}{e_p}\Bigr)^{2/3} (\ln\nu)^{2/3}.
\end{align}
Dropping the numerical under the logarithm in Eq.~\eqref{eq:j_dp_s}, we can
express this condition in the form $j\lesssim j_\mathrm{dp}(T)$.

\begin{figure}
\includegraphics[scale=1]{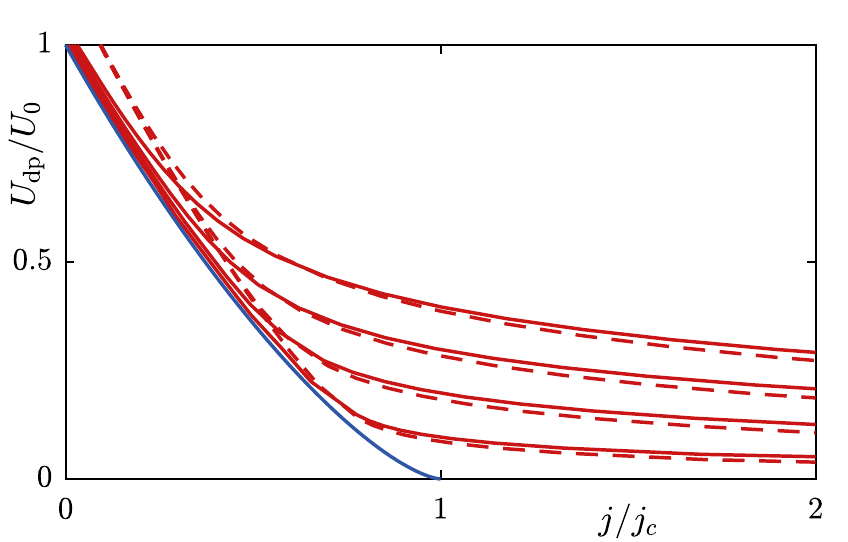}
\caption{Depinning barrier as function of the driving current at temperatures
$T/e_p = (0.5,\,1.0,\,1.5,\,2.0)\times 10^{-2}$ and a small density of defects
$n_p a_0\xi^2 = 10^{-4}$ for a Lorentzian pinning potential with $\kappa = 5$.
The blue curve vanishing at $j = j_c$ corresponds to the limit $T\to 0$, while
higher temperatures produce non-vanishing barriers growing with temperature
beyond $j_c$. The solid lines track the numerical results for the average
barrier value, see Eq.~\eqref{eq:barrier_numerics}. Dashed curves are the
analytical approximation of the barriers as provided by Eqs.\
\eqref{eq:U_small_j}--\eqref{eq:U_large_j2}.} \label{Fig:j_U}
\end{figure}

Beyond depinning, the vortex motion is characterized by a steep rise in
velocity and the dissipative forces cannot be ignored any longer. The vortex
motion is slowed down as compared to the pure thermally activated situation
and the resulting barrier attains larger values than described by
Eq.~\eqref{eq:U_small_j}, see Fig. \ref{Fig:j_U}. In this regime, the
current-velocity characteristic is approximately linear and joins the $T = 0$
excess-current characteristic at a current $(1+\mathcal{A})j_c$ corresponding
to the velocity $v_\mathrm{th}$. The vortex velocity then can be approximated
by
\begin{align}\label{eq:v_linear}
   v(j)\approx v(j_c) + \frac{1}{\mathcal{A}}(j/j_c-1) \,[v_\mathrm{th} - v(j_c)].
\end{align}
Using the expression for $v(j_c)$ from Eq.~\eqref{eq:v_jc}, we find the barrier
\begin{align}\label{eq:U_large_j1}
   U(j) \approx U(j_c) \! - \! T \ln\Bigl[1+\frac{1}{\mathcal{A}}
   \Bigl(\frac{\nu}{(\ln\nu)^{2/3}}\! -\! 1\Bigr) \frac{j\! -\! j_c}{j_c}\Bigr]
\end{align}
with the activation barrier at the critical drive $U(j_c)\approx T
\ln[\nu/(\ln\nu)^{2/3}]$. Assuming $\nu\gg 1$, comparing with the definition
of the depinning current density, and ignoring numerical factors inside the
logarithm, allows us to rewrite Eq.~\eqref{eq:U_large_j1} as
\begin{align}\label{eq:U_large_j2}
   U(j) \approx U(j_c) - T \ln\Bigl[1+\frac{j-j_c}{j_c-j_\mathrm{dp}(T)}\Bigr].
\end{align}
As shown in Sect.~\ref{sec:IVchar}, the effects of thermal depinning and creep
persist well above the critical current $j_c$. In accordance with this result,
the corresponding activation barrier vanishes at large currents $j\sim
(1+\mathcal{A})j_c\gg j_c$. The slow logarithmic decay of the barrier
predicted by Eq.\ \eqref{eq:U_large_j2} then corresponds to the linear
current-velocity characteristic.

In order to test the quality of our approximations, we compare the above
analytical results with a computational scheme that relies on the numerical
solution of the rate equation and the current--voltage characteristic.  We
exploit the insight that the transitions from pinned to free states go
together with a smooth drop of the occupation probability $p(x,y)$ from $p =
1$ to $p = 0$ when increasing $x$ across the depinning jump point $x^\mathrm{jp}_{\scriptscriptstyle
+}$; it is this region that defines the relevant barriers in the depinning
process. The derivative $\partial_x p$ is sharply peaked around $x^\mathrm{jp}_{\scriptscriptstyle
+}$ and serves as a convenient measure (of total weight unity) to define the
average depinning barrier,
\begin{align}\label{eq:barrier_numerics}
   \langle U_\mathrm{dp} \rangle = \frac{1}{2x_{\scriptscriptstyle -}}\! \int_{-x_-}^{x_-}\!\! dy\,
   \int_{x_-}^{x_+} \!\! dx\,U_\mathrm{dp}(x,y)[-\partial_x p(x,y)].
\end{align}
The $y$-integration is cut by $|y|=\pm x_{\scriptscriptstyle -}$ since vortices passing at
larger transverse distances from the defect are not pinned. The average
pinning barrier $\langle U_\mathrm{p}\rangle$ is defined in a similar way, with the
$x$-integration ranging from $-x_{\scriptscriptstyle +}$ to $-x_{\scriptscriptstyle -}$ and with the
derivative replaced via $-\partial_x p(x,y) \to \partial_x p(x,y)$ (during
pinning, the branch occupation grows from $p=0$ to $p=1$). This scheme
provides us with the average barrier at a particular temperature $T$ and
velocity $v$ (since the solution of the rate equation is obtained at fixed $T$
and $v$); in order to obtain the activation barriers as function of drive $j$,
we have to combine this result with the numerical predictions for the
current--velocity characteristic.  The result of this numerical procedure and
its comparison with the analytical prediction are displayed on
Fig.~\ref{Fig:j_U}, with very satisfactory agreement between the two.

Against common expectations, our activation barrier grows with temperature in
the region beyond depinning (cf.\ the expression for $U(j_c)$ and
Fig.~\ref{Fig:j_U}).  Thermal activation enhances the vortex motion and the
magnitude of the pinning force $F_\mathrm{pin}(v,T)$ decreases with increasing
temperature.  Eq.~\eqref{eq:F_pin_expansion_barrier} then relates the
activation barrier $U$ and the pinning-force density $F_\mathrm{pin}$ and shows that
decreasing pinning forces indeed correspond to an increasing barrier.  On a
more technical level, within the strong pinning paradigm, the relevant barrier
at a given drive $j$ and temperature $T$ is selected by the jump positions
$\pm x^\mathrm{jp}_\pm$; increasing $T$ pushes $\pm x^\mathrm{jp}_\pm$ further away from
$\pm x_\pm$ where the barriers are larger, see Figs.\ \ref{Fig:energy_landscape}
and \ref{Fig:occupation_high_v}.

The current-dependence of the activation barrier $U(j)$ is directly related to
the magnetic relaxation rate \cite{Blatter1994} and, conversely, the
measurement of creep rates can be used to reconstruct the barrier, a topic we
are going to analyse in the next paragraph.

\subsubsection{Magnetic relaxation through creep}\label{sec:current_relaxation}

Magnetic relaxation measurements \cite{Maley1990,Yeshurun1996,Civale2017}
represent a~convenient way to study thermal vortex creep. The sample is
typically cooled in zero field, then a magnetic field is applied generating a
Bean critical state with a vortex density gradient. Subsequently, this density
gradient is relaxed as vortices move further into the sample due to creep,
what results in a decay of the diamagnetic moment with time. The magnetization
is linearly proportional to the persistent current $j(t)$ flowing in the
sample and the role of thermal fluctuations can be quantified via the creep
rate $S = -d \ln j/d \ln t$, a quantity that is closely related to the
activation barrier $U(j)$. Within the Anderson-Kim flux creep theory
\cite{Anderson1962}, barriers are linear in the current, which results in the
creep rate $S = T/U_c$. However, the observed creep rate often exhibits a
finite value when extrapolated to small temperatures \cite{Civale2017}, a
feature that is usually ascribed to the phenomenon of quantum creep
\cite{Blatter1991, Blatter1994}. Below, we discuss creep rates in the
framework of strong pinning theory with non-linear barriers near $j_c$ and
show that such a scenario may provide an alternative explanation of the
apparent low-$T$ saturation phenomenon of creep.

Specifically, we consider a superconducting slab of thickness $d$ in the
presence of an external magnetic field. Variations in the magnetic induction
$\mathbf{B}$ inside the slab lead to a transport current density $\mathbf{j} =
(c/4\pi)\nabla\times\mathbf{B}$. The resulting Lorentz force density $\mathbf{F}_L$
moves vortices in the direction of the gradient of $\mathbf{B}$, what diminishes
the variation in the magnetic field and leads to the decay of the observed
current density. Assuming activation barriers independent on the magnetic
field, the time evolution of the current density $j$ follows from 
\begin{align}\label{eq:current_relaxation}
   \frac{\partial j}{\partial t} = -\frac{j_c}{\tau_0}e^{-U(j)/T},
\end{align}
where $\tau_0$ denotes a macroscopic timescale, see Ref.\
[\onlinecite{Blatter1994}] (we use Eqs.\ \eqref{eq:jump_U},
\eqref{eq:v_th_v_c} and $\eta = BH_{c2}/c^2\rho_n$ in the second equation)
\begin{align}\label{eq:jt}
   \tau_0 &= \frac{\pi j_c d^2}{2c v_\mathrm{th} H} 
   = \frac{\pi}{2\mathcal{A}} \frac{H_{c2}}{H}\frac{d^2}{\rho_n c^2}.
\end{align}
Using typical values $H/H_{c2}\sim 10^{-1}$, $\rho_n\sim
10^{-4}\,\Omega\,\mathrm{cm}$, $d\sim 0.1\,\mathrm{cm}$, we estimate the timescale
$\tau_0\sim 10^{-5}/\mathcal{A} [s]$, where the parameter $\mathcal{A}$
depends on temperature $T$, density of defects $n_p$, and the Labusch
parameter $\kappa$ according to Eq.~\eqref{eq:v_th_v_c}.

\begin{figure}
\includegraphics[width=8cm]{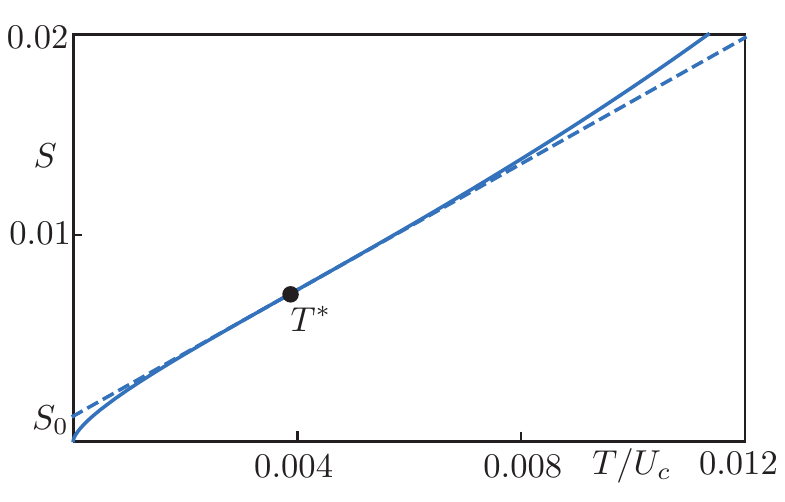}
\caption{ Normalized creep rate $S = -d\log (j/j_c)/d \log (t/t_0)$ as
predicted by Eq.~\eqref{eq:creep_rate} for $n_p a_0 \xi^2 = 10^{-3}$, $\kappa
= 5$ and a measurement time $t = 10^3 \, \mathrm{s}$.  The initial non-linear increase
$S\propto (T/U_c)^{2/3}$ crosses over to the approximately linear regime near
the inflection point $T^*$ defined by $\partial_T^2 S(T^*) = 0$, $T^*/U_c=
1/[5^{3/2}\log (t/t_0)]\approx 4\times 10^{-3}$. Extrapolating the creep rate
around the inflection point to zero temperatures yields a finite creep value
$S_0\approx 1/[36 \log (t/t_0)] \approx 1.2\times
10^{-3}$.}\label{Fig:creep_rate}
\end{figure}

The differential equation \eqref{eq:current_relaxation} describing the creep
induced decay of $j(t)$ is easily transformed to one describing the evolution
of the barrier $U(t)$ and its integration produces the well known
result\cite{Blatter1994}
\begin{align}\label{eq:U_t}
   U(t\gg t_0) \approx T\log(t/t_0),
\end{align}
with the new timescale defined self-consistently as $t_0 = \tau_0 T / (j_c
|\partial_j U|)$. From Eq.~\eqref{eq:U_small_j}, one finds that $|\partial_j
U| = (2/3)(U_c/j_c)(U/U_c)^{1/3}$, that reduces the self-consistent
relation to $t_0 = (2\tau_0/3)(T/U_c)^{2/3}(\log t/t_0)^{1/3}$. Ignoring
logarithmic corrections, we obtain
\begin{align}\label{eq:t_0}
   t_0 = \frac{2}{3}\tau_0 g(\kappa)\Bigl(\frac{T}{e_p}\Bigr)^{2/3} =
   \frac{\pi}{3\nu}\frac{H_{c2}}{H}\frac{d^2}{\rho_n c^2}
   \sim \frac{10^{-5}\,\mathrm{s}}\nu
\end{align}
with $\nu$ given in Eq.\ \eqref{eq:tau}.  Using a typical value $\nu\sim 10^2$
then gives $U(t = 10^3\, \mathrm{s})\approx 23 \, T$, confirming the
assumption $U\gg T$. The
non-linear barrier \eqref{eq:U_small_j} translates \eqref{eq:U_t} into a slow
time-decay of the screening current (or magnetization) that depends
non-linearly on $\log (t/t_0)$,
\begin{align}\label{eq:current_relaxation2}
   \frac{j(t)}{j_c} = 1-\Bigl(\frac{T}{U_c}\Bigr)^{\! 2/3}\!
   \bigl[\log(t/t_0) \bigr]^{2/3}.
\end{align}
Neglecting the temperature dependence of $\log (t/t_0)$ (note that $t_0$
depends on temperature through the parameter $\nu$), we arrive at the
expression for the normalized creep rate $S = -d \log (j/j_c)/d
\log(t/t_0)$,
\begin{align}\label{eq:creep_rate}
   S = \frac{2}{3}\frac{(T/U_c)^{2/3}[\log (t/t_0)]^{-1/3}}
                       {1-[(T/U_c) \log (t/t_0)]^{2/3}}.
\end{align}
This result predicts an initial non-linear and convex increase of the creep
rate $S\propto (T/U_c)^{2/3}$ at small temperatures, followed by a concave
growth at higher temperatures beyond the inflection point at $T^* =
U_c/[5^{3/2}\log (t/t_0)]$, see Fig.~\ref{Fig:creep_rate}. Such a creep rate
with a temperature dependence indicative for an inflection point has been
observed in recent experiments \cite{Civale2017}. As the limit $T \to 0$ is
not accessible, the extrapolation of such a line shape below $T^*$ then
reaches a finite value $S_0 = S(T=0)$; a linear extrapolation at the
inflection point $T^*$ produces the result $S_0\approx 1/36 \log(t/t_0)$. Such
an extrapolation to a finite $T=0$ creep rate within strong pinning theory
then is in competition with the interpretation of a finite value $S_0 >0$
generated by quantum creep\cite{Blatter1991}.

Increasing the temperature beyond the inflection point $T^*$, the
creep rate increases further and formally diverges when $U_c = T \log(t/t_0)$,
see Eq.\ \eqref{eq:creep_rate}; under these conditions, the finite barriers
have effectively vanished. In reality, this result is expected to be modified
due to collective pinning effects appearing at small drive.  Such collective
effects will lead to diverging barriers and a saturation of the creep rate at
fixed decay time $t$ \cite{Blatter1994}.

\subsection{Small drives, low velocities}\label{sec:low_velocities}

As the velocity of the vortex system approaches values of order  $v_{\rm
\scriptscriptstyle TAFF} \simeq v_\mathrm{th} e^{-U_0/T}$, the activation
barrier $U(v,T)$ reaches its maximum possible height $U_0$ and the jump points
$x^\mathrm{jp}_{\scriptscriptstyle \pm}$ approach the location $x_0$ where the branches cross. The
result Eq.~\eqref{eq:F_pin_large_v} suggesting a vanishing pinning force (due
to the vanishing energy jumps $\Delta e_{\mathrm{p}}(- x_0) = \Delta e_\mathrm{dp} (x_0) =
0$) is no longer applicable. We then have to go back and solve the rate
equation \eqref{eq:rate} for the present situation which involves both terms
$\propto p$ and $\propto (1-p)$ in order to find the shape of $p(x)$ near
$x_0$.

We start from the equilibrium occupation $p_\mathrm{eq}(x)$ and calculate the
corrections due to a finite but weak drive $j$ or small velocity $v$.
Rewriting the rate equation \eqref{eq:rate} with the help of the equilibrium
distribution \eqref{eq:p_eq1}, we can cast it into the form
\begin{align}
   \frac{d p}{d x} &= (p_\mathrm{eq} -p)\left(\omega_\mathrm{p}\,e^{-U_\mathrm{dp}/T}
   +\omega_\mathrm{f}\,e^{-U_\mathrm{p}/T}\right)\\
   &=\frac{p_\mathrm{eq}(x)-p}{\ell_\mathrm{eq}(x)},
\label{eq:rate_tau_rep}
\end{align}
with the local equilibrium relaxation length
\begin{align}
   \ell_\mathrm{eq}(x) = [\ell_\mathrm{p}(x)^{-1} + \ell_\mathrm{dp}(x)^{-1}]^{-1}.
   \label{eq:ell_eq_rep}
\end{align}

\subsubsection{Equilibrium properties}

In order to solve the rate Eq.\ \eqref{eq:rate_tau}, we need to analyze the
equilibrium distribution $p_\mathrm{eq}(x)$ and relaxation length
$\ell_\mathrm{eq}(x)$.  We restrict the discussion to the interval
$[x_{\scriptscriptstyle -},x_{\scriptscriptstyle +}]$; a similar analysis
applies to the region $[-x_{\scriptscriptstyle +},-x_{\scriptscriptstyle -}]$.
Using the definition of $\ell_\mathrm{dp}(x)$, Eq.~\eqref{eq:ell_dp} (and
analogous for $\ell_\mathrm{p}(x)$) and expressing the jump in energy $\Delta
e_\mathrm{pin}$ through the difference of the barriers, $\Delta e_\mathrm{pin}
(x) = U_\mathrm{p}(x)-U_\mathrm{dp}(x) >0 $, we can rewrite the equilibrium
distribution Eq.\ \eqref{eq:p_eq1} in the form
\begin{align}\label{eq:p_eq2}
   p_\mathrm{eq}(x) &= \frac{1}{1 + (\omega_\mathrm{p}/\omega_\mathrm{f})\,
   e^{\Delta e_\mathrm{pin}/T}}\\
   \label{eq:p_eq3}
   &= \frac{1}{1 + (\lambda_\mathrm{p}/\lambda_\mathrm{f})^{1/2}\,
   e^{\Delta e_\mathrm{pin}/T}}.
\end{align}
The equilibrium distribution formulated through the Eq.~\eqref{eq:p_eq3} is
expressed purely in terms of equilibrium properties of the energy
landscape: the energy difference of branches $\Delta e_\mathrm{pin}$ and the
curvatures $\lambda_\mathrm{p}$, $\lambda_\mathrm{f}$ at the local minima of the double-well
pinning energy $e_\mathrm{pin}(x;r)$. The ratio of curvatures $\lambda_\mathrm{p}/\lambda_\mathrm{f}$
plays an important role in the equilibrium probability distribution. The two
minima, pinned and free, come with a different geometrical shape: e.g., at
very strong pinning and $x_{\scriptscriptstyle -}\ll x\ll x_{\scriptscriptstyle +}$, we can estimate the curvatures
(see Fig.\ \ref{Fig:metastable_states}) $\lambda_\mathrm{p}\sim e_p/\xi^2$ and
$\lambda_\mathrm{f}\sim \bar{C}$. As a result, the branches are not populated equally at
the branch crossing point $x_0$ where $\Delta e_\mathrm{pin}(x_0) = 0$. For $\kappa\gg
1$, we find instead $p_\mathrm{eq}(x_0)\sim 1/\sqrt{\kappa}$, i.e., the shallower well
of the free branch is more strongly populated since it accommodates more
states.

\begin{figure}
\centering \includegraphics[width = 8 truecm]{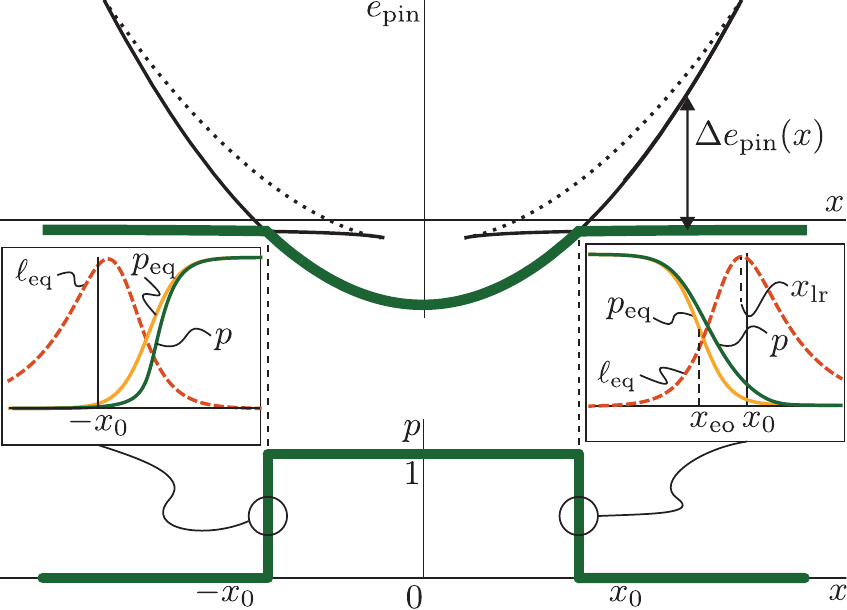}\vspace{20pt}
\caption{Occupied branches (thick) of the multi-valued energy landscape for a
Lorentzian pin with $\kappa = 5$ in the low velocity limit $v\ll
v_\mathrm{th}e^{-U_0/T}$. The occupation changes sharply in the vicinity of
the branch crossing points $\pm x_0$ and $p(x)$ takes the form of a shifted
equilibrium occupation, $p(x)\approx p_\mathrm{eq}[x-\ell_\mathrm{eq}(x)]$;
the solid curves in the insets show the distribution functions $p(x)$ and
$p_\mathrm{eq}(x)$ near $\pm x_0$ on an expanded scale. The dashed curves show
the equilibirum relaxation distance $\ell_\mathrm{eq}(x)$ with the maxima
attained at $\pm x_\mathrm{lr}$, $x_\mathrm{lr} \approx x_0 + [T/2(n+1) \Delta
f_\mathrm{pin}(x_0)] \ln\kappa$. The points of equal equilibrium branch
occupation are further shifted to $\pm x_\mathrm{eo}$, with $x_\mathrm{eo}
\approx x_0 + [T/2\Delta f_\mathrm{pin}(x_0)]\ln\kappa$.}
\label{Fig:occupation_small_v}
\end{figure}

This implies that the point $x_\mathrm{eo}$ of equal branch occupation is
shifted to the left from $x_0$ (see Fig.\ \ref{Fig:occupation_small_v} and the
Appendix \ref{sect:APP_low_v} for a more elaborate discussion). A similar
analysis of the equilibrium relaxation length, Eq.~\eqref{eq:ell_eq_rep},
shows that the point $x_\mathrm{lr}$ where the relaxation length is maximal is
also shifted to the left from $x_0$ and that the three points, $x_0$,
$x_\mathrm{eo}$, and $x_\mathrm{lr}$ are arranged in the sequence
$x_\mathrm{eo} < x_\mathrm{lr} < x_0$. For marginally strong pinning, the
energy landscape becomes symmetric and the three positions join up.

\subsubsection{Solution of the rate equation}\label{sect:solution_rate}

We use the ansatz $\delta p = p-p_\mathrm{eq}$ in the rate equation
\eqref{eq:rate_tau} what takes us to the linear problem
\begin{align}\label{eq:rate_tau_lin}
   \delta p'+\frac{\delta p}{\ell_\mathrm{eq}(x)}=-p'_\mathrm{eq}
\end{align}
with the boundary condition $\delta p(x_{\scriptscriptstyle -}) = 0$. The
Green's function $G(x,x')$ satisfying the differential equation
\begin{align}\label{eq:Green_rate}
  &\bigl[\partial_x + \ell_\mathrm{eq}(x)^{-1}\bigr] G(x,x') = \delta(x-x')
\end{align}
takes the form
\begin{align}\label{eq:Green_rate1}
   G(x,x') = \Theta(x-x')\exp\left[-\int_{x'}^x 
   \frac{d x''}{\ell_\mathrm{eq}(x'')}\right],
\end{align}
in terms of which the solution for \eqref{eq:rate_tau_lin} reads 
\begin{align}\label{eq:delta_p}
   \delta p(x) &= -\int_{x_{\scriptscriptstyle -}}^x \!\!\! 
   d x'\, G(x,x')p'_\mathrm{eq}(x').
\end{align}
Fixing $x$ and varying $x'<x$, the function $G(x,x')$ grows on a scale
$\ell_\mathrm{eq}(x) \lesssim \ell_\mathrm{eq}(x_\mathrm{lr})$, which is small
compared to $T/|U_\mathrm{dp}'|$ or $T/|U_\mathrm{p}'|$. As a result, the
change in $p_\mathrm{eq}'(x)$ or $\ell_\mathrm{eq}(x)$ is small on this scale
and we can approximate $G(x,x')\approx e^{-(x-x')/\ell_\mathrm{eq}(x)}$ as
well as $p_\mathrm{eq}'(x')\approx p_\mathrm{eq}'(x)$ in Eq.\
\eqref{eq:delta_p}. Moving the lower integration bound in Eq.\
\eqref{eq:delta_p} from $x_{\scriptscriptstyle -}$ to $-\infty$ provides us
with the final result
\begin{align}\nonumber
   p(x) &\approx p_\mathrm{eq}(x) -\int_{-\infty}^x \!\!\!\! 
   d x'\, \exp\Bigl[-\frac{x-x'}
   {\ell_\mathrm{eq}(x)}\Bigr]p'_\mathrm{eq}(x)\\ \label{eq:delta_p2} 
   &= p_\mathrm{eq}(x)-\ell_\mathrm{eq}(x) p'_\mathrm{eq}(x) 
   \approx p_\mathrm{eq}[x-\ell_\mathrm{eq}(x)]. 
\end{align}
Hence, we find that the branch occupation is shifted from its equilibrium
distribution by a small distance $\ell_\mathrm{eq}(x)$. This shift is small
compared to the scale where the equilibrium distribution $p_\mathrm{eq}(x)$
changes.

\subsubsection{Pinning force}

The average pinning force $\langle f_\mathrm{pin}\rangle$ involves an
integration over the asymptotic vortex positions $x$.  Using the symmetry $x
\leftrightarrow -x$ of the integrand $\ell_\mathrm{eq}p'_\mathrm{eq}\Delta
f_\mathrm{pin}$, we can restrict the integration in Eq.\
\eqref{eq:f_pin_small_v1} to the interval $[x_{\scriptscriptstyle
-},x_{\scriptscriptstyle +}]$,
\begin{align}\label{eq:f_pin4}
   \langle f_\mathrm{pin}\rangle = \frac{2}{a_0}
   \int_{x_{\scriptscriptstyle -}}^{x_{\scriptscriptstyle +}} \!\!\!\!
   d x\,\ell_\mathrm{eq}(x)\, p'_\mathrm{eq}(x)\, \Delta f_\mathrm{pin}(x).
\end{align}
The functions $p'_\mathrm{eq}(x)$ and $\ell_\mathrm{eq}(x)$ are sharply peaked
around $x_\mathrm{eo}$ and $x_\mathrm{lr}$ on the scale $T/|\Delta
f_\mathrm{pin}|$. We therefore expand $\Delta e_\mathrm{pin}(x_0 + \delta x) =
-\Delta f_\mathrm{pin} \delta x$ in Eq.~\eqref{eq:p_eq2} and neglect the
$x$-dependence of the frequency factors $\omega_{\mathrm{p},\mathrm{dp}}$. We
find that
\begin{align}
   p_\mathrm{eq}'(x_0+\delta x) = \frac{\Delta f_\mathrm{pin}/4T}{\cosh^2
   \Bigl[\frac{-\Delta f_\mathrm{pin}}{2T}\delta x
   +\frac{1}{2}\ln\frac{\omega_\mathrm{p}}{\omega_\mathrm{f}}\Bigr]}
   \label{eq:p_eq'}
\end{align}
and similarly,
\begin{align}
   \ell_\mathrm{eq}(x_0+\delta x) = \frac{ve^{U_0/T}}{\sqrt{\omega_\mathrm{p} 
   \omega_\mathrm{f}}} \frac{e^{(U_\mathrm{p}'+U_\mathrm{dp}')\delta x/2T}}
   {\cosh\Bigl[\frac{-\Delta f_\mathrm{pin}}{2T}
   \delta x+ \frac{1}{2}\ln\frac{\omega_\mathrm{p}}{\omega_\mathrm{f}}\Bigr]}.
   \label{eq:ell_eq2}
\end{align}
The limits of integration in Eq.\ \eqref{eq:f_pin4} can be shifted to $\pm
\infty$ since the contributions from the region $|\delta x|\gg T/|\Delta
f_\mathrm{pin}|$ are negligible. We abbreviate $[(-\Delta f_\mathrm{pin}/T)\,
\delta x+\ln(\omega_\mathrm{p}/\omega_\mathrm{f})]/2 = z$ and define the ratio
\begin{align}\label{eq:Gamma}
   \Gamma = \left.-\frac{U_\mathrm{p}'+U_\mathrm{dp}'}
   {\Delta f_\mathrm{pin}}\right|_{x_0} 
   = \left.\frac{U_\mathrm{p}'+U_\mathrm{dp}'}{U_\mathrm{p}'
   -U_\mathrm{dp}'}\right|_{x_0}
\end{align}
ranging between $\Gamma = 0$ for marginally strong pinning where ${U_p'
\approx -U_\mathrm{dp}'}$ and $\Gamma = 1$ for very strong pinning with
${U_\mathrm{p}' \gg |U_\mathrm{dp}'|}$. Furthermore, we ignore variations of
$\Delta f_\mathrm{pin}$ that appear on the large scale $\kappa\xi$. With these
approximations, the average pinning force is given by
\begin{align}\label{eq:f_pin_final}
   \langle f_\mathrm{pin} \rangle =
   \frac{v|\Delta f_\mathrm{pin}|(\omega_\mathrm{p}/\omega_\mathrm{f})^{-\Gamma/2}
   e^{U_0/T}}{2a_0 \sqrt{\omega_\mathrm{p}\, \omega_\mathrm{f}}} \!\!
   \int_{-\infty}^\infty \!\!\!\!\!\! d z\, \frac{e^{\Gamma z}}{\cosh^3 z}.
\end{align}
The integral can be evaluated by deforming the contour in the complex plane
around the poles ${z_n = (n+\frac{1}{2})\pi i},\,n\in\mathbb{Z}_0^+$.  The
residues $\mathrm{Res}(z_n)$ derive from the prefactor of the term $\propto \delta
z^{-1}$ in the expansion
\begin{align}\label{eq:residuum_expansion}
   &\frac{e^{\Gamma(z_n+\delta z)}}{\cosh^3(z_n+\delta z)} = 
   e^{\Gamma z_n}\frac{1+\Gamma \delta z+\Gamma^2\delta z^2/2+
   \mathcal{O}(\delta z^3)}{-i(-1)^n\left[\delta z+\delta z^3/6+
   \mathcal{O}(\delta z^5)\right]^3}\nonumber\\
   &=-\frac{e^{\Gamma z_n}}{i\delta z^3}
   \Bigl[1+\Gamma\delta z+\frac{\Gamma^2-1}{2}
   \delta z^2+\mathcal{O}(\delta z^3)\Bigr],
\end{align}
and provide us with the following result for the integral in Eq.\
\eqref{eq:f_pin_final}
\begin{align}\label{eq:geometry_factor}
   2\pi i\sum_{n = 0}^\infty \mathrm{Res}(z_n)&=
   \pi(1-\Gamma^2)\sum_{n=0}^\infty (-1)^ne^{-\Gamma(n+\frac{1}{2})\pi i}\\
   &= \frac{\pi}{2}\frac{1-\Gamma^2}{\cos(\pi\Gamma/2)}
\end{align}
ranging between $\pi/2$ at $\Gamma = 0$ and $2$ for $\Gamma = 1$.

Again, we have to consider the change in the average pinning force $\langle
f_\mathrm{pin}\rangle$ when vortices impact the defect at a finite distance $y$. The
pinning force exerted on a vortex in branch $i$ passing the defect at an
arbitrary transverse distance $y$ is obtained from Eq.\ \eqref{eq:e_pin_xy},
\begin{align}\nonumber
   f_\mathrm{pin}^i(x,y) &= -\frac{\partial}{\partial x}e_\mathrm{pin}^i(x,y)\\
   \nonumber
   &=f_\mathrm{pin}^i\left(\sqrt{x^2+y^2},0\right)\frac{x}{\sqrt{x^2+y^2}}.
\end{align}
The position of the branch crossing point with equal energies of the pinned
and free branches is given by $x_0(y) = \sqrt{x_0^2-y^2}$; the force
exerted at this point is
\begin{align}\label{eq:f_pin_x0y}
   f_\mathrm{pin}^i[x_0(y),y] = f_\mathrm{pin}^i(x_0,0)\frac{\sqrt{x_0^2-y^2}}{x_0}
\end{align}
and vanishes at the transverse distance $y = x_0$. 

The quantity $\Delta f_\mathrm{pin}(x_0)$ is the only term in
Eq.~\eqref{eq:f_pin_final} that depends on the transverse distance. The
frequency factors are independent of $y$ (they are determined by the 
asymptotic distance $|\mathbf{R}|$ of the vortex from the defect, see
discussion in Sect.~\ref{subsect:formalism_pinning_force}). The derivatives
$U_\mathrm{p}'$, $U_\mathrm{dp}'$ acquire the same $y$-dependent factor
$\sqrt{x_0^2-y^2}/x_0$ as the pinning force in Eq.~\eqref{eq:f_pin_x0y} above,
however, this factor is cancelled in the ratio $\Gamma$.

Averaging the pinning force $\Delta f_\mathrm{pin}(x_0(y),y)$ over the impact
parameter $|y|<x_0$ results in an additional numerical factor $\pi/4$
in the expression for the pinning-force density (the fraction of trapped
trajectories is $2 x_0/a_0$),
\begin{align} \nonumber
   F_\mathrm{pin} = n_p\frac{\langle f_\mathrm{pin}\rangle}{a_0}\!\! 
   \int_{-x_0}^{x_0}\!\!\!\!  dy\, \frac{\sqrt{x_0^2-y^2}}{x_0}
    = n_p\frac{\pi}{4}\frac{2x_0}{a_0}\langle f_\mathrm{pin}\rangle,
\end{align}
where $\langle f_\mathrm{pin} \rangle $ is the (thermally) averaged pinning force
exerted on vortices passing at $y = 0$ and calculated through Eq.\
\eqref{eq:f_pin4}.

Finally, we substitute for the attempt frequencies in Eq.\
\eqref{eq:f_pin_final} using Eq.\ \eqref{eq:omega} and collect the various
contributions from above to arrive at the pinning-force density in the form
\begin{align}\label{eq:F_pin_small_v}
   F_\mathrm{pin} = \eta v \, h(\kappa)\,(n_p \xi^2 a_0)\,e^{U_0(\kappa)/T},
\end{align}
with the dimensionless scaling function $h(\kappa)$
\begin{align}\label{eq:h}
   h(\kappa) &= \frac{\pi^3}{4} \frac{x_0|\Delta f_\mathrm{pin}(x_0)| 
   (\lambda_\mathrm{p}/\lambda_\mathrm{f})^{-\Gamma/4}} {\xi^2
   |\lambda_\mathrm{us}|^{1/2}(\lambda_\mathrm{f}\lambda_\mathrm{p})^{1/4}}
   \frac{1-\Gamma^2}{\cos\bigl(\frac{\Gamma\pi}{2}\bigr)}
\end{align}
accounting for the dependence on the Labusch parameter. At very strong
pinning, we use $\Gamma \approx 1$, $|\Delta f_\mathrm{pin}(x_0)| \approx
\bar{C} x_0$, and $e_p\approx 1/2 \bar{C} x_0^2$, which simplifies $h(\kappa)$
to
\begin{align}
   h(\kappa)\approx \pi^2\frac{e_p}{2\xi^2(|\lambda_\mathrm{us}|
   \lambda_\mathrm{f})^{1/2}} \sim \kappa^{(n+2)/4(n+1)},
\end{align}
while for marginally strong pinning, we find that $h(\kappa)\sim
(\kappa-1)^{-1/2}$. Hence, the function $\tilde{h}(\kappa) =
{h(\kappa)(\kappa-1)^{1/2} \kappa^{-(3\alpha+1)/(4\alpha)}}$ is roughly
constant and ranges between $\tilde{h}(\infty)\approx 10$ and
$\tilde{h}(0)\approx 22$ (see Fig.~\ref{Fig:scaling_functions}).

The scaling form of the result \eqref{eq:F_pin_small_v} can be obtained quite
straightforwardly: Using $p_\mathrm{eq}'(x)\approx
{\delta(x+x_0)-\delta(x-x_0)}$ to integrate Eq.\ \eqref{eq:f_pin_small_v1}
provides the estimate $\langle f_\mathrm{pin}\rangle \sim
[\ell_\mathrm{eq}(x_0)/a_0]\Delta f_\mathrm{pin}(x_0)$.  With the transverse
trapping distance $t_\perp \sim x_0$, the pinning force density can be
estimated as
\begin{align}\label{eq:F_pin_TAFF_scaling}
   F_\mathrm{pin} \sim n_p\frac{x_0}{a_0}\frac{\ell_\mathrm{eq}(x_0)}{a_0}
   |\Delta f_\mathrm{pin}(x_0)|.
\end{align}
With the further approximations $x_0 \sim \xi$, $\Delta f_\mathrm{pin}(x_0)
\sim e_p/\xi$ and $\omega_\mathrm{f} \sim (e_p/\xi^2)/\eta a_0^3$, see Eq.\
\eqref{eq:omega}, we find the result \eqref{eq:F_pin_small_v} up to the
scaling function $h(\kappa)$.

The analytical and numerical results for the pinning-force density at low
velocities are compared in Fig.~\ref{Fig:vF_comparison}. Note that the
pinning-force density Eq.~\eqref{eq:F_pin_small_v} can be written through the
ratio $v/v_{\rm \scriptscriptstyle TAFF}(T)$ using Eq.~\eqref{eq:v_th_v_c},
$F_\mathrm{pin}/F_c = (T/e_p)a(\kappa)h(\kappa)(v/v_{\rm \scriptscriptstyle
TAFF})$.  Plotting the pinning-force density as a function of $v/v_{\rm
\scriptscriptstyle TAFF}(T)$, see Fig.~\ref{Fig:vF_comparison}, the additional
factor $T/e_p$ in this expression implies that the upper-most curve
corresponds to the highest temperature (this is due to the $T$-dependence of
$v_{\rm \scriptscriptstyle TAFF}$ itself; if plotting the result as a function
of $v/v_c$, where $v_c$ is temperature-independent, the highest pinning force
corresponds to the lowest temperature). The numerical results show excellent
agreement with the analytic formula for low velocities, while for $v\sim
v_{\rm \scriptscriptstyle TAFF}(T)$, the force dependence crosses over to the
logarithmic behaviour as described by Eq.~\eqref{eq:F_pin_interm_v}.

Finally, we compare our results to those of Brazovskii, Larkin and Nattermann
(BLN, Refs.\ \cite{BrazovskiiLarkin1999,BrazovskiiNattermann2004}). In
their study of charge density wave pinning, the pinning-energy landscape is
symmetric around the branch crossing point (as is the case for marginally
strong vortex pinning, see Appendix \ref{sect:APP_mod_strong}); furthermore,
their pinning is effectively one-dimensional, involving no transverse
dimensions. Therefore the results of BLN are to be compared to the pinning
force $\langle f_\mathrm{pin} \rangle$ calculated for $\kappa\to 1$ and $y = 0$. BLN
neglected the variations of $p_\mathrm{eq}$ about the equal occupation point $x_0$.
Using $p_\mathrm{eq}(x)\approx \Theta(x_0-x)$, we find their solution for the branch
occupation directly by integrating Eq.~\eqref{eq:delta_p}. If $x<x_0$, then
$p(x) = 1$ while for $x>x_0$
\begin{align}\label{eq:p_BN}
   p_{\scriptscriptstyle \mathrm{BN}}(x) = \exp\Bigl[-\!\!\int_{x_0}^x\!\!
   \frac{d x'}{\ell_\mathrm{eq}(x')}\Bigr]
   \approx \exp\!\left[-\dfrac{x-x_0}{\ell_\mathrm{eq}(x_0)}\right]\! ,
\end{align}
where in the second step, we used the fact that the scale of variations of
$\ell_\mathrm{eq}(x)$ is small compared to $\ell_\mathrm{eq}(x_0)$ itself. The solution of BLN thus
decays on the short scale $\ell_\mathrm{eq}(x_0)$ while the step in our solution is
governed by the variation of $p_\mathrm{eq}(x)$ changing on the larger scale $T/|\Delta
f_\mathrm{pin}|$, see Eq.~\eqref{eq:p_eq4}. Adopting the approximation of BLN, the
average pinning force Eq.~\eqref{eq:f_pin4} becomes
\begin{align}
   \langle f_\mathrm{pin} \rangle =\nu\frac{|\Delta f_\mathrm{pin}(x_0)|}{a_0}\ell_\mathrm{eq}(x_0),
\end{align}
with the numerical factor $\nu = 2$ (the result in Ref.\
[\onlinecite{BrazovskiiNattermann2004}] is reduced by half since it is calculated per
one multivalued interval), however, our result \eqref{eq:f_pin_final} with
$\Gamma(\kappa\to 1) = 0$ and $\omega_\mathrm{dp} = \omega_\mathrm{p}$ results in a different
factor $\nu = \pi/2$. As the parametric dependence of the area between $p(x)$
and $p_\mathrm{eq}(x)$ turns out to be the same in BLN and in our case, the results
differ only by a numerical factor. Our result \eqref{eq:f_pin_final} for
$\langle f_\mathrm{pin} \rangle$ thus provides a more accurate and universal (as it
deals with non-symmetric pinning landscapes) result than BLN.

\subsubsection{Current--velocity characteristic}

Inserting the result \eqref{eq:F_pin_small_v} for the pinning-force density
$F_\mathrm{pin}$ back to the equation of motion \eqref{eq:force_balance} provides the
current--voltage characteristic at small velocities $v \ll v_{\rm
\scriptscriptstyle TAFF} = v_\mathrm{th} e^{-U_0/T}$,
\begin{align}
   \frac{v}{v_c}\left[1+h(\kappa)n_p a_0\xi^2e^{U_0/T}\right] = \frac{j}{j_c}.
\end{align}
The exponential term dominates for low temperatures, $T \ll U_0/
|\ln\,[h(\kappa)\, n_p a_0 \xi^2]|$ and the slope of the characteristic is
exponentially suppressed compared to the slope $v_c/j_c$ describing free
flux-flow,
\begin{align} \label{eq:TAFF_char}
   v = \frac{v_c}{j_c}\frac{\exp(-U_0/T)}{h(\kappa)\,n_p a_0\xi^2}\,j.
\end{align}
The activation barrier $U_0$ does not depend on $j$ but remains constant, see
Eq.\ \eqref{eq:U0}, different from the weak collective pinning scenario where
the creep barriers $U(j) \propto j^{-\mu}$ diverge as $j \to 0$. As a result,
the glassy response of a true superconductor in the weak collective pinning
framework is replaced by a resistive normal metallic behavior in the strong
pinning setting. The exponential reduction of the normal resistance is known
under the name TAFF, thermally assisted flux flow \cite{Kes1989}.  The
crossover to high velocities with a non-linear characteristic at $v_{\rm
\scriptscriptstyle TAFF} = v_\mathrm{th}e^{-U_0/T}$ is realized at the
driving current $j_{\rm \scriptscriptstyle TAFF} = (v_\mathrm{th}/v_c)
h(\kappa) (n_p a_0\xi^2)\, j_c = a(\kappa) h(\kappa) (T/e_p)\, j_c$.

\section{Summary and conclusion}

Including thermal fluctuations into the strong pinning paradigm, we have
advanced this quantitative theory of pinning by a further important step.  As
a result, we have arrived at a rather comprehensive picture of strong pinning
that provides us with the critical current density \cite{Blatter2004}, the
current--voltage characteristic \cite{Thomann2012,Thomann2017}, and its
smoothing due to thermal fluctuations. Furthermore, the parametric conditions
for the validity of strong pinning theory have been identified
\cite{Blatter2004}. An important result of the present study is the obtained
insight on the form of the current--voltage characteristic which is of the
excess-current form---thermal fluctuations leave this simple form essentially
unchanged. From this finding, we conclude that pinning and creep are preserved
when driving the system above the critical current density $j_c$. This is in
contrast to the usual perception of a steep rise in dissipation appearing
above $j_c$, an expectation that may be understood in terms of avalanche
formation. The strong pinning paradigm, though, is rather in agreement with
Coulomb's law of dry friction, telling that the friction force remains
unchanged when motion sets in.

The present work provides us with a precise prediction for the
current--voltage characteristic at finite temperatures. We have found that the
role of $j_c$ is taken over by the depinning current density $j_\mathrm{dp}(T)$
separating flat and steep regions of the characteristic, with $j_\mathrm{dp}(T)$
reduced with respect to $j_c$ by a term $\propto (T/e_p)^{2/3}$, see Eq.\
\eqref{eq:j_dp_s}. Below $j_\mathrm{dp}(T)$, vortex motion is determined by creep
over barriers $U(j) \approx U_c (1-j/j_c)^{3/2}$, see Eq.\
\eqref{eq:U_small_j}.  The exponent $3/2$ is universal for smooth pinning
potentials and derives from the exponent describing the vanishing of depinning
and pinning barriers at the boundaries of the bistable region $[x_{\scriptscriptstyle -},
x_{\scriptscriptstyle +}]$, see Eq.\ \eqref{eq:U_onset_gen}.  The motion above $j_\mathrm{dp}(T)$ is
flux-flow like until joining the $T=0$ characteristic at $v_\mathrm{th} \sim
(T/e_p)\kappa v_p$, $v_p \sim f_p/a_0^3 \eta$ the velocity scale for vortex
motion within a pinning well. Formulating this flow-type motion through
barriers, the latter exhibit a weak logarithmic dependence on $j$, see Eq.\
\eqref{eq:U_large_j2}, and hence a linear rise in dissipation.  

Creep barriers are conveniently measured through relaxation experiments; our
result with the $3/2$ exponent in the activation barriers provides us with the
normalized creep rate $S(T,t)$ in Eq.\ \eqref{eq:creep_rate} with an initial
non-linear and convex increase with temperature $T$ that transforms into a
concave shape at higher temperatures; as a result, the relaxation rate $S(T)$
exhibits an inflection point.  Finally, the temperature scale where pinning
stops suppresssing the flow-type vortex motion is given by the thermal
depinning temperature $T_\mathrm{dp}$ that is of order $e_p$, see Eq.\
\eqref{eq:T_dp}.  Focussing the discussion on low drives $v <
v_{\rm\scriptscriptstyle TAFF} = v_\mathrm{th}e^{-U_0/T}$ with $U_0$ the
maximal barrier appearing at the branch cutting point, we have found a
quantitative result for the thermally assisted flux-flow characteristic $v/v_c
\propto e^{-U_0/T} (j/j_c)$, see Eq.\ \eqref{eq:TAFF_char}.

Several findings in the present paper are amenable to experimental
verification, foremost, the thermal modifications of the excess-current
characteristics predicted by strong pinning theory. Although systematic data
is scarce, we have found that transport experiments on NbSe$_2$
\cite{Xiao2002, Xiao_unpublished} and on MoGe \cite{Roy2019, Pratap_unpublished} can be
successfully analyzed in terms of our strong pinning theory; we will devote a
separate publication to this topic \cite{Buchacek2019}.  Regarding creep rates,
indications for an inflection point have been reported in experiments on
pnictide and cuprate superconductors \cite{Civale2017}. Finally, thermally assisted
flux-flow has been experimentally observed and quantitatively analyzed in high
temperature superconductors \cite{Iye1987,Palstra1988,Tinkham1988}.

Within the strong pinning paradigm, pinning due to individual defects is
finite.  As $\kappa$ drops below unity, individual pins cannot hold a vortex
any longer (since the energy landscape is single valued and hence the jumps
$\Delta e_\mathrm{pin} = 0$ vanish). As a result, collective pinning effects
due to multiple defects have to be included in the pinning analysis around
$\kappa \simeq 1$. This is also the case when the current drive vanishes or
approaches its critical value---in both cases corrections from neighboring
defects become important and hence our results will get modified in these
regimes. Other topics of interest are the decay of lattice order
\cite{Larkin1970} and the physics of one-dimensional strong pinning, a regime
enclosed between 3D strong pinning and 1D weak collective pinning within the
pinning diagram of Ref.~[\onlinecite{Blatter2004}], and future work will
address several of these issues.

\acknowledgements
We thank Eva Andrei, Alexei Koshelev, Zhi-Li Xiao, and Eli Zeldov for
illuminating discussions and acknowledge financial support of the Swiss
National Science Foundation, Division II and the PostDoc Mobility fellowship (R.W.).

\appendix

\section{Marginally strong pinning}\label{sect:APP_mod_strong}

\begin{figure}
\includegraphics[width = 8truecm]{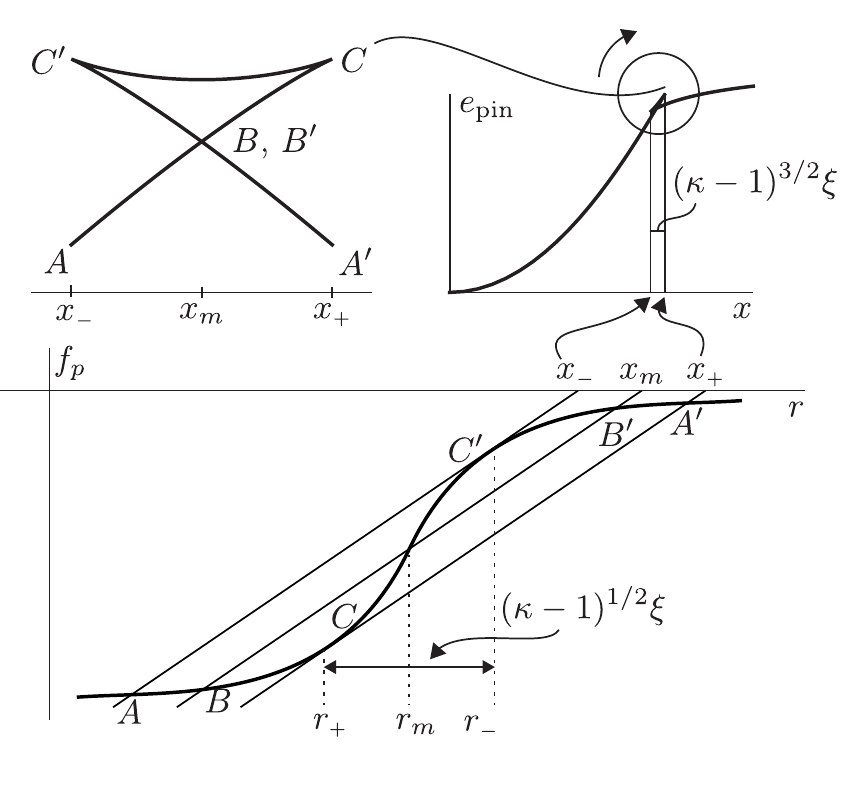}
\caption{In the marginally strong pinning regime, the energy landscape is
fully characterized by an expansion of the bare pinning force (bottom) around
the inflection point $r_m$ up to the order $(r_m-r)^3$. The pinning force is
locally symmetric around the inflection point $r_m$, which gives rise to the
symmetric pinning energy landscape $e_\mathrm{s,pin}^i(x)$ (top left) and
implies that $x_m = r_m - f_p(r_m)/\bar{C}$ corresponds to the branch crossing
point $x_0$, $x_m = x_0$.}
\label{Fig:weak}
\end{figure}

At marginally strong pinning, the slope $\bar{C}$ is close to (but smaller than)
the maximum slope of the bare pinning force $f_p(r)$ and the multivalued
solutions for the tip position $r(x)$ reside in the vicinity of the
inflection point $r_m$, ${f_p''(r_m)=0}$, see Fig.\ \ref{Fig:weak}. The
characteristics of the pinning energy landscape then can be obtained from an
expansion of the pinning force around $r_m$,
\begin{align}\label{eq:f_p_expansion}
   f_p(r_m+\delta r) = f_p(r_m)+\kappa \bar{C}\delta r
   -\frac{1}{3}\gamma \,\delta r^3+\cdots
\end{align}
with $\kappa \bar{C} = f_p'(r_m)$, see Eq.\ \eqref{eq:Labusch}, and the third derivative
${\gamma = -f_p'''(r_m)/2>0}$, $\gamma \sim f_p/\xi^3$. As long as the higher
order terms in \eqref{eq:f_p_expansion} can be neglected, the bare pinning
force is antisymmetric about the inflection point,
\begin{align}\label{eq:f_symmetry}
   f_p(r_m+\delta r)-f_p(r_m)=-[f_p(r_m-\delta r)-f_p(r_m)].
\end{align}
The line with slope $\bar{C}$ passing through the inflection point defines the
asymptotic vortex position $x_m$ related to $r_m$, $x_m = r_m-f_p(r_m)/\bar{C}$,
see Eq.\ \eqref{eq:equilibrium} and Fig.\ \ref{Fig:weak}.  The symmetry
property \eqref{eq:f_symmetry} implies that $x_m = (x_{\scriptscriptstyle +}+x_{\scriptscriptstyle -})/2$.

Making use of the relation \eqref{eq:f_symmetry}, we determine the derivatives of
the pinning energy for the branches $i = \mathrm{p,f}$ of the multi-valued
energy landscape, see Eq.\ \eqref{eq:effective_energy},
\begin{align}\label{eq:e_pin_derivative}
   &\frac{d}{d x}e_\mathrm{pin}^i(x) = \frac{d}{d x}e_\mathrm{pin}[x,r_i(x)] = -f_p[r_i(x)]r_i'(x)\\
   \nonumber
   &\quad +\bar{C}[r_i(x)-x][r_i'(x)-1] = -f_p[r_i(x)] = -f_\mathrm{pin}^i(x).
\end{align}
Subtracting the slope $(x-x_m)f_p(r_m)$ from $e_\mathrm{pin}^i(x)$, we obtain the
symmetrized energy landscape
\begin{align}\label{e_pin_symmetrized}
   e_\mathrm{s,pin}^i(x) \equiv e_\mathrm{pin}^i(x)+(x-x_m)f_p(r_m),
\end{align}
symmetric around the branch crossing point at $x_m$, see Fig.\ \ref{Fig:weak}.
In particular, the difference in the tilted energies $e_\mathrm{s,pin}^i(x)$
between the points $B'$ and $B$ (as defined in the force diagram of Fig.\
\ref{Fig:weak}) can be expressed as an integral from $r_{\scriptscriptstyle +}$ to $r_{\scriptscriptstyle -}$; this
quantity vanishes due to the symmetry \eqref{eq:f_symmetry},
\begin{align} \nonumber
   \begin{split}
   &e_\mathrm{pin}^\mathrm{p}(x_m)-e_\mathrm{pin}^\mathrm{f}(x_m) = 
   \int_{B}^{B'}\!\!\! d x\, \left\lbrace -f_p[r(x)]+f_p(r_m) \right\rbrace =0,
   \end{split}
\end{align}
implying that $x_m$ indeed corresponds to the branch crossing point, $x_m =
x_0$. Differentiating Eq.~\eqref{eq:equilibrium} and evaluating the second
derivative $d^2 e_\mathrm{pin}^i(x)/d x^2$ gives
\begin{align}
   \frac{d^2 e_\mathrm{pin}^i(x)}{d x^2} 
   = \frac{\bar{C}}{1-\bar{C}/f_p'[r_i(x)]}.
\end{align}
We conclude that the energy landscape is concave for the pinned and free
branches (for which the curvature $\lambda_{\mathrm{p},\mathrm{f}} = \bar{C} -
f_p'[r_i(x)] > 0$) and convex for the unstable branch for which
$\lambda_{\mathrm{us}} < 0$.

All quantities characteristic for strong pinning, such as the bistable region
$[x_{\scriptscriptstyle -},x_{\scriptscriptstyle +}]$, the jumps $\Delta
e_\mathrm{pin}(x_{\scriptscriptstyle \pm})$, the energy barriers
$U_{\mathrm{p},\mathrm{dp}}(x)$, etc., must vanish at the crossover point
$\kappa=1$; they scale with a power of the small parameter $\kappa-1$. In
particular, as a straightforward calculation shows, we find the extremal tip
locations $r_{\scriptscriptstyle \pm} = r_m \mp \delta r_{\max}$ from the
condition $f_p'(r_{\scriptscriptstyle \pm}) = \bar{C}$,
\begin{align}\label{eq:delta_r_max}
   \delta r_{\max} = \sqrt{\frac{\bar{C}}{\gamma}}(\kappa-1)^{1/2}
   \sim (\kappa-1)^{1/2}\xi
\end{align}
from which the end-points $x_{\scriptscriptstyle\pm} = x_m \pm \delta
x_{\max}$ of the multi-valued interval appear at
\begin{align}
   \delta x_{\max} =
   \frac{2}{3}\sqrt{\frac{\bar{C}}{\gamma}}(\kappa-1)^{3/2}\sim(\kappa-1)^{3/2}\xi,
\end{align}
see \eqref{eq:end_points}, with the dimensional estimate $\gamma \sim
f_p/\xi^3$ and $\kappa \sim 1$ used in the last relation.

The results for the curvatures, barriers, and frequency factors near the
multi-valued interval end points $x_{\scriptscriptstyle \pm}$ are obtained
using $f_p''(r_{\scriptscriptstyle +}) = -f_p''(r_{\scriptscriptstyle -}) =
2\gamma\delta r_{\max}$,
\begin{align}\label{eq:lambda_U_weak}
   \lambda_{\mathrm{f},\mathrm{us}}(x_{\scriptscriptstyle -} + \delta x) 
   &= \pm\sqrt{\frac{8}{3}}(\kappa-1)\bar{C}
   \left(\frac{\delta x}{\delta x_{\max}}\right)^{1/2},\\
   U_\mathrm{p}(x_{\scriptscriptstyle -} +\delta x) 
   &= \frac{4\bar{C}^2}{3\gamma}(\kappa-1)^2
   \left(\frac{2\delta x}{3\delta x_{\max}}\right)^{3/2},\\
   \omega_{\mathrm{f}}(x_{\scriptscriptstyle -}+\delta x)
   &=\frac{\bar{C}}{2\pi\eta a_0^3}
   \sqrt{\dfrac{8}{3}}(\kappa-1)
   \left(\dfrac{\delta x}{\delta x_{\max}}\right)^{1/2},
\end{align}
and corresponding results hold for $\lambda_{\mathrm{p},\mathrm{us}}$,
$U_\mathrm{dp}$, and $\omega_\mathrm{p}$ evaluated at $x_{\scriptscriptstyle
+} - \delta x$.

In order to find the force jumps $\Delta f_\mathrm{pin}(x_{\scriptscriptstyle
+}) = -\Delta f_\mathrm{pin}(x_{\scriptscriptstyle -})$, we first determine
the tip location $r_\mathrm{f}$ of the free solution at the interval end
point, $r_\mathrm{f}(x_{\scriptscriptstyle +}) = r_m+2\delta r_{\max}$; then
\begin{align}\nonumber
   \Delta f_\mathrm{pin}(x_{\scriptscriptstyle +}) &= f_p(r_m-\delta r_{\max})
   -f_p(r_m+2\delta r_{\max})\\
   \label{eq:Delta_f_weak}
   =& -3\bar{C}\sqrt{\frac{\bar{C}}{\gamma}}(\kappa-1)^{1/2}
   \sim -f_p(\kappa-1)^{1/2}.
\end{align}
A more lengthy calculation yields the energy jumps \cite{Labusch1969,Blatter2004}
\begin{align}
   \nonumber
   &\Delta e_\mathrm{pin}(x_{\scriptscriptstyle +}) 
   = e_p(r_{\scriptscriptstyle +})+\frac{1}{2}
   \bar{C}(x_{\scriptscriptstyle +}-r_{\scriptscriptstyle +})^2 
   -e_p[r_\mathrm{f}(x_{\scriptscriptstyle +})]\\
   &-\frac{1}{2}\bar{C}(x_{\scriptscriptstyle +}
   -r_\mathrm{f}(x_{\scriptscriptstyle +}))^2 
   = \frac{9\bar{C}^2}{4\gamma}(\kappa-1)^2
   \label{eq:delta_ec_weak}
\end{align} 
and $\Delta e_\mathrm{pin}(x_{\scriptscriptstyle -}) = -\Delta
e_\mathrm{pin}(x_{\scriptscriptstyle +})$.

Next, we determine various quantities at the branch crossing point $x_0 =
x_m$.  The three solutions for the tip locations $r_i(x)$ at the branch
crossing point are $r_\mathrm{us}(x_m) =r_m$ and $r_{\mathrm{p},
\mathrm{f}}(x_m) = r_m \mp \sqrt{3} \delta r_{\max}$.  Using
Eq.~\eqref{eq:lambda}, we find the curvatures and attempt frequencies
\begin{align}\label{eq:Delta_e_pin_xplus_moderate}
   &\lambda_\mathrm{us}(x_0)=-\bar{C}(\kappa-1),\quad \lambda_{\mathrm{p},\mathrm{f}}(x_0) =
   \frac{\bar{C}}{2}(\kappa-1),\\ &\omega_\mathrm{f}(x_0) = \omega_\mathrm{p}(x_0) =
   \frac{\bar{C}(\kappa-1)}{2\pi\sqrt{2}\eta a_0^3}.
\end{align}
The force difference between pinned and free branches is
\begin{align}
   \Delta f_\mathrm{pin}(x_0) = -2\bar{C}\sqrt{\frac{3\bar{C}}{\gamma}}(\kappa-1)^{1/2},
\end{align}
the energy difference $\Delta e_\mathrm{pin}(x_0)$ vanishes. The maximum energy
barrier $U_0 = U_\mathrm{dp}(x_0)=U_\mathrm{p}(x_0)$ is
\begin{align}\label{U_max_weak}
   U_0 &= e_\mathrm{pin}(x_0,r_m)-e_\mathrm{pin}(x_0,r_m+\delta r_0)\nonumber\\
   &= \frac{3\bar{C}^2}{4\gamma}(\kappa-1)^2.
\end{align}

We further calculate the various scaling functions discussed in the text. The
coefficients $\kappa_{\scriptscriptstyle \pm}$ appearing in the curvatures and barrier
expansions \eqref{eq:lambda_onset_gen}--\eqref{eq:U_onset_gen2} read
\begin{align}
  \kappa_{\scriptscriptstyle \pm} = \xi \sqrt{\frac{\gamma}{\bar{C}}} (\kappa-1)^{1/2} 
  \sim  (\kappa-1)^{1/2}.
\end{align}
The scaling function $g(\kappa)$ appearing in the pinning-force density
$F_\mathrm{pin}$, see Eq.\ \eqref{eq:g}, takes the form
\begin{align}
   g(\kappa) = \left[\frac{4\gamma^2e_p^2}{3(\kappa-1)^4\bar{C}^4}\right]^{1/3}
   \sim \frac{1}{(\kappa-1)^{4/3}},
\end{align}
and the function $a(\kappa)$ entering the characteristic at high velocities,
see Eq.~\eqref{eq:a}, is given by
\begin{align}
   a(\kappa) = \frac{1}{9\pi} \frac{e_p\xi^2}{x_m\bar{C}}
   \left(\frac{\gamma/\bar{C}}{\kappa-1}\right)^{3/2}
   \sim \frac{1}{(\kappa-1)^{3/2}}.
\end{align}
The pinning-force density $F_\mathrm{pin}$ at low velocities involves the scaling
function $h(\kappa)$ of Eq.\ \eqref{eq:h},
\begin{align}
   h(\kappa)\approx
   \frac{\pi^2\sqrt{6}}{4}\frac{x_0}{\xi^2}\left(\frac{\bar{C}/\gamma}{\kappa-1}
   \right)^{1/2} \sim \frac{1}{(\kappa-1)^{1/2}}.
\end{align}

\section{Very strong pinning}\label{sect:APP_very_strong}

\begin{figure}
\centering \includegraphics[width = 8truecm]{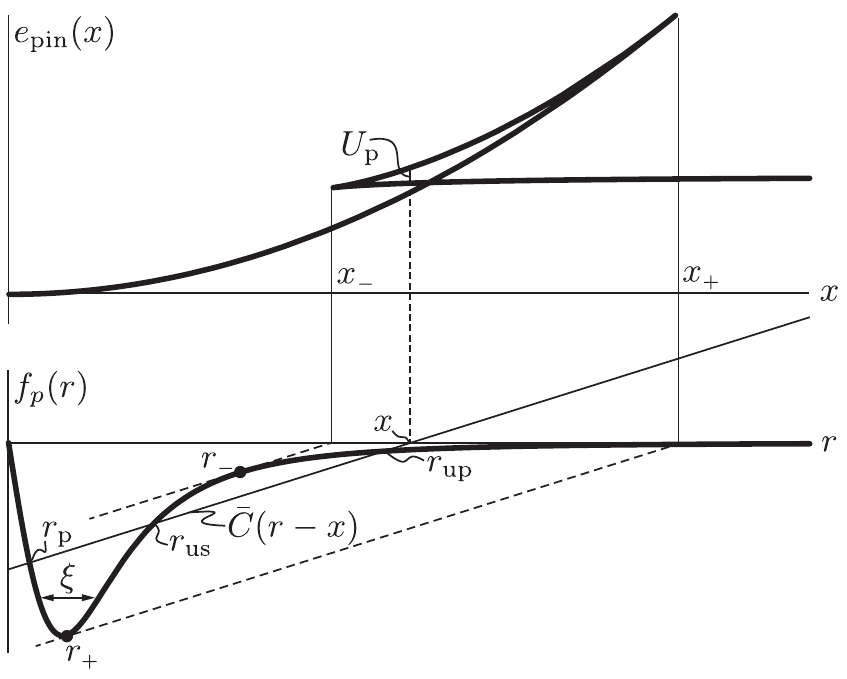}
\caption{For very strong pinning with $\kappa\gg 1$, the slopes $f'_p(r)$ are
steep compared to the slope $\bar{C}$ of the elastic restoring force. This results
in an extended multistable region between $x_{\scriptscriptstyle -} \sim \xi
\kappa^{1/(1+n)}$ and $x_{\scriptscriptstyle +} \sim \kappa \xi$, wherein the tip
positions of the free and pinned states assume values $r_\mathrm{f}\approx x$ and
$r_p\sim x/\kappa$, while the unstable solution lies between $r_{\scriptscriptstyle +} \sim \xi$
and $r_{\scriptscriptstyle -}\sim \kappa^{1/(1+n)}\xi$. The extent of the multivalued
region is $x_{\scriptscriptstyle +} - x_{\scriptscriptstyle -} \sim \kappa\xi$.} \label{Fig:vs_pinning}
\end{figure}

When pinning is very strong, i.e., the slope $\bar{C}$ is small compared to the
maximum slope of $f_p(r)$, see Fig.\ \ref{Fig:vs_pinning}, the relevant
expansions of the pinning force are around the origin when investigating the
pinned branch and in the tail when dealing with the free branch. Here, we
discuss the pinning energy for bare pinning potentials $e_p(r)$ decaying on a
scale $\xi$ from its minimal value $-e_p = e_p(0)$. The induced pinning force
$f_p(r) = -e_p'(r)$ quickly reaches its maximal value $f_p \sim e_p/\xi$ on
the distance of the vortex core size $r \sim \xi$. We assume an algebraically
decaying tail of the bare pinning potential, $e_p(r)\sim -e_p (r/\xi)^{-n}$,
$r\gg \xi$, for instance, $n = 2$ for the commonly-used Lorentzian pinning
potential
\begin{align}\label{eq:Lorenzian}
   e_p(r) = -\frac{e_p}{1+r^2/2\xi^2}.
\end{align}

Whenever convenient, we will make use of the parametric dependence $\kappa\sim
f_p/\bar{C}\xi\sim e_p/\bar{C}\xi^2$. The pinning energy profile $e_\mathrm{pin}(x;r)$ in Fig.\
\ref{Fig:metastable_states} involves the superposition of a shallow (elastic)
parabola $\bar{C}(r-x)^2/2$ centered at $r = x$ and a narrow bare pinning potential
$e_p(r)$ centered at the origin. For $x = 0$ the vortex is placed directly at
the pinning center and $e_\mathrm{pin}(r)$ has only one minimum $r_\mathrm{p} = 0$
corresponding to the pinned branch. A second minimum with a tip position at
$r_\mathrm{f}(x)$ appears at the point $x_{\scriptscriptstyle -}$. Since the point $r_{\scriptscriptstyle -}  = r_\mathrm{f}(x_{\scriptscriptstyle -})$
lies on the tail of the bare pinning force $f_p(r)$, evaluating the condition
$f_p'(r_{\scriptscriptstyle -})=\bar{C}$ gives $r_{\scriptscriptstyle -} \sim \xi \kappa^{1/(1+n)}$. The corresponding
asymptotic position of the vortex is $x_{\scriptscriptstyle -} = r_{\scriptscriptstyle -}-f_p(r_{\scriptscriptstyle -})/\bar{C}\sim \xi
\kappa^{1/(1+n)}$.

The new (free) minimum becomes deeper as the asymptotic position $x$ is further
increased away from the defect. Both minima are of equal energy at the
branch-crossing point when $\bar{C} x_0^2/2 \approx  e_p$ and hence $x_0\approx
\sqrt{2 e_p/\bar{C}} \sim \kappa^{1/2}\xi$.  Finally, the original (pinned) minimum
disappears for very large $x$ when the slope of the parabola balances the
maximum pinning force $\bar{C} x_{\scriptscriptstyle +}\approx f_p$ and we obtain $x_{\scriptscriptstyle +}\sim
\kappa\xi$.

To evaluate the scaling form of curvatures, barriers, and frequencies near the
end points $x_{\scriptscriptstyle +}$ and $x_{\scriptscriptstyle -}$ of the
multivalued interval, we use Eqs.\
\eqref{eq:lambda_onset_gen}--\eqref{eq:U_onset_gen2} with
$f_p''(r_{\scriptscriptstyle -})\sim (f_p/\xi^2) \kappa^{-\nu}$, $\nu =
(n+2)/(n+1)$ and $f_p''(r_{\scriptscriptstyle +})\sim (f_p/\xi^2)$. As a
result, we find a different behavior of the barriers and curvatures near the
two end points $x_{\scriptscriptstyle \pm}$,
\begin{align}\label{eq:lambda_onset_strong}
   \begin{split}
      \lambda_\mathrm{f,us}(x_{\scriptscriptstyle -}+\delta x_{\scriptscriptstyle -})
      &\sim \pm\frac{e_p}{\kappa^{\nu/2}\xi^2}
      \left(\frac{\delta x_{\scriptscriptstyle -}}{\kappa \xi}\right)^{1/2},\\
       \lambda_\mathrm{p,us}(x_{\scriptscriptstyle +}-\delta x_{\scriptscriptstyle +})
      &\sim \pm \frac{e_p}{\xi^2}
      \left(\frac{\delta x_{\scriptscriptstyle +}}{\kappa \xi}\right)^{1/2},
   \end{split}
\end{align}
and
\begin{align}
  \begin{split}
    U_\mathrm{p}(x_{\scriptscriptstyle -}+\delta x_{\scriptscriptstyle -}) &\sim e_p\kappa^{\nu/2}
    \left(\frac{\delta x_{\scriptscriptstyle -}}{\kappa\xi}\right)^{3/2},\\
    U_\mathrm{dp}(x_{\scriptscriptstyle +}-\delta x_{\scriptscriptstyle +}) &\sim e_p
    \left(\frac{\delta x_{\scriptscriptstyle +}}{\kappa\xi}\right)^{3/2}.
  \end{split}
\end{align}
Note the additional large factor $\kappa^{\nu/2}$ that appears in connection
with quantities evaluated near $x_{\scriptscriptstyle -}$.

For positions $x$ far away from the boundaries $x_{\scriptscriptstyle \pm}$,
the curvatures are dominated either by the shallow parabolic well with
$\lambda_\mathrm{f}(x\gg x_{\scriptscriptstyle -})\approx \bar{C}\sim
e_p/\kappa\xi^2$ on the free branch or by the narrow pinning potential well on
the pinned branch, $\lambda_\mathrm{p}(x\ll x_{\scriptscriptstyle +})\approx
e_p''[r_\mathrm{p}(x)]\sim e_p/\xi^2$. For the unstable solution, we have
$r_\mathrm{us}(x)<r_{\scriptscriptstyle -} \ll x$ and hence $r_\mathrm{us}$
resides in the tail of $f_p(x)$ for $x\ll x_{\scriptscriptstyle +}$. Then, the
equilibrium Eq.\ \eqref{eq:equilibrium} reads $\bar{C} x \sim
f_p(r_\mathrm{us}/\xi)^{-n}$ and $r_\mathrm{us} \sim \xi\,
(x/\kappa\xi)^{-1/n}$.  Evaluating the curvature in this situation, we find
that $\lambda_\mathrm{us}(x) \sim -(e_p/\xi^2) (x/\kappa\xi)^{(n+1)/n}$. The
scaling forms for the frequency factors follow from Eq.\ \eqref{eq:omega} and
are summarized in Table \ref{table:omega}.
%

In order to find the barrier near the branch crossing point $x_0$, we make use
of Eq.\ \eqref{eq:e_pin_derivative} and integrate away from
$x_{\scriptscriptstyle -}$,
\begin{align}
   U_\mathrm{p}(x) &= \int_{x_{\scriptscriptstyle -}}^{x}d x'\, 
   \left\lbrace f_p[r_\mathrm{f}(x)]-f_p[r_\mathrm{us}(x)]\right\rbrace.
\end{align}
Using the equilibrium condition \eqref{eq:equilibrium} for the tip position
and $r_\mathrm{f}(x)\approx x$, this simplifies to 
\begin{align}\label{eq:U_middle_strong}  
   U_\mathrm{p}(x) &= \bar{C} \int_{x_{\scriptscriptstyle -}}^{x}
   d x'\, [r_\mathrm{f}(x')-r_\mathrm{us}(x')]\\
   \nonumber
   &\approx \frac{1}{2}\bar{C} (x^2-x_{\scriptscriptstyle -}^2)
   -\bar{C}\int_{x_{\scriptscriptstyle -}}^x d x'\, r_\mathrm{us}(x').
\end{align}
We make use of the scaling form of $r_\mathrm{us}(x)$ derived above and
note that the integral is dominated by its upper boundary, resulting
in 
\begin{align}
   \frac{1}{2}\bar{C} (x^2-x_{\scriptscriptstyle -}^2) - U_\mathrm{p}(x)
   =\mu \, e_p\left( \frac{x}{\kappa \xi} \right)^{(n-1)/n},
\end{align}
where $\mu$ is a $\kappa$-independent numerical. The barrier $U_0 =
U_\mathrm{p}(x_0)$ at the branch crossing point $x_0 \sim \kappa^{1/2} \xi$
then is given by
\begin{align}\label{U_max_strong2}
   U_0 \sim \bar{C} \kappa \xi^2 \sim e_p,
\end{align}
with corrections of order $\sim -e_p/ \kappa^{(n-1)/2n}$.

\section{Equilibrium properties}\label{sect:APP_low_v}

We discuss the properties of the equilibrium occupation $p_\mathrm{eq}$ and
the relaxation length $\ell_\mathrm{eq}$ and expand both quantities around the
branch crossing point $x_0$.  Expanding $\Delta e_\mathrm{pin}(x_0+\delta
x)\approx -\Delta f_\mathrm{pin}(x_0)\delta x$ in the exponential and
neglecting the change in the frequency factors provides a good approximation
of $p_\mathrm{eq}(x)$ in the entire multivalued interval; when the change in
the frequency factors becomes significant ($\omega_\mathrm{p,up}$ vary on the
scale $\kappa\xi$ of the multivalued interval), the occupation is already
completely dominated by the exponential (which changes on a short scale
$T/|\Delta f_\mathrm{pin}(x_0)|$). We thus can rewrite Eq.~\eqref{eq:p_eq2} as
\begin{align}\label{eq:p_eq4}
   p_\mathrm{eq}(x_0+&\delta x)\approx\frac{1}{2}
   +\frac{1-(\omega_\mathrm{p}/\omega_\mathrm{f})
   e^{-\Delta f_\mathrm{pin} \delta x/T}} {2[1+(\omega_\mathrm{p}/\omega_\mathrm{f})
   e^{-\Delta f_\mathrm{pin}\delta x/T}]}\nonumber \\ 
   &=\frac{1}{2}-\frac{1}{2}\tanh
   \Bigl[\frac{1}{2}\ln \frac{\omega_\mathrm{p}}{\omega_\mathrm{f}} 
       - \frac{\Delta f_\mathrm{pin}}{2T}\delta x\Bigr].
\end{align}
The force difference between the branches scales as $\Delta f_\mathrm{pin}\sim
-f_p(\kappa-1)^{1/2},\,-f_p\kappa^{-1}$ in the limit of marginally and very
strong pinning. The ratio of frequency factors is $\omega_\mathrm{p} /
\omega_\mathrm{f}=\lambda_\mathrm{p}/\lambda_\mathrm{f}\sim \sqrt{\kappa}$ for
very strong pinning and reaches unity if pinning is marginally strong.
Determining the point of equal branch occupation
$p_\mathrm{eq}(x_\mathrm{eo})=1/2$, we then find it shifted to the left from
the branch crossing point,
\begin{align}\label{eq:x_eo}
   x_\mathrm{eo} \approx x_0 + \frac{T}{\Delta f_\mathrm{pin}}
  \ln\left.\frac{\omega_\mathrm{p}}{\omega_\mathrm{f}}\right|_{x_0} \!\!\!
   \approx x_0+\frac{T \ln\kappa}{2\Delta f_\mathrm{pin}(x_0)}.
\end{align}
This shift is comparable to the scale $T/|\Delta f_\mathrm{pin}|$ of
variations in $p_\mathrm{eq}(x)$ but is small (at low temperatures) compared
to the extension of the pinning landscape.


Next, we wish to understand the local behavior of $\ell_\mathrm{eq}(x)$ around
$x_0$.  We expand the depinning and pinning barriers in Eq.~\eqref{eq:ell_eq}
around $x_0$, $U_\mathrm{p,dp} = U_0+U_\mathrm{p,dp}'(x_0)\delta x$ and
neglect variations of the attempt frequencies
$\omega_{\mathrm{p},\mathrm{f}}(x)$,
\begin{align}\label{eq:ell_eq3}
   \ell_\mathrm{eq}(x_0+\delta x) \approx \frac{ ve^{U_0/T}}
   {\omega_\mathrm{p}\, e^{-U_\mathrm{dp}'\delta x/T} 
   +\omega_\mathrm{f}\, e^{-U_\mathrm{p}'\delta x/T}}.
\end{align}
The maximal barrier $U_0$ at $x_0$ is given by 
\begin{align}\label{eq:U0}
   U_0 = e_\mathrm{pin}(x_0,r_\mathrm{us})-e_\mathrm{pin}(x_0,r_\mathrm{p})
\end{align}
and vanishes as $e_p (\kappa-1)^2$ for marginally strong pinningi, while
asssuming a value of order $e_p$ with corrections of order $e_p/\kappa^{n
/(2n+2)}$ at very strong pinning, see Appendix~\ref{sect:APP_very_strong}.
For marginally strong pinning, the slopes of the barriers have equal
magnitude, $U_\mathrm{p}'(x_0) = -U_\mathrm{dp}'(x_0)$ while for very strong
pinning, we have $r_\mathrm{p}\sim\xi/\kappa$, $r_\mathrm{f} \sim x$, and
$r_\mathrm{us} \sim \xi(x/\kappa \xi)^{-1/\alpha}$ with $x = x_0\sim
\kappa^{1/2}\xi$. This provides us with the ratio
\begin{align}\label{eq:U_ratio}
   \frac{|U_\mathrm{dp}'|}{U_\mathrm{p}'} = \frac{f_\mathrm{pin}^\mathrm{us} 
   -f_\mathrm{pin}^\mathrm{p}}{f_p^\mathrm{f} -f_p^\mathrm{us}}
   = \frac{\bar{C} (r_\mathrm{us} -r_\mathrm{p})}{\bar{C}(r_\mathrm{f} 
   -r_\mathrm{us})} \sim \frac{\kappa^{1/2(n+1)}}{\kappa^{1/2}}
\end{align}
with $n$ the exponent describing the tails of the pinning energy relevant for
very strong pinning.

Maximizing the equilibrium relaxation length \eqref{eq:ell_eq3} with respect
to $\delta x$ then provides us with the location $x_{\mathrm{lr}}$ of the
longest relaxation length,
\begin{align}\label{eq:x_lr}
   x_{\mathrm{lr}} - x_0 \approx \frac{T}{\Delta f_\mathrm{pin}}
   \ln\frac{\omega_\mathrm{p}}{\omega_\mathrm{f}} \frac{|U_\mathrm{dp}'|}
   {U_\mathrm{p}'}\bigg|_{x_0}
   \!\!\!\!\!\sim  \frac{T\ln\kappa}{2(n \! + \! 1)\Delta f_\mathrm{pin}},
\end{align}
with $n$ the exponent describing the tails of the pinning energy relevant for
very strong pinning. Note that the maximum relaxation length
$\ell_\mathrm{eq}(x_\mathrm{lr})$ does not differ significantly from its value
at the point $x_0$. Substituting Eq.~\eqref{eq:x_lr} to Eq.~\eqref{eq:ell_eq3}
provides the estimate $\ell_\mathrm{eq}(x_\mathrm{lr})\sim
\ell_\mathrm{eq}(x_0)\kappa^{\alpha}$ with a small exponent $\alpha =
{|U_\mathrm{dp}'|/[2(n+1)|\Delta f_\mathrm{pin}|]}\sim
\kappa^{1/2(n+1)-1/2}/2(n+1)$ at very strong pinning (we use
$\omega_\mathrm{p}/\omega_\mathrm{f}\sim\sqrt{\kappa}$).  As a result, we find
that the various positions $x_0$, $x_\mathrm{eo}$, and $x_{\mathrm{lr}}$ are
arranged in the sequence $x_\mathrm{eo} < x_{\mathrm{lr}} < x_0$, see Fig.\
\ref{Fig:occupation_small_v} and note that $\Delta f_\mathrm{pin} < 0$; for
marginally strong pinning $\kappa\to 1$, the energy landscape becomes
symmetric and $x_\mathrm{eo} = x_\mathrm{lr} = x_0$.

Finally, we analyze the decay of $\ell_\mathrm{eq}$ away from its maximum. We
note that for $x > x_0$, $U_\mathrm{dp}(x) < U_\mathrm{p}(x)$ and therefore at
small temperatures $\ell_\mathrm{dp}(x) \ll \ell_\mathrm{p}(x)$, resulting in
$\ell_\mathrm{eq}(x)\approx \ell_\mathrm{dp}(x)\propto e^{U_\mathrm{dp}(x)/T}$
and thus $\ell_\mathrm{eq}(x)$ decays on the scale $T/|U_\mathrm{dp}'|$ to the
right of its maximum. Similarly if $x < x_0$, $\ell_\mathrm{eq}(x) \approx
\ell_\mathrm{p}(x) \propto e^{U_\mathrm{p}(x)/T}$ and thus $\ell_\mathrm{eq}$
grows on approaching $x_0$ from the left on the different scale
$T/U_\mathrm{p}'$, see the inset of Fig.\ \ref{Fig:occupation_small_v}.  For
very strong pinning, the ratio of growth and decay is $|U'_\mathrm{dp}|
/U_\mathrm{p}' < 1$, see Eq.~\eqref{eq:U_ratio}, while for marginally strong
pinning the growth and decay scales are identical.

\section{Current-velocity characteristic}\label{sect:APP_IV}

\subsection{Iteration scheme}
We solve the equation \eqref{eq:motion_expanded_2}, 
\begin{align}\label{eq:motion_expanded_2A}
   \frac{v}{v_\mathrm{th}} = \frac{1}{\mathcal{A}}\frac{\delta j}{j_c}
   +\frac{1}{\nu} \Bigl[\ln \frac{v_\mathrm{th}}{v}\Bigr]^{2/3}
\end{align}
for the current--velocity characteristic at the point $\delta j = 0$
corresponding to the critical drive. The iteration procedure for the solution
$x = v(j_c)/v_\mathrm{th}$ is given by
\begin{align}\label{eq:iteration_2}
    x_0 = 1/\nu,\qquad x_{n+1} = \frac{[\log (1/x_n)]^{2/3}}{\nu}.
\end{align}
The condition $\nu>1$ is not sufficient to ensure positivity of all
logarithms. For instance, if we consider $\nu = 1+\varepsilon$, with
$\varepsilon$ a small correction, we find that the logarithm in $x_3$ becomes
negative,
\begin{align}\label{eq:iteration_3}
   x_0&\approx 1-\varepsilon,\\
   x_1&\approx \varepsilon^{2/3},\\
   x_2&\approx \Bigl(\frac{2}{3}\log \frac{1}{\varepsilon}\Bigr)^{2/3},\\
   x_3&\approx \Bigl[\frac{2}{3}\log\frac{1}
   {(\frac{2}{3}\log 1/\varepsilon)^{2/3}}\Bigr]^{2/3}.
\end{align}
In fact, the iterative procedure can converge to the true solution $x^*$ of $x
= (1/\nu)[\ln (1/x)]^{2/3}$ only if
\begin{align}
   \left|\frac{\partial }{\partial x}\left.
   \frac{[\ln(1/x)]^{2/3}}{\nu}\right|_{x = x^*}\right|&<1,\\
   \frac{2}{3\nu x^* [\ln(1/x^*)]^{1/3} } = \frac{2}{3 \ln(1/x^*)}&<1,
\end{align}
i.e., the solution must satisfy $x^*<e^{-2/3}\equiv x_\mathrm{lim}$.
Substituting back to the equation fixing $x^*$, we find the limiting value of
$\nu$,
\begin{align}\label{eq:tau_lim}
   \nu_\mathrm{lim} = \frac{1}{x_\mathrm{lim}}\Bigl(\ln\frac{1}{x_\mathrm{lim}}\Bigr)^{2/3}
   = \Bigl(\frac{2 e}{3}\Bigr)^{2/3}\approx 1.49.
\end{align}
Results for Eq.\ \eqref{eq:motion_expanded_2A} at large and small values of
$\nu$ are given in the main text, see Eqs.\ \eqref{eq:v_jc} and
\eqref{eq:approx_sol}.

\subsection{Depinning current}

The definition of the depinning current density $j_\mathrm{dp}(T)$ as the
point of steepest change in the differential resistivity leads to the
condition $\partial^3 v/\partial j^3 = 0$. Assuming that we know the
expression $j(v)$, we need to use the chain rule repeatedly to arrive at
\begin{align}\label{eq:chain_rule_repeated}
\begin{split}
   \frac{\partial v}{\partial j} &= \Bigl(\frac{\partial j}{\partial v}\Bigr)^{-1},\\
   \frac{\partial ^2 v}{\partial j^2} &= \frac{\partial}{\partial v}
   \Bigl[\Bigl(\frac{\partial j}{\partial v}\Bigr)^{-1}\Bigr]
   \Bigl(\frac{\partial j}{\partial v}\Bigr)^{-1}=
   -\Bigl(\frac{\partial j}{\partial v}\Bigr)^{-3}\frac{\partial^2 j}{\partial v^2},\\
   \frac{\partial^3 v}{\partial j^3} &= -\frac{\partial}{\partial v}
   \Bigl[\Bigl(\frac{\partial j}{\partial v}\Bigr)^{-3}
   \frac{\partial^2 j}{\partial v^2}\Bigr]
   \Bigl(\frac{\partial j}{\partial v}\Bigr)^{-1}\\
   &= 3\Bigl(\frac{\partial j}{\partial v}\Bigr)^{-5}
   \Bigl(\frac{\partial^2 j}{\partial v^2}\Bigr)^2-
   \Bigl(\frac{\partial j}{\partial v}\Bigr)^{-4}\frac{\partial^3 j}{\partial v^3}.
\end{split}
\end{align}
The condition $\partial^3 v/\partial j^3 = 0$ is thus equivalent to
\begin{align}
   \frac{\partial^3 j}{\partial v^3}\frac{\partial j}{\partial v}
   = 3\Bigl(\frac{\partial^2 j}{\partial v^2}\Bigr)^2.
\end{align}
Substituting Eq.~\eqref{eq:motion_expanded_2} leads to the condition ($x =
v_\mathrm{dp}/v_\mathrm{th}$)
\begin{align}\nonumber
& 54\nu x[\ln(1/x)]^{7/3}-27\nu x[\ln(1/x)]^{4/3}+12\nu x[\ln(1/x)]^{1/3}\\ 
&\qquad -18[\ln(1/x)]^2+18\ln(1/x)+2 = 0. \label{eq:APP_v_dp}
\end{align}
We expect $x\lesssim 1/\nu$, with $\nu\gg 1$. In this limit, the approximate
solution to the above equation is found by balancing the term $54\nu x
(\ln(1/x)]^{7/3}$ against $18[\ln(1/x)]^2$. This simplifies the previous
equation to $3\nu x(\ln(1/x)]^{1/3} = 1$, which we solve iteratively to obtain
$x \approx 1/[3\nu (\ln 3\nu)^{1/3}]$.

To find the slope of the characteristic and the differential resistivity at
depinning, we need to evaluate the quantity $y = 3\nu x[\ln(1/x)]^{1/3}$ (see
Eq.\ \eqref{eq:rho}).  Rewriting Eq.~\eqref{eq:APP_v_dp} gives
\begin{align}\nonumber
   &y - \frac{y}{2 \ln(1/x)} + \frac{2}{9}\frac{y}{[\ln(1/x)]^2}-1\\
   \nonumber
   & \qquad\qquad +\frac{1}{\ln(1/x)}+\frac{1}{9[\ln(1/x)]^2} = 0
\end{align}
Treating the small parameter $[\ln(1/x)]^{-1}\approx (\ln 3\nu)^{-1}$
perturbatively, we arrive at the solution $y\approx 1- (2\ln 3\nu)^{-1}$.
Substituting further to the expression for differential resistivity gives
$\rho(j_\mathrm{dp})/\rho_\mathrm{ff} \approx (1/3)[1-(3\ln 3\nu)^{-1}]$.

\section{Scaling functions}\label{sect:APP_scal_fun}

The scaling functions $g(\kappa)$, $a(\kappa)$, and $h(\kappa)$ for marginally
strong pinning have been calculated in Appendix \ref{sect:APP_mod_strong} and
its asymptotic scaling for very strong pinning has been discussed in Secs.\
\ref{sec:high_velocities} and \ref{sec:low_velocities}. It remains to
determine the function $\varphi(\kappa)$ defined in Eq.\
\eqref{eq:F_pin_interm_v}; this can be obtained from an expansion of the
barriers and energy jumps around the branch crossing point $x_0$. With the
jumps realized at the points $x^\mathrm{jp}_{\scriptscriptstyle \pm} =
x_0+\delta x_{\scriptscriptstyle \pm}$, we use the expansions
\begin{align}
   U_\mathrm{dp}(x^\mathrm{jp}_{\scriptscriptstyle +}) 
   &= U_0+U_\mathrm{dp}'(x_0)\delta x_{\scriptscriptstyle +},\\
   \Delta e_\mathrm{pin}(x^\mathrm{jp}_{\scriptscriptstyle +}) 
   &= -\Delta f_\mathrm{pin}(x_0)\delta x_{\scriptscriptstyle +},\\
   U_\mathrm{p}(x^\mathrm{jp}_{\scriptscriptstyle -}) 
   &= U_0+U_\mathrm{p}'(-x_0)\delta x_{\scriptscriptstyle -},\\
   \Delta e_\mathrm{pin}(x^\mathrm{jp}_{\scriptscriptstyle -}) 
   &= -\Delta f_\mathrm{pin}(-x_0)\delta x_{\scriptscriptstyle -},
\end{align}
and the symmetries $U_\mathrm{p}'(-x_0) = -U_\mathrm{p}'(x_0)$ and $\Delta
f_\mathrm{pin}(-x_0)=-\Delta f_\mathrm{pin}(x_0)$. Setting
$U_\mathrm{dp}(x^\mathrm{jp}_{\scriptscriptstyle +}) =
U_\mathrm{p}(x^\mathrm{jp}_{\scriptscriptstyle -}) = U$, we rewrite the total
energy jump as
\begin{align}
   \Delta e_\mathrm{pin}^\mathrm{tot} 
   &= \Delta e_\mathrm{pin}(x^\mathrm{jp}_{\scriptscriptstyle +})
   -\Delta e_\mathrm{pin}(x^\mathrm{jp}_{\scriptscriptstyle -})\\
   &=\Delta f_\mathrm{pin}(x_0)\left[\frac{U_0-U}{U_\mathrm{dp}'(x_0)}
   -\frac{U_0-U}{U_\mathrm{p}'(x_0)}\right].
\end{align}
The last expression is simplified using $(U_\mathrm{p}-U_\mathrm{dp})' =
(e_\mathrm{pin}^\mathrm{p} - e_\mathrm{pin}^\mathrm{f})' = -\Delta
f_\mathrm{pin}$. To leading order, we can assume a constant trapping length
$x^\mathrm{jp}_{\scriptscriptstyle -} = -x_0$ and obtain
\begin{align} \label{eq:G_kp}
   \varphi(\kappa) &= \frac{x_0}{x_{\scriptscriptstyle -}\Delta e_c}
   \frac{\partial \Delta e_\mathrm{pin}^{\mathrm{tot}}}{\partial (U_0-U)/e_p}\\
   \nonumber
   &=\frac{x_0}{x_{\scriptscriptstyle -}}\frac{e_p}{\Delta e_c}
   \frac{\Delta f_\mathrm{pin}(x_0)^2}{|U_\mathrm{dp}'(x_0)|U_\mathrm{p}'(x_0)}.
\end{align}
In the marginally strong pinning regime, we use $x_{\scriptscriptstyle
-}\approx x_0$, $|U_\mathrm{dp}'(x_0)| = |f_p[r_\mathrm{p}(x)]-f_p(r_m)| =
|\Delta f_\mathrm{pin}(x_0)|/2$, and similarly $U_\mathrm{p}'(x_0) = |\Delta
f_\mathrm{pin}(x_0)|/2$, and hence
\begin{align}
   \varphi(\kappa)\approx \frac{8e_p\gamma}{9\bar{C}^2}(\kappa-1)^{-2},
\end{align}
where we have made use of Eq.\ \eqref{eq:delta_ec_weak}.  For very strong
pinning $\kappa\gg 1$, we use $x_0\sim \kappa^{1/2}\xi$,
$x_{\scriptscriptstyle -}\sim \kappa^{1/(1+n)}\xi$, $\Delta e_c\sim \kappa
e_p$, $|U_\mathrm{dp}'(x_0)| = \bar{C}[r_\mathrm{us}(x_0) -
r_\mathrm{p}(x_0)]\sim \bar{C}\xi \kappa^{1/2n}$, $U_\mathrm{p}' =
\bar{C}[r_\mathrm{f}(x_0) - r_\mathrm{us}(x_0)]\sim \bar{C}\xi\kappa^{1/2}$,
and $|\Delta f_\mathrm{pin}(x_0)| = \bar{C}[r_\mathrm{f}(x_0) -
r_\mathrm{p}(x_0)]\sim \bar{C}\xi\kappa^{1/2}$, and therefore
$\varphi(\kappa)\sim \kappa^{-\nu'}$ with the power $\nu' =
(3n+4)/2(n+1)(n+2)$.

\begin{figure}[h]
\includegraphics[width = 8truecm]{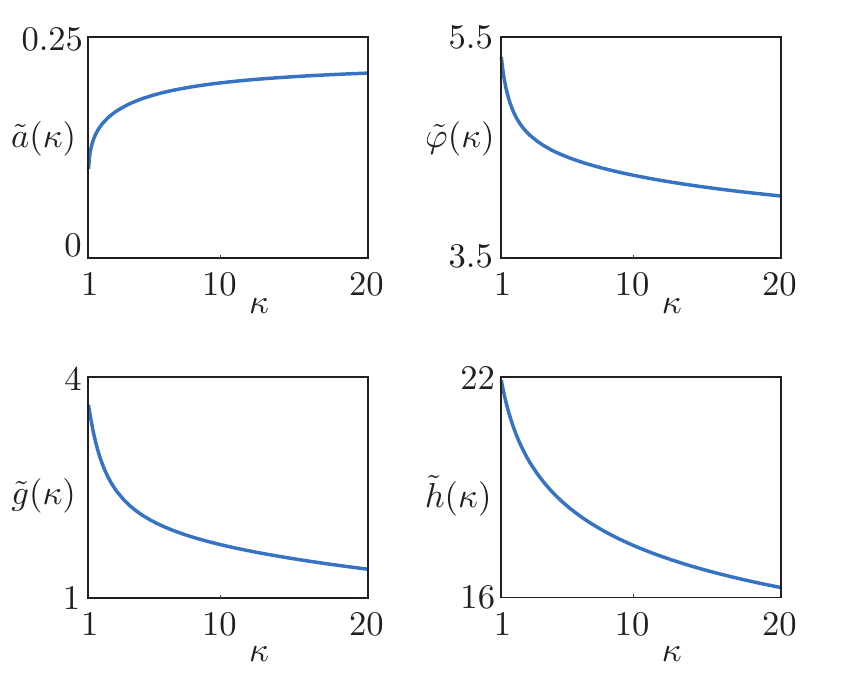}
\caption{Rescaled functions characterising the properties of pinning and creep
calculated for Lorentzian bare pinning potential. They depend on the pinning
strength as described by the Labusch parameter $\kappa$. Note the different
scales of vertical axes.}\label{Fig:scaling_functions}
\end{figure}

Finally, we plot in Fig.\ \ref{Fig:scaling_functions} the properly scaled
factors $\varphi(\kappa) (\kappa-1)^2 \kappa^{{\nu'}-2}$, $g(\kappa)(\kappa -
1)^{4/3}$, $a(\kappa)(\kappa-1)^{3/2} \kappa^{-(3n+4) /(2n+4)}$, and
$h(\kappa)(\kappa -1)^{1/2} \kappa^{-(3n+4)/(4n+4)}$.

\FloatBarrier

\section{Table of results}\label{sect:APP_table}
\begin{widetext}

\begin{table}
\caption{Properties of the pinning energy landscape for a defect with
asymptotic decay $e_p\sim r^{-n}$, $r\gg \xi$, in the limit of marginally
strong and very strong pinning.}
\centering
\setlength\extrarowheight{10pt}
\begin{tabular}{p{3cm}|p{4cm}|p{5cm}|p{4cm}}
Quantity & Exact formula & Marginally strong pinning & Very strong pinning\\
\hline
turning point & $f_p''(r_m) = 0$ & &\\
instability points & $f_p'(r_{\scriptscriptstyle \pm}) = \bar{C}$ & $r_\pm-r_m \sim \mp (\kappa-1)^{1/2}\xi$
   & $r_{\scriptscriptstyle -}\sim \kappa^{1/(n+2)}\xi,\,r_{{\scriptscriptstyle +}}\sim \xi$\\
end points & $x_{\scriptscriptstyle \pm} = r_{\scriptscriptstyle \pm}-\dfrac{f_p(r_{\scriptscriptstyle \pm})}{\bar{C}}$ 
   &  $x_{\scriptscriptstyle \pm} - x_m \sim \pm (\kappa - 1)^{3/2} \xi$ 
      & $x_{\scriptscriptstyle -} \sim \kappa^{1/(1+n)} \xi,\,x_{\scriptscriptstyle +} \sim \kappa \xi$\\
critical force density & $F_c = n_p\dfrac{2x_{\scriptscriptstyle -}}{a_0}\dfrac{\Delta e_{c}}{a_0}$ 
   & $(\xi/a_0)^2 n_p f_p (\kappa-1)^2$ & $(\xi/a_0)^2 n_p f_p \kappa^{(n+3)/(n+2)}$\\
frequency factor & $\omega_\mathrm{p}(x) = \dfrac{\sqrt{\lambda_\mathrm{p}|\lambda_\mathrm{us}|}}{2\pi\eta a_0^3}$ 
   & $\dfrac{v_p}{\xi} (\kappa - 1)^{1/4} \sqrt{\delta x_{\scriptscriptstyle +}/\xi},\qquad x\to x_{\scriptscriptstyle +}$
      \vspace*{5pt}\newline $ \dfrac{v_p}{\xi}(\kappa-1)$,\qquad $x\approx x_0$ & $\dfrac{v_p}{\xi} 
           \sqrt{\delta x_{\scriptscriptstyle {\scriptscriptstyle +}}/\kappa\xi}$,\qquad $x\to x_{\scriptscriptstyle +}$\vspace*{5pt}
                \newline $\Bigl(\dfrac{x}{\kappa\xi}\Bigr)^{(n+2)/2(n+1)}$, $x \ll x_{\scriptscriptstyle +}$\\ 
frequency factor & $\omega_\mathrm{f}(x) = \dfrac{\sqrt{\lambda_\mathrm{f}|\lambda_\mathrm{us}|}}{2\pi\eta a_0^3}$ 
    & $\dfrac{v_p}{\xi} (\kappa - 1)^{1/4} \sqrt{\delta x_{\scriptscriptstyle -}/\xi}$,\qquad $x\to x_{\scriptscriptstyle -}$\vspace{5pt}
       \newline $\dfrac{v_p}{\xi}(\kappa-1)$,\qquad $x \approx x_0$& $\dfrac{v_p}{\kappa^{\nu/2}\xi} 
            \sqrt{\delta x_{\scriptscriptstyle {\scriptscriptstyle +}}/\kappa\xi}$, $x\to x_{\scriptscriptstyle -}$,\vspace*{5pt}
                 \newline $\Bigl(\dfrac{x}{\kappa\xi}\Bigr)^{(n+2)/2(n+1)}\kappa^{-1/2}$, $x\gg x_{\scriptscriptstyle -}$\\
Labusch parameter & $\kappa_{\scriptscriptstyle \pm} = \dfrac{\xi|f_p''(r_{\scriptscriptstyle \pm})|}{\bar{C}}$ 
    & $\sqrt{\kappa-1}$ & $\kappa_{\scriptscriptstyle +} \sim \kappa$, $\kappa_{\scriptscriptstyle -}\sim \kappa^{-1/(n+2)}$\\
depinning barrier & $U_\mathrm{dp}(x) = e_\mathrm{pin}^\mathrm{us}(x)-e_\mathrm{pin}^\mathrm{p}(x)$ & $e_p(\kappa-1)^2\Bigl[\dfrac{\delta x_{\scriptscriptstyle +}}
   {(\kappa-1)^{3/2}}\Bigr]^{3/2}$, $x\to x_{\scriptscriptstyle +}$\vspace{5pt}\newline$e_p(\kappa-1)^2$, $x=x_0$ 
       & $e_p\Bigl(\dfrac{\delta x_{\scriptscriptstyle +}}{\kappa\xi}\Bigr)^{3/2}$, $x\to x_{\scriptscriptstyle +}$,\vspace{5pt}
           \newline $e_p[1-\mathcal{O}(\kappa^{n/2(n+1)})]$, $x = x_0$\\
pinning barrier & $U_\mathrm{p}(x) = e_\mathrm{pin}^\mathrm{us}(x)-e_\mathrm{pin}^\mathrm{f}(x)$ & $e_p(\kappa-1)^2\Bigl[\dfrac{\delta x_{\scriptscriptstyle -}}
    {(\kappa-1)^{3/2}}\Bigr]^{3/2}$, $x\to x_{\scriptscriptstyle -}$\vspace{5pt}\newline$e_p(\kappa-1)^2$, $x=x_0$ 
         & $e_p\Bigl(\dfrac{\delta x_{\scriptscriptstyle +}}{\kappa\xi}\Bigr)^{3/2}\kappa^{(n+3)/2(n+2)}$, $x\to x_{\scriptscriptstyle -}$,\vspace{5pt}
             \newline $e_p[1-\mathcal{O}(\kappa^{n/2(n+1)})]$, $x = x_0$\\
thermal velocity & $v_\mathrm{th} = \left.\dfrac{\omega_p T}{U_\mathrm{dp}'}\right|_{x_{\scriptscriptstyle +}}$ 
    & $\dfrac{T}{e_p}(\kappa-1)^{1/4}v_p$ & $\dfrac{T}{e_p}\kappa v_p$\\
scaling factor & $\varphi(\kappa)$ [Eq.~\eqref{eq:F_pin_interm_v}] 
   & $(\kappa-1)^{-2}$ & $\kappa^{-\nu'}$, $\nu' = \dfrac{3n+4}{2(n+1)(n+2)}$\\
scaling factor & $g(\kappa)$ [Eq.~\eqref{eq:g}] & $\mathcal{O}(1)$ & $(\kappa-1)^{-4/3}$\\
scaling factor & $a(\kappa) = \dfrac{\kappa_{\scriptscriptstyle +}}{4\pi}\dfrac{\xi}{x_{\scriptscriptstyle -}}\dfrac{e_p}{\Delta e_c}$ 
   & $(\kappa-1)^{-3/2}$ & $\kappa^{-1/(n+2)}$\\
scaling factor & $h(\kappa)$ [Eq.~\eqref{eq:h}] & $(\kappa-1)^{-1/2}$ & $\kappa^{(n+2)/4(n+1)}$\\
\end{tabular}
\label{table:omega}
\end{table}

\end{widetext}

\newpage
\bibliography{bibliography_thermal_creep}

\end{document}